\documentclass[12pt]{article}
\usepackage{mathtools}
    \allowdisplaybreaks
\usepackage{amssymb}
\usepackage{bm}
\usepackage{physics}
\usepackage[mathscr]{euscript}
\usepackage{multirow}
\usepackage{enumitem}
\usepackage{xcolor}
\usepackage[colorlinks,citecolor=blue,linkcolor=blue,urlcolor=blue,pagebackref]{hyperref}
\usepackage{amsthm}
    \theoremstyle{plain} %
        \newtheorem{theorem}{Theorem}
        \newtheorem{corollary}{Corollary}
        \newtheorem{lemma}{Lemma}
        \newtheorem{proposition}{Proposition}
    \theoremstyle{definition}
        \newtheorem{assumption}{Assumption}
        \newtheorem{definition}{Definition}
        
    \theoremstyle{remark}
        
        \newtheorem{remark}{Remark}
\usepackage[margin=1in]{geometry}
\usepackage{booktabs}
\usepackage{microtype}
\usepackage{setspace}
\usepackage[ruled]{algorithm2e}
\usepackage[labelfont=bf, font=small]{caption}
\usepackage{natbib}
    \bibpunct[, ]{(}{)}{,}{a}{}{,}%

\DeclareMathOperator{\E}{\mathbb{E}}
\DeclareMathOperator{\Var}{Var}
\DeclareMathOperator{\Cov}{Cov}

\DeclareMathOperator{\ind}{\mathbb{I}}

\DeclareMathOperator*{\argmax}{argmax}
\def\Real{\mathbb{R}}

\def\BFd{\boldsymbol{d}}

\def\BFG{\boldsymbol{G}}
\def\BFg{\boldsymbol{g}}

\def\BFI{\boldsymbol{I}}

\def\BFM{\boldsymbol{M}}

\def\BFN{\boldsymbol{N}}

\def\BFR{\boldsymbol{R}}

\def\BFu{\boldsymbol{u}}
\def\BFV{\boldsymbol{V}}
\def\BFv{\boldsymbol{v}}

\def\BFx{\boldsymbol{x}}

\def\BFZ{\boldsymbol{Z}}
\def\BFz{\boldsymbol{z}}

\def\BFzero{\boldsymbol{0}}

\def\BFalpha{\boldsymbol{\alpha}}

\def\BFdelta{\boldsymbol{\delta}}

\def\BFvarepsilon{\boldsymbol{\varepsilon}}

\def\BFSigma{\boldsymbol{\Sigma}}

\def\BFOmega{\boldsymbol{\Omega}}

\def\calA{\mathcal{A}}

\def\calH{\mathcal{H}}

\def\calN{\mathcal{N}}

\def\calP{\mathcal{P}}

\def\scrA{\mathscr{A}}
\def\scrB{\mathscr{B}}
\def\scrC{\mathscr{C}}
\def\scrD{\mathscr{D}}
\def\scrE{\mathscr{E}}
\def\scrF{\mathscr{F}}
\def\scrG{\mathscr{G}}

\def\scrS{\mathscr{S}}

\def\SFc{\mathsf{c}}

\def\SFp{\mathsf{p}}

\def\SFs{\mathsf{s}}

\def\Real{\mathbb{R}}
\def\ttR{\mathtt{T}_1}
\def\ttU{\mathtt{T}_2}

\def\Reg{\mathsf{Regret}}

\def\calA{\mathcal{A}}
\def\calH{\mathcal{H}}
\def\calN{\mathcal{N}}
\def\calP{\mathcal{P}}
\def\scrA{\mathscr{A}}
\def\scrB{\mathscr{B}}
\def\scrC{\mathscr{C}}
\def\scrD{\mathscr{D}}
\def\scrE{\mathscr{E}}
\def\scrF{\mathscr{F}}
\def\scrG{\mathscr{G}}

\def\scrS{\mathscr{S}}

\def\SFc{\mathsf{c}}
\def\SFp{\mathsf{p}}
\def\SFs{\mathsf{s}}

\def\Cbern{\mathsf{C_{Bern}}}

\begin{document}

\begin{titlepage}
\title{Dynamic Selection in Algorithmic Decision-making\thanks{Contact information: Jin Li (\url{jli1@hku.hk}), Ye Luo (\url{kurtluo@hku.hk}), Faculty of Business and Economics, The University of Hong Kong, Pokfulam Road, Hong Kong SAR.
Xiaowei Zhang (\url{xiaoweiz@ust.hk}),
Department of Industrial Engineering and Decision Analytics, The Hong Kong University of Science and Technology, Clear Water Bay, Hong Kong SAR.
We are grateful for helpful conversations with Chunrong Ai, Xi Chen, Pingyang Gao, Bob Gibbons, John Klopfer, Danielle Li, Whitney Newey, Michael Song, Wing Suen, Chang Sun, Brian Viard, and seminar participants at Causal Data Science Meeting 2021, University of Hong Kong, UC Riverside, and CUHK Workshop: ``Advances in Econometrics, Machine Learning, and Big Data.'' All remaining errors are ours.}}
\author{Jin Li \and Ye Luo \and Xiaowei Zhang}
\date{}
\maketitle
\begin{abstract}
This paper identifies and addresses dynamic selection problems in online learning algorithms with endogenous data. In a contextual multi-armed bandit model, a novel bias (\emph{self-fulfilling bias}) arises because the endogeneity of the data influences the choices of decisions, affecting the distribution of future data to be collected and analyzed. We propose an instrumental-variable-based algorithm to correct for the bias. It obtains true parameter values and attains low (logarithmic-like) regret levels.
We also prove a central limit theorem for statistical inference. To establish the theoretical properties, we develop a general technique that untangles the interdependence between data and actions.



\end{abstract}
\setcounter{page}{0}
\thispagestyle{empty}
\end{titlepage}
\pagebreak \newpage

\section{Introduction}

This paper studies the dynamic selection problem in algorithmic decision-making. The use of algorithms has become increasingly popular among business firms. The algorithms analyze data continuously and allow firms to adjust their decisions in response to new information constantly. This feature, known as \emph{online} learning, implies that algorithms are both consumers and producers of data. They take data as inputs and analyze them to make decisions. They also \emph{produce data} by choosing which product---movies, songs, or articles---to recommend.

Because an online algorithm uses prior data analysis to generate data, the data it generates are not random. As a result, the data the algorithm uses for its analysis is also non-random. The non-randomness of the data naturally leads to a data selection problem. Importantly, this selection problem is dynamic due to the online-learning nature of algorithms: past selection problems affect the data the algorithm generates, creating new data selection problems.

To demonstrate and address this dynamic selection problem, we focus on a class of algorithms known as contextual multi-armed bandits (MAB).
Contextual bandit algorithms are commonly used across various domains due to their ability to balance exploration and exploitation while incorporating contextual information; see, for example, applications in product recommendation \citep{LiChuLangfordSchapire10}, online advertising \citep{TangRosalesSinghAgarwal13}, and mobile healthcare \citep{TewariMurphy17}. News outlets such as the New York Times utilize contextual bandit algorithms to learn about their readers' behavior and recommend articles tailored to their interests more effectively \citep{Coenen19}.

To illustrate how a contextual bandit algorithm works, consider a website that recommends news articles to its users. The website aims to maximize user engagement, measured by the number of articles read or the time users spend reading articles. The website can base its recommendation on the information it has on the users, known as the \emph{contextual information}, which may include features like user demographics (age, gender, location) and user behavior (browsing history, time of the visit).

When a user visits the website, the contextual bandit algorithm gathers information about the user. Using this context and its current knowledge, it selects a news article to recommend, also known as which arm to pull, by using a model (such as linear regression) to predict user engagement. After the recommendation, the algorithm observes a reward, such as whether the user read the article or the time spent reading. The algorithm then updates its knowledge based on the context, chosen arm, and reward to improve its future recommendations.

A potential issue with the contextual bandit algorithm is that its knowledge-updating process may be impacted by an endogeneity problem. For instance, the news website might estimate the effectiveness of recommending sports articles by regressing their click rates on the reader's age. Click rates depend on both the reader's age and their free time available. But the website only observes the reader's age and not the free time. If the reader's free time is correlated with his age, an omitted-variable problem arises, leading to a biased estimate of the effectiveness of sports articles.

The omitted-variable problem causes a dynamic selection problem when the size of the omitted-variable bias varies among subsamples of data. In this case, the estimated effectiveness of sports articles depends on which subsamples are used for the regression. Because the algorithm recommends the article based on its estimated effectiveness, this implies that different past subsamples can lead to different recommendation rules. The past subsample therefore affects the new data generated, creating the dynamic selection problem.

Because of the dynamic selection problem, the decision rule (i.e., policy) in the long run under the contextual bandit algorithm is a fixed point: the optimal policy derived from analyzing the subsample must coincide with the long-run policy that generates the subsample. For example, suppose that the algorithm recommends the article to readers aged 60 and above. In this case, the regression analysis performed on the subsample of readers aged 60 and above must indicate that the optimal policy involves recommending the article to readers aged 60 and above.

The fixed-point property of the long-run policy leads to two problems. First, the \emph{oracle} policy---that is, the true optimal policy had the effectiveness of each arm been known in advance---may not be a fixed point. This implies that the long-run policy is suboptimal. Second, and, relatedly, because the past policys affect the choice of future policys, the long-run policy may not be unique, implying that there are multiple fixed points.

Our first result shows that the non-uniqueness problem can be severe. Depending on the distribution of the omitted variable, there can be an arbitrarily large number of fixed points. Furthermore, these fixed points can be arbitrarily away from the optimal policy induced by the true parameter of interest. We refer to the difference between true parameter and its long-run estimate as the \emph{self-fulfilling bias}. Our first result states that the self-fulling bias can be arbitrarily large and can take arbitrarily many values.

We then propose algorithms that correct for the self-fulfilling bias and generate actions with low levels of regret (loss in payoff compared to optimal actions).
Our method integrates the instrumental variable (IV) approach into a greedy-type online learning algorithm designed for contextual bandit problems.
Our proposed algorithms consist of three phases.
In Phase~1, arms are randomly selected.
At the end of Phase~1, the two-stage least squares (2SLS) method is implemented on the data associated with each arm to compute the arm-specific coefficient estimates. In Phase~2, the arm with the highest expected reward (using the coefficient estimates in Phase~1) is selected in each period. The covariance matrix of the estimates are calculated at the end of Phase~2. In Phase~3, arms are selected in a greedy fashion and the coefficients and covariance matrices are updated using ``joint-2SLS'' in each period.
Under this algorithm, the coefficients converge to the true value in the long run.
The regret of our algorithm is of order $\mathcal{O}(\log(T))$.
In particular, the two special features of the algorithm---the existence of Phase~2 and joint-2SLS in Phase~3---are critical. Without either of the two, the coefficient estimates can fail to converge to the true value.

Our third contribution is to develop a technique that facilitates the theoretical analysis of online learning algorithms with dynamic selection problems.
As mentioned above, a key feature of online learning algorithms is that data and actions are intertwined. This complicates the analysis because the effects of past actions can be long-lasting.
Mathematically, in our proposed algorithm and in online learning algorithms in general, this complication manifests in a two-way dependence between (i) the estimates of the coefficients and (ii) the estimates of the covariance matrix between the IVs and an augmented vector of the covariates.
On the one hand, in 2SLS, the coefficient estimates depend on the covariance estimates.
On the other hand, each covariance estimate depends on \emph{all} the past
coefficient estimates because they affect the past actions and the past realization of data. This latter dependence is new and central to the analysis of online learning algorithms with dynamic selection problems. Moreover, the dependence is complicated both because the coefficient estimates are serially correlated and because they affect the estimates of the covariance matrix in a nonlinear manner.

Our technique deals with this two-way dependence by performing an induction in a zig-zag matter.
In each induction step, we first derive the properties of the covariance estimates (zig), and then use this property to establish the induction hypothesis on coefficient estimates (zag). The heart of the induction is the zig-part, which employs a special technique to eliminate the dependence of the covariate estimates on all past coefficient estimates. The theoretical analysis of this dependence also indicates that, for the coefficient estimates to converge to the true value, the covariance estimates  at the beginning of Phase 3 must be sufficiently close to the covariance matrix induced by the oracle policy. This motivates Phase 2 in our algorithm, which stabilizes the covariance estimates.

Finally, this technique establishes the rate of convergence of the estimates.
These rates are used to analyze the regret of the algorithm.
They are also used to establish the asymptotic distribution of the parameter estimators at the terminal period of the algorithm and, therefore, allow for the construction of confidence intervals and hypothesis testing.

Our paper contributes to the literature on dynamic learning.
Within this literature, a key emphasis has been the trade-off between exploration and exploitation \citep{March91}.
The workhorse for analyzing this trade-off is the multi-armed bandit models.
Economists have extensively studied their theoretical and empirical implications  (see \citealt{BergemannValimaki08} for a review).
Recently, there is a growing sub-literature that studies algorithmic decision-making based on multi-armed bandit models
(\citealt{PerchetRigolletChassangSnowberg16,Caria_etal20,CurrieMacLeod20,KasySautmann21,Kawaguchi21}, \citealt{KawaguchiUetakeWatanabe21})\footnote{While we focus on online learning environments, there is a rich line of work that studies offline learning of optimal policies \emph{after} the experimental or observational data is collected (\citealt{KitagawaTetenov18,NaritaYasuiYata19,AtheyWager21,KallusZhou21}, \citealt{ShiWanChernozhukovSong21}, \citealt{ZhanHadadHirshbergAthey21}, \citealt{ChenAndrews23}).}.
The closest one to our study is that of \citet{LiRaymondBergman20}, which examines application of contextual bandit algorithms for making decisions in the hiring process.
We add to this sub-literature by both studying the biases that arise in algorithmic decision-making when there is an endogeneity problem and showing how to correct for it.

Our paper also contributes to the vast literature of casual inference; see, for example, \cite{ImbensRubin15} for a review.
Most of this literature does not examine the environment of online learning, where the actions influence and are influenced by data. Two notable exceptions are \citet{NambiarSimchiWang19} and  \citet{LiLuoZhang21_fullversion}.
\citet{NambiarSimchiWang19} consider a dynamic pricing problem with learning, where the endogeneity problem arises from mis-specification of demand functions.
\citet{LiLuoZhang21_fullversion} consider a Markov
decision process setup with learning and show that online learning environments may exacerbate the estimation bias.
Different from these papers, our focus is on the dynamic selection problem where current data, through affecting the choice of actions, affects how future data is generated. This dynamic selection problem also differs from selection problems in \cite{Heckman79,Heckman90}, \cite{rubinpropensity}, \citet{Altonji2005Selection00},  and \cite{Oster19}, which do not feature how past selection problems affect future data selection.

The remainder of the paper is organized in the following manner. Section \ref{sec:formulation} introduces the background of the contextual bandit problem. Section \ref{sec:double-endog} discusses the case where there exists at least one endogenous variable in the contextual bandit problem, which leads to the dynamic selection problem and self-fulfilling bias. Section \ref{sec:IV-UCB} presents a set of regularity conditions for IVs and the contextual bandit setup as well as proposes a class of IV-based bandit algorithms. Section \ref{sec:theo} establishes the main results including regret bounds and asymptotic distribution for the proposed algorithms. Section  \ref{sec:simulation} demonstrates the performance of these algorithms in a series of simulations. Section \ref{sec:conclusions} presents the conclusions. Technical proofs are included in the Appendix.

\section{Background}\label{sec:formulation}

Consider an agent who follows an algorithm to make a sequence of decisions over a time horizon of $T$ periods.
At each time $t=1,\ldots,T$, the agent observes a vector of covariates
$\BFv_t\in\Real^p$,
takes an action (i.e., pulls an arm) $a_t\in \{1,\ldots,K\}$ with $K\geq 2$,
and receives a \emph{random} reward $R_t$:
\begin{equation}\label{eq:rand-reward}
    R_t =  \sum_{i=1}^K \ind(a_t=i) \mu_i(\BFv_t) +\varepsilon_t,
\end{equation}
where $\mu_i(\BFv)$ is the \emph{unknown} reward function associated with arm $i$, and $\varepsilon_t$ is the noise.
At each time point $t$, the algorithm specifies a policy that maps $\BFv_t$ to $a_t$, based on the historical data $\calH_{t-1}\coloneqq (\BFv_1,a_1,R_1,\ldots,\BFv_{t-1},a_{t-1},R_{t-1})$.
The goal is to maximize the expected cumulative reward $\E[\sum_{t=1}^T \mu_{a_t}(\BFv_t)]$, where the expectation is taken with respect to the distribution of $\calH_T$ induced by the particular algorithm.

This sequential decision-making problem is commonly known as the contextual bandit problem \citep[chapter 8]{Slivkins19}.
Evidently,
if each reward function $\mu_i$ is known in advance,
the agent would employ the \emph{oracle policy} at each time $t$, pulling the optimal arm associated with $\BFv_t$, that is, $\pi^*(\BFv_t)\coloneqq \argmax_{ i =  1,\ldots,K}\mu_i(\BFv_t)$.
In practice, however, $\mu_i$ is unknown and must be estimated from data.
Hence, the arm that is actually pulled at each time $t$ is in general not the optimal arm,  leading to an
\emph{instantaneous regret}: $\max_{i=1,\ldots,K} \mu_i(\BFv_t) - \mu_{a_t}(\BFv_t)$.
A standard metric for measuring the performance of an algorithm is \emph{expected cumulative regret}:
\begin{equation}\label{eq:regret-def}
\Reg(T) \coloneqq \sum_{t=1}^T \E\biggl[\max_{i=1,\ldots,K}\mu_i(\BFv_t) - \mu_{a_t}(\BFv_t)\biggr].
\end{equation}
We simply refer to it as ``regret'' for the remainder of this paper.
The regret quantifies the difference in expected cumulative reward between an algorithm and the oracle policy.
As such, the agent's goal is to approximate the performance of the oracle by progressively learning the reward functions.
For simplicity, we will henceforth assume that for each arm $i$, the reward function $\mu_i$ is linear in the covariate vector (i.e., $\mu_i(\BFv) = \BFv^\intercal \BFalpha_i$), where $\BFalpha_i\in\Real^p$ is an unknown parameter vector. Consequently,
the problem is reduced to learning these linear coefficients over time.

When the covariates are exogenous, that is, $\mathbb{E}[\varepsilon_t|\BFv_t]=0$, numerous effective algorithms have been proposed to learn the coefficients.
While these algorithms vary in their finite-time and asymptotic properties, they all converge to the true value of the coefficients over time and, therefore, exhibit a \emph{sublinear} regret: that is,  $T^{-1}\Reg(T)\to 0$ as $T\to\infty$.
This indicates that as the horizon increases, the agent is increasingly likely to take the optimal action in most periods.

\section{Self-fulfilling Bias: An Example}\label{sec:double-endog}

We now study an example in which the covariates are endogenous---that is, $\mathbb{E}[\varepsilon_t|\BFv_t]\neq 0$.
We show that the endogeneity of covariates leads to distortion in actions, which generates another type of endogeneity; that is, it creates ``endogeneity spillover''. The additional endogeneity generates a dynamic selection problem and leads to a new type of bias, which we refer to as self-fulfilling bias.

Consider the following special case of our setup.
Assume
that there exist two arms, and between the two arms $\calA=\{1,2\}$,
one is the safe arm, having a \emph{known} expected reward that is independent of the covariates,
while the other is the risky arm, whose expected reward has an unknown linear dependence on the covariates.
Further, assume that the covariates are one-dimensional.
In this simple example, the reward functions are
\[
\mu_1(v) \equiv c \qq{and} \mu_2(v) = \alpha v,\quad \forall v\in\Real.
\]
It follows that, if $\alpha$ is known and positive, then the oracle policy for the agent is to pull arm~2 if $v> c/\alpha$,
and pull arm~1 otherwise.

However, because $\alpha$ is unknown, the agent needs to  estimate its value over time and makes her decisions accordingly.
In contrast to the standard regression analysis in which data is taken as given, in the dynamic environment considered here, the data used for estimating $\alpha$ is only available when arm~2 is pulled
(pulling arm~1 generates noisy observations of $c$, which are irrelevant to $\alpha$).
Now, suppose the agent uses the ordinary least squares (OLS) method to estimate $\alpha$
and the estimates converge to a limit, say $\widehat \alpha$.
Then, in the long run,
\begin{equation}\label{eq:OLS-est}
\widehat\alpha = \frac{\Cov[v_t, R_t|R_t\mbox{ is generated from arm~2}]}{\Var[v_t|R_t\mbox{ is generated from arm~2}]}.
\end{equation}

Note that given $\widehat \alpha$, the agent's policy  in the long run is to pull arm~2 when $v>c/\widehat \alpha$.
Thus, the conditioning event in equation \eqref{eq:OLS-est} is identical to $\{v_t > c/\widehat\alpha\}$. Using the fact that $R_t = \alpha v_t + \varepsilon_t$ for arm~2, we have
\begin{equation}\label{eq:bias}
\widehat\alpha = \alpha + \frac{\Cov[v_t, \varepsilon_t|v_t > c/\widehat\alpha]}{\Var[v_t|v_t > c/\widehat\alpha]}.
\end{equation}

It is evident from equation \eqref{eq:bias} that $\widehat{\alpha}$ is a fixed point as it appears on both sides of the equation.
It also shows that the bias of the estimate, $\widehat\alpha-\alpha$, has an expression that is similar to, but different from, that of the usual OLS bias, which is given by $ \frac{\Cov[v_t, \varepsilon_t]}{\Var[v_t]}$.
We refer to $\widehat \alpha - \alpha$ as the \emph{self-fulfilling bias} to reflect that the limit policy of the agent is induced by
her limit belief (i.e., the limit estimate $\widehat{\alpha}$),
and the limit belief is confirmed by the data generated from the limit policy.

To unpack the self-fulfilling bias, note that  there are two---rather than one---types of endogeneity problems. The first type results from the correlation between the noise and the covariates. This directly creates a bias in the estimate in a manner that is similar to the usual OLS bias. The second type results from sample selection, which arises because the estimated coefficients affect the actions of the agents. This sample selection is dynamic in nature because the policy that determines agent's actions changes over time.

To illustrate the two types of endogeneity formally, we rewrite equation \eqref{eq:rand-reward} as
\begin{equation}\label{eq:regression-double-endo}
R_t  =  c + \alpha v_t \ind(a_t = 2) - c\ind(a_t = 2)  +  \varepsilon_t.
\end{equation}
When $\varepsilon_t$ and $v_t$ are correlated, the first type of endogenous variable in this regression is $ v_t \ind(a_t = 2)$ if the arm $a_t$ is selected \emph{independent} of $v_t$. But because $a_t$ depends on $v_t$---through the relation $\ind(a_t=2) = \ind(v_t > c/\widehat\alpha)$ in the long run---the correlation between $\varepsilon_t$ and $v_t$ leads to a correlation between $\ind(a_t=2)$ and $\varepsilon_t$. In other words, the endogeneity problem in $v_t$ spills over to $\ind(a_t = 2)$, creating another endogenous variable.

The additional endogenous variable complicates the dependence of the value of the limit coefficient estimate $\widehat\alpha$ on the joint distribution of $v_t$ and $\varepsilon_t$. In particular, unless
the term
$\frac{\Cov[v_t, \varepsilon_t|v_t > s]}{\Var[v_t|v_t>s]}$ is the same for all cutoff $s$---this happens only when $\mathbb{E}[\varepsilon_t|v_t]$ is a linear function of $v_t$\footnote{$\mathbb{E}[\varepsilon_t|v_t]$ is a linear function of $v_t$ when $(v_t,\varepsilon_t)$ has a joint normal distribution.}---$\widehat\alpha$ is different from the OLS estimate. In other words, an additional bias arises whenever $\mathbb{E}[\varepsilon_t|v_t]$ is a nonlinear function of $v_t$.

\begin{figure}[ht]
    \begin{center}
    \caption{Multiplicity of the Self-fulfilling Bias \label{fig:equilibria}}
    \includegraphics[height=0.34\textwidth]{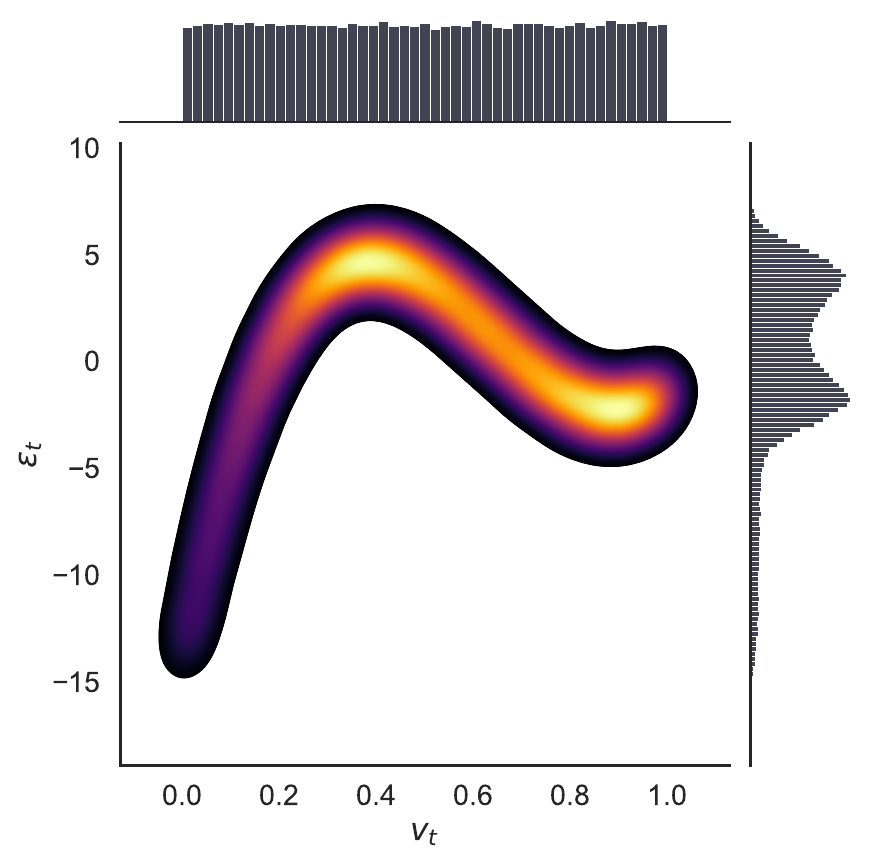}
    \includegraphics[height=0.34\textwidth]{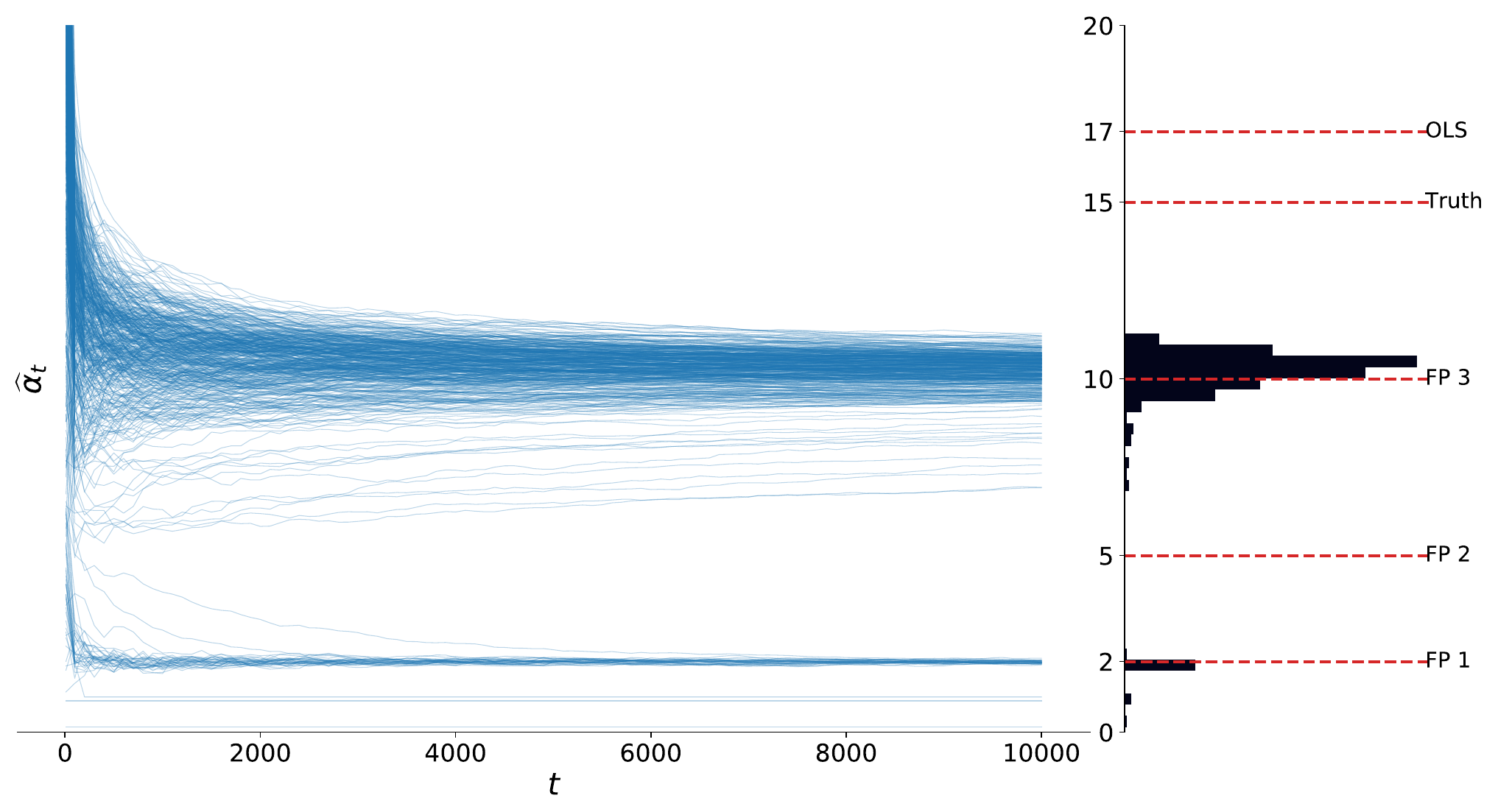}
    \end{center}
\footnotesize{\emph{Note.} Suppose that
$c=1$, $\alpha=15$,
$v_t$ is uniformly distributed on $[0,1]$, and
$\varepsilon_t = \sum_{k=0}^3 \beta_k v_t^k + \eta_t$, where
$\beta_k$'s are some constants and
$\eta_t$ is an independent standard normal random variable.
Then, (i) the limit of the OLS estimate is 17 and (ii) equation~\eqref{eq:bias} is reduced to a cubic equation in terms of $\widehat{\alpha}$, thereby
yielding three distinct real fixed points (FPs)---2, 5, and 10---if $\beta_k$'s are appropriately selected (see the Appendix for details).
The left plot shows the joint distribution of $(v_t, \varepsilon_t)$.
The middle plot shows 500 sample paths of estimates of $\alpha$ following a greedy policy.
The right plot shows the histogram of the terminal values of these sample paths, indicating that the greedy estimates mostly converge to either 2 or 10.
}
\end{figure}

When there is the additional bias, the value of $\widehat{\alpha}$ (and hence the value of the self-fulling bias) may not be unique, which happens when equation~\eqref{eq:bias} contains multiple solutions. Figure~\ref{fig:equilibria} illustrates one such case and shows that the agents' beliefs converge to two different values in the long run. With multiple self-fulfilling biases, this example shows that agents can adopt different policies in the long run, even if the underlying production environment is the same. This provides one explanation for why seemingly similar enterprises may choose different practices and result in persistent differences in performance, a central question in organizational economics and strategy
(see, e.g., \citealt{GibbonsHenderson12} for a review).

\begin{remark} \label{multiplicity}
There are few restrictions on the relationship between the true value $\alpha$ and the agent's long-run belief. For a given $\alpha$, the beliefs regarding $\alpha$ may converge in the long run to \emph{any} number of \emph{arbitrarily chosen} values in the long run. In other words, for a group of agents with the same initial beliefs, they can hold many very different beliefs in the long run. We formalize this in the following proposition, and the proof is given in the Appendix.
\end{remark}

\begin{proposition}\label{prop:multiplicity}
Let $v_t$ be a uniform random variable on $[0,1]$ and
$\eta_t$ be an independent standard normal random variable.
Then, for any integer $n\geq 1$ and real numbers $\{r_k:k=1,\ldots,n\}$,
there exist real numbers $\{\beta_k:k=0,1,\ldots,n\}$ such that
$\{r_k:k=1,\ldots,n\}$ are the roots of Equation~\eqref{eq:bias}, provided that
$\varepsilon_t = \sum_{k=0}^n \beta_k v_t^k + \eta_t$.
\end{proposition}

We conclude this section by making two comments. First, we have illustrated the self-fulfilling bias and its severity using a simple example. Despite its simplicity, the example points to a general problem: in any dynamic environment where the agent's actions depend on some covariates, the endogeneity in the covariates can generate additional endogeneity with the actions. Because of the self-fulfilling bias, the policy is suboptimal in the long run. The number of periods a wrong action is taken is proportional to the time horizon $T$. Thus, the regret of the policy with self-fulfilling bias is in the order of $\mathcal{O}(T)$.

Second, the self-fulfilling bias cannot be corrected by the usual exploration of the agent, i.e., randomly choosing actions occasionally. This is because as long as $v_t$ is endogenous, OLS bias remains. Moreover, when the data from the non-random actions is used in the regression, the endogeneity in $v_t$ spills over to $\ind(a_t=2)v_t$, creating additional bias.
The solution for the endogeneity in $v_t$ is not to carry out exploration in the conventional sense.
Instead, the solution is to do  \emph{causal exploration}---that is, to perturb the data generating process, and, through doing so, generating IVs accordingly. Unlike the randomization of actions in the literature (conventional exploration), which we denote as \emph{ex-post} exploration, the generation of IVs through perturbation of data generating process is ex-ante to decision making, thus we call it \emph{ex-ante} exploration. We discuss how to correct the self-fulfilling bias with IVs in the next section.

\section{IV-Greedy Algorithm}\label{sec:IV-UCB}

We now proceed to illustrate how to integrate IVs into bandit algorithms.
The algorithm we propose not only corrects the self-fulfilling bias, thereby providing consistent estimation of the reward functions and progressively improving a policy $\pi(\cdot | v)$ towards the oracle policy over time, but also attains a sublinear regret of order $\mathcal{O}(\log(T))$.
Simultaneously addressing the two concerns---endogeneity and regret---is crucial for online learning environments.

As illustrated in the example above, we can rewrite observed rewards in Equation~\eqref{eq:rand-reward} as:
\begin{align}
R_t =  \widetilde{\BFv}_t^\intercal \BFalpha +\varepsilon_t,\quad \mbox{for all } t=1,\ldots,T,\label{eq:main2}
\end{align}
where
\begin{equation}\label{eq:augmented-covariates}
\BFalpha \coloneqq
\begin{pmatrix}
\BFalpha_1 \\
\vdots \\
\BFalpha_K
\end{pmatrix}\in\Real^{Kp}
\qq{and}
\widetilde{\BFv}_{t}\coloneqq
\begin{pmatrix}
\ind(a_t=1)\BFv_t  \\
\vdots\\
\ind(a_t=K)\BFv_t
\end{pmatrix}\in\Real^{Kp}.
\end{equation}
Equation~\eqref{eq:main2} represents a \emph{joint} linear regression model in that it integrates reward observations across \emph{all} time periods, as opposed to partitioning them based on the selected arm.
The use of $\widetilde{\BFv}_t$ as the regressors in formulation \eqref{eq:main2} underscores the endogeneity spillover issue discussed in Section~\ref{sec:double-endog}. Specifically, the action $a_t$ may become endogenous due to its potential dependence on $\BFv_t$ and the endogeneity of $\BFv_t$, hence making $\ind(a_t=i)\BFv_t$ endogenous for each $i$.
As a result, the OLS method leads to a biased estimation of the coefficient vector $\BFalpha$.

Assuming a set of IVs, as denoted by $z_t$, are available,
we can use the 2SLS method to correct for this bias, as is typically done in a static data analysis environment.
However, our online learning environment introduces substantial challenges due to the intertemporal dependencies among the regressors $\{\widetilde{\BFv}_t : t = 1, \ldots, T\}$, which are a result of the adaptive nature intrinsic to any effective algorithm.
To address the challenges, we propose to employ a \emph{joint-2SLS} approach, as defined later in Equation~\eqref{eq:joint-2SLS}, to estimate $\BFalpha$ using the joint linear regression model \eqref{eq:augmented-covariates},
rather than estimating for each arm separately.

We introduce our algorithm in Section~\ref{sec:algo-detail}.
The algorithm's two key features---the joint-2SLS and three-phase structure---are discussed in detail in Section~\ref{sec:joint-2sls} and Section~\ref{sec:three-phases}, respectively.

\subsection{Structure}\label{sec:algo-detail}

In this subsection, we propose an IV-based algorithm for linear contextual bandits and name it \emph{IV-Greedy}.
This algorithm consists of three phases, which are formally demonstrated in Algorithm \ref{algo:IV-bandits}.

\textbf{Phase 1: Coefficient Stabilization.}
In this phase, we seek a stable estimate of the coefficient vector $\BFalpha$,
meaning that the estimate should stay close to the true value with a high probability.
For this purpose,
our policy is to select each arm randomly---irrespective of the observed covariates---with equal probability.
At the end of Phase~1, we run \emph{arm-specific-2SLS} to obtain an estimate of the coefficient vector $\BFalpha$.
Specifically,
depending on the arm pulled in each period---whether $a_t=1$, $2$, etc.---we divide the periods into $K$ groups and then conduct 2SLS separately on the data (i.e., rewards, covariates, and IVs) in each group.
Note that the estimate is unbiased because the actions are  generated randomly, and the IVs are used to correct for the endogeneity in the covariates.
But the absence of bias is achieved
at the cost of a regret that is linear in the number of periods $\ttR$.
To balance the regret and the purpose of coefficient stabilization, we choose the duration of Phase~1 to be $\ttR = \mathcal{O}(\log(T))$.

\textbf{Phase 2: Covariance Stabilization.}
In this phase, the policy takes the form of a greedy policy---that is, it uses the estimate of $\BFalpha$ from Phase 1 to evaluate the expected payoff of each arm for a given value of the covariates, and then
selects an arm that maximizes the expected payoff.
During this phase, which lasts from period $\ttR+1$ to $\ttU$, the estimate of $\BFalpha$ is \emph{not} updated.
Fixing the coefficient estimate helps initialize Phase~3 of the algorithm with a good estimate of the covariance matrix $\BFOmega^*$, as defined in Equation~\eqref{eq:cov-matrix}.
As discussed in Section~\ref{sec:joint-2sls},
the presence of Phase~2 is critical for the estimate of $\BFalpha$ to converge to the true value.
Similar to Phase~1, we set the duration of this phase to be $\ttU-\ttR=\mathcal{O}(\log(T))$ so that sufficient data is generated without causing excessive regret.

\textbf{Phase 3: Policy Improvement.}
In this phase, the policy is again greedy: we select the arm that maximizes the estimated expected reward in each period.
However, in contrast to Phase~2, the criterion for evaluating the expected reward is constantly updated.
In each period, the expected reward for each arm is calculated from the most updated estimate of $\BFalpha$.
Note that in this phase, the estimator that we propose to use is the \emph{joint-2SLS} estimator.
Namely, we consider to estimate a 2SLS model jointly for all arms $i=1,\ldots,K$,
This is critical for the algorithm to produce consistent estimator asymptotically. We discuss this point in more detail in Section \ref{sec:joint-2sls}.

\begin{algorithm}[ht]
\caption{IV-Greedy}
\label{algo:IV-bandits}
\SetAlgoLined
\DontPrintSemicolon
\SetKwInOut{Init}{Input}
\SetKwInOut{CoefStab}{Phase 1}
\SetKwInOut{CovStab}{Phase 2}
\SetKwInOut{Impr}{Phase 3}
\Init{Set $\ttR$, $\ttU$, and $T$}
\CoefStab{
\For{$t=1,\ldots,\ttR$}{
Observe $(\BFv_t, \BFz_t)$,
pull arm $a_t=1,\ldots,K$ uniformly at random, and
collect $R_t$\;
}
\For{$i=1,\ldots,K$}{
Run arm-specific-2SLS on the data corresponding to arm $i$ (that is,
$\{(\BFv_t, \BFz_t, R_t):t\leq \ttR, a_t=i\}$)
to compute $\widehat{\BFalpha}_{i,\ttR}$, as defined by Equation~\eqref{eq:arm-spec-2SLS}

}
}
\CovStab{
\For{$t=\ttR+1,\ldots, \ttU$}
{
Observe $(\BFv_t, \BFz_t)$,
pull arm $\displaystyle a_t = \argmax_{i=1,\ldots,K} \bigl\{\BFv_t^\intercal \widehat{\BFalpha}_{i,\ttR} \bigr\}$,
and collect $R_t$\;
}
Set    $\widehat\BFalpha_{i,\ttU} = \widehat\BFalpha_{i,\ttR}$ for all $i=1,\ldots,K$\;
}
\Impr{
\For{$t = \ttU + 1,\ldots, T$}
{
Observe $(\BFv_t, \BFz_t)$,
pull arm
$\displaystyle
a_t=\argmax_{i=1,\ldots,K} \left\{\BFv_t^\intercal\widehat{\BFalpha}_{i,t-1}
\right\} $,
and collect $R_t$\;
Run joint-2SLS on all data collected after Phase~1 (that is, $\{(\widetilde{\BFv}_s, \BFz_s, R_s):\ttR+1 \leq s\leq t\}$)
to compute $\widehat{\BFalpha}_t$, as defined by Equation~\eqref{eq:joint-2SLS}
}
}
\end{algorithm}

\medskip

To describe the arm-specific-2SLS method used in Algorithm~\ref{algo:IV-bandits}, we introduce the following notation.
Let $\BFV_{i,\ttR}$ denote the observed covariates in those time periods of Phase 1 when arm $i$ is selected.
Specifically, $\BFV_{i,\ttR}$ is a matrix formed of row vectors $\BFv_t^\intercal$ for all $t$ such that $t \leq \ttR$ and $a_t = i$.
Similarly, we define $\BFZ_{i,\ttR}$ as the matrix composed of row vectors $\BFz_{i,t}^\intercal$, and $\BFR_{i,\ttR}$ as the column vector composed of elements $R_{i,t}$, both corresponding to the same time periods as those utilized in defining $\BFV_{i,\ttR}$.
The number of rows in these three matrices (or in the case of the vector) corresponds to the number of times that arm $i$ is pulled during Phase~1.
Then, the arm-specific-2SLS estimator for the coefficient vector $\BFalpha_i$ of the reward function associated with arm~$i$ is given by
\begin{equation}\label{eq:arm-spec-2SLS}
    \widehat{\BFalpha}_{i,\ttR} \coloneqq \bigl(\BFV_{i,\ttR}^\intercal   \calP[\BFZ_{i,\ttR}]\BFV_{i,\ttR}\bigr)^{-1}\BFV_{i,\ttR}^\intercal  \calP[\BFZ_{i,\ttR}]\BFR_{i,\ttR},
\end{equation}
where $\calP$ denotes the projection operator, defined as $\calP[\BFZ] \coloneqq \BFZ(\BFZ^\intercal \BFZ)^{-1} \BFZ^\intercal$ for any matrix $\BFZ$.
The formula provided above is derived from standard calculations associated with the 2SLS method.

To define the joint-2SLS estimator used in Phase 3, for any $t \geq \ttR+1$, we introduce $\widetilde{\BFV}_{\ttR:t}$ as the matrix composed of row vectors $\widetilde{\BFv}_s^\intercal$ for all $s = \ttR+1, \ldots, t$,
where $\widetilde{\BFv}_{s}$ is the augmented covariate vector as defined in \eqref{eq:augmented-covariates}.
This matrix has $(t-\ttR)$ rows because we use all data that starts from the beginning of Phase 2.
Likewise, we can define $\BFZ_{\ttR:t}$ and $\BFR_{\ttR:t}$.
These three represent the data accumulated from the start of Phase~2 through time $t$.
The joint-2SLS estimator for the coefficient vector $\BFalpha$ is then given by
\begin{equation}\label{eq:joint-2SLS}
\widehat{\BFalpha}_t \coloneqq  (\widetilde{\BFV}_{\ttR:t}^\intercal \calP[\BFZ_{\ttR:t}]\widetilde{\BFV}_{\ttR:t})^{-1} \widetilde{\BFV}_{\ttR:t}^\intercal \calP[\BFZ_{\ttR:t}] \BFR_{\ttR:t},
\end{equation}

Lastly, the need for inference calls for a characterization of the limiting covariance matrix linked with the joint-2SLS estimate of $\BFalpha$.
Intuitively, if the policy induced by the proposed algorithm converges to the oracle policy, the data will, in the long run, be generated under this oracle policy.
Following standard calculations for the 2SLS method, it becomes evident that the covariance matrix should converge to
\begin{equation}\label{eq:cov-matrix}
\BFOmega^* \coloneqq  \bigl(\BFSigma_{vz}^*\BFSigma_{zz}^{-1}(\BFSigma_{vz}^*)^\intercal \bigr)^{-1},
\end{equation}
where
\[\BFSigma^*_{vz}\coloneqq \mathbb{E}[\BFv_t^*\BFz_t^\intercal],\quad
\BFSigma_{zz}\coloneqq \mathbb{E}[\BFz_t \BFz_t^\intercal], \qq{and}
\BFv_t^*\coloneqq
\begin{pmatrix}
\ind(a_t^*=1)\BFv_t \\
\vdots \\
\ind(a_t^*=K)\BFv_t
\end{pmatrix} \in\Real^{Kp}.\]

\begin{remark}
The IV-Greedy algorithm assumes that the time horizon $T$ is known in advance. However, we can adapt the algorithm by applying the standard ``doubling trick'' from the bandit literature \citep{BessonKaufmann18}.
This modification allows the algorithm to function without prior knowledge of $T$, while preserving the same order of regret.
\end{remark}

\subsection{Key Features of the Algorithm}

The IV-Greedy algorithm has two distinctive features that merit further exploration.
The first feature is the use of the joint-2SLS approach, and the second is the three-phase structure, with a particular emphasis on the role of Phase 2.

\subsubsection{The Necessity of the Joint-2SLS}\label{sec:joint-2sls}

During Phase 3, the main phase of the algorithm, the joint-2SLS approach is utilized. As we will demonstrate, several seemingly obvious alternate design choices for the algorithm are available. However, they fail to achieve consistent estimation with a sublinear regret, which is the dual objective in our online learning environment.

Contrary to the 2SLS estimator \eqref{eq:joint-2SLS},
a simple approach to estimating $\BFalpha$ in Phase~3 involves using the standard OLS method in an arm-specific way.
Specifically,
\begin{equation}\label{eq:OLS}
\widehat{\BFalpha}_{i,t}^{\mathsf{OLS}} :=  (\BFV^\intercal_{i,\ttR:t}\BFV_{i,\ttR:t})^{-1}\BFV_{i,\ttR:t}^\intercal\BFR_{i,\ttR:t},
\end{equation}
where $\BFV_{i,\ttR:t}$ denotes the  covariates $\BFv_s$ (instead of the augmented covariates $\widetilde{\BFv}_s$) that are observed in the time periods $s=T_1+1,\ldots, t$ when arm $i$ is selected, and $\BFR_{i,\ttR:t}$ is similarly defined.
This estimator is separately constructed for the parameter $\BFalpha_{i,t}$ for each arm $i$ based on the data available up to time $t$.

While the use of the OLS method in an arm-specific manner is standard practice in linear contextual bandit problems, the presence of endogeneity clearly results in a biased estimate of each $\BFalpha_i$.
Moreover, as illustrated in the example in Section \ref{sec:double-endog}, a self-fulfilling bias will emerge, leading to sub-optimal policies in the long run.

A second approach to estimating $\BFalpha$ in Phase~3 is the arm-specific application of the 2SLS method.
This approach is a natural choice if a set of IVs are available and can be considered as a generalization of \eqref{eq:OLS}:
\begin{equation}\label{eq:Arm-2SLS}
\widehat{\BFalpha}_{i,t}^{\mathsf{2SLS}} := \bigl(\BFV_{i,\ttR:t}^\intercal   \calP[\BFZ_{i,\ttR:t}]\BFV_{i,\ttR:t}\bigr)^{-1}\BFV_{i,\ttR:t}^\intercal  \calP[\BFZ_{i,\ttR:t}]\BFR_{i,\ttR:t}.
\end{equation}
Here, $\BFZ_{i,\ttR:t}$ is defined in the same way as $\BFV_{i,\ttR:t}$.
However, this estimator still does not yield consistent estimates of $\BFalpha_i$.
This inconsistency stands in contrast to the traditional use of the 2SLS method in static data analysis settings, where it is employed to correct bias caused by endogeneity issues.
As discussed in Section \ref{sec:double-endog}, this failure is attributed to the endogeneity spillover effect.
In other words, the selected arms, which form part of the data, can also become endogenous, in addition to the endogenous covariates.
The arm-specific 2SLS approach fails to address the part of endogeneity that spills over to the selected arms.
This issue arises from the dynamic sample selection, which is a consequence of the policy's adaptive nature in arm selection.

A straightforward way to circumvent the dynamic sample selection issue is to render the policy non-adaptive.
This leads to a third alternative approach to estimating $\BFalpha$ in our endogenous online learning environment. Specifically, one could implement the so-called ``randomize-then-commit'' (RTC) strategy as follows.
Initially, arms are selected uniformly at random for a series of consecutive time periods, say $t=1,\ldots,\ttR$, similar to what is done in Phase 1 of the IV-Greedy algorithm.
Then, the estimate $\widehat\BFalpha_{\ttR}$ is constructed using the arm-specific 2SLS.
As $\ttR\to\infty$, this becomes a consistent estimator because all arms have been randomly selected, thus avoiding the endogeneity spillover effect.
Finally, one commits to the greedy policy that selects the seemingly optimal arm based on the estimate $\widehat\BFalpha_{\ttR}$, without updating the estimates in all subsequent periods.
In other words, $a_t = \argmax_i \{\widehat\BFalpha_{\ttR}^\intercal \BFv_t \}$ for all $t=\ttR+1,...T$.

While the RTC strategy avoids the endogeneity spillover effect and ensures consistent estimation, it does so at the expense of a linear regret---a drawback not shared by the IV-Greedy algorithm.
The greedy mechanism, or any adaptive mechanism aiming to emulate the decisions of the oracle policy in an online learning environment, is indispensable if the objective is to achieve a sublinear regret.
Consequently, the dynamic selection issue is likely to be generated by such an adaptive mechanism, necessitating an adjustment in the estimation approach accordingly.

\subsubsection{The Necessity of the Covariance Stabilization Phase} \label{sec:three-phases}

The second key feature of the IV-Greedy algorithm is the inclusion of Phase~2.
This phase aids in stabilizing the matrix $\mathbb{E}[\widetilde{\BFv}_t \BFz_t^\intercal]$ towards the limit $\mathbb{E}[\BFv_t^* \BFz_t^\intercal]$, resulting in an accurate estimate of the covariance matrix $\BFOmega^*$, as defined in Equation~\eqref{eq:cov-matrix}.
An intuitive alternative approach might involve carrying out Phase 1 first, which provides a sufficiently accurate estimate of $\alpha_t$, and then moving directly to the joint-2SLS.
However, due to the online nature of our problem, this approach fails to yield a consistent estimate over the long term.
Poor estimates of either the coefficients or the covariance matrix in the past have lasting effects through their impact on the chosen actions, leading to biased long-term estimates.

Suppose one arm (for the sake of discussion, let's say arm~1) encounters a bad draw, which results in a lower estimate of $\BFalpha$.
Without the covariance stabilization phase, the magnitude of this downward estimate could be substantially large.
In such a scenario, a direct application of Phase~3 of the algorithm implies that arm~1 may not be chosen again throughout the entire sequence of iterates.
This leads to a permanent underestimate of the reward function of arm~1, hence resulting in a sub-optimal policy.
Technically, a bad draw could imply that the indicator $\ind(a_t = 1)$ approaches zero as $t\rightarrow \infty$.
This leads to the near singularity of the matrix $\mathbb{E}[ \widetilde{\BFv}_t \BFz_t^\intercal]$, causing unstable estimates of $\BFOmega^*$ and inconsistent estimates of $\BFalpha$.
We formally demonstrate the necessity of Phase 2 in the proposition below.

\begin{proposition}[Non-convergence  Without Phase~2]\label{prop:Non-Convergence}
    Suppose there are $K=2$ arms. If the Covariance Stabilization Phase takes $t=T_1+1,...,T_2$ with $T_2-T_1 = C>0$ being an absolute constant that does not depend on $T$, then, for any $T$ large enough, there exists $\underline{p}>0$ such that with probability at least $\underline{p}$, one of the two arms does not converge in the IV-Greedy algorithm, leading to a sub-optimal policy and a linear regret.
\end{proposition}

\section{Theoretical Results}\label{sec:theo}

In this section, we analyze the theoretical properties of the IV-Greedy algorithm.
We show that this algorithm produces a regret of order  $\mathcal{O}(\log(T))$.
We also show that it delivers consistent and asymptotically normal estimates of $\BFalpha$.

\begin{assumption}\label{assump:IV}
Let $\{(\BFv_t,\BFz_t,\varepsilon_t):t=1,\ldots,T\}$ be an i.i.d. sequence, where $\BFz_t\in \Real^q$ is a vector of IVs with $q\geq Kp$.
\begin{enumerate}[label=(\roman*)]
\item
$\varepsilon_t$ is sub-Gaussian with variance proxy $\varsigma^2$,
that is, $\mathbb{E}[e^{\lambda \varepsilon_t}] \leq e^{\lambda^2\varsigma^2/2}$ for all $\lambda\in\Real$.
\label{cond:subG}
\item $\mathbb{E}[\BFz_t\varepsilon_t]=0$ and $\Var(\varepsilon_t)=\sigma^2<\infty$. \label{cond:valid-IV}
\item There exist positive constants $\bar{v}$ and $\bar{z}$, such that
$\|\BFz_t\|_\infty\leq \bar{z}$ and $\|\BFv_t\|_\infty \leq \bar{v}$ almost surely, where $\|\cdot\|_\infty$ denotes the $L^\infty$ norm of a vector. \label{cond:bounded}

\item Let $\BFdelta_{i,j}\coloneqq \BFalpha_i-\BFalpha_j$, $\BFSigma^*_{vz}:=\mathbb{E}[\BFv_t^*\BFz_t^\intercal]$, and $\BFSigma_{zz}:=\mathbb{E}[\BFz_t \BFz_t^\intercal]$, where
\[\BFv_t^*\coloneqq
\begin{pmatrix}
\ind(a_t^*=1)\BFv_t \\
\vdots \\
\ind(a_t^*=K)\BFv_t
\end{pmatrix} =
\begin{pmatrix}
\ind(\min_{j\neq 1}\BFv_t^\intercal \BFdelta_{1j} > 0)\BFv_t \\
\vdots\\
\ind(\min_{j\neq K}\BFv_t^\intercal \BFdelta_{Kj}  > 0)\BFv_t
\end{pmatrix} \in\Real^{Kp}.\]
Then, $\BFSigma_{vz}^*$ is a full-rank matrix and $\BFSigma_{zz}$ is a positive definite matrix. \label{cond:full-rank}

\item
There exists a constant $L>0$, such that
$\Pr(|\BFv_t^\intercal \BFdelta_{i,j}|\leq c)\leq Lc$  for all $c > 0$ and $1\leq i\neq j \leq K$.\label{cond:margin}

\end{enumerate}
\end{assumption}

Condition~\ref{cond:subG}  is  common in the  bandit literature and is key for establishing a $\log(T)$-like regret bound. For example, \cite{GoldenshlugerZeevi13} assume normality on $\varepsilon_t$, and \citet{ZhongHongLiu21} assume boundedness,
both of which are special cases of Condition~\ref{cond:subG}.
\citet{BastaniBayatiKhosravi21} impose the same sub-Gaussian condition as we do.
If Condition~\ref{cond:subG} is replaced with  moment restrictions on $\varepsilon_t$,
one might still obtain a sublinear regret, but not a $\log(T)$ regret. Conditions~\ref{cond:valid-IV}  and \ref{cond:bounded} are again common assumptions in the IV and contextual bandit literature, respectively.

Condition~\ref{cond:full-rank}
implies that $\BFSigma_{vz}^*\BFSigma_{zz}^{-1}(\BFSigma_{vz}^*)^\intercal \in \Real^{(Kp)\times (Kp)} $
is a  positive definite matrix.
This is a key condition for parameter identification.
It requires that the oracle policy must be dependent on $\BFv_t$; more specifically, $\ind(a_t^*=1)\BFv_t, \ldots,\ind(a_t^*=K)\BFv_t$ should not be co-linear when projected to the $\BFz_t$ space.
Condition~\ref{cond:full-rank}  holds in general if $\BFdelta_{i,j} \neq \BFzero$ for all $1\leq i\neq j\leq K$.

Further,
Condition~\ref{cond:margin} states that the density of $\BFv_t$ is Lipschitz continuous  in the vicinity of the hyperplance $\{\BFv:\BFv^\intercal \BFdelta_{i,j} = 0\}$.
This condition is often referred to as the ``margin condition'' in statistical learning literature \citep{Tsybakov04}. It is also a standard condition in the literature on contextual bandits.
Moreover, Conditions~\ref{cond:full-rank}
and \ref{cond:margin} together imply that  $ \Pr(\BFv_t^\intercal \BFdelta_{i,j} <0)>0$ and  $ \Pr(\BFv_t^\intercal \BFdelta_{i,j}>0)>0$,
which is referred to as the ``diversity condition'' in  \cite{GoldenshlugerZeevi13}.

\begin{remark}\label{comment:decomp}
The covariate vector $\BFv_t\in\Real^p$ may be decomposed into
two parts: $\BFx_t$ and $\BFd_t$, where
$\BFx_t\in\Real^{\ell}$ is exogenous, $\BFd_t\in\Real^m$ is potentially endogenous, and $p=\ell+m$.
Then,
there are $((K-1)\ell+Km)$ number of endogenous variables in \eqref{eq:main2}, with
$m$ of them from the term $\BFv_t^\intercal \BFalpha_1$ and the other $(K-1)(l+m)$ of them from the term
$\ind(a_t = i)\BFv_t^\intercal(\BFalpha_i - \BFalpha_1)$, $i=2,\ldots,K$.
Thus, the minimal number of IVs that we need is $ (K-1)\ell+K m = Kp - \ell$.
However, for the convenience of exposition, we assume
there are a total of $q$ IVs with $q\geq Kp$.
\end{remark}

\begin{remark}\label{comment:multipe-IVs}
To construct a sufficient number of IVs,
following the decomposition of $\BFv_t$
in Remark~\ref{comment:decomp}, practitioners may begin with a set of IVs for $\BFd_t$, denoted as $\check{\BFz}_t\in\Real^{\check{q}}$ with $\check{q} \geq m$.
The exogenous variables $\BFx_t$ can serve as IVs for themselves.
A natural means
to construct additional IVs for $\ind(a_t=i)\BFv_t$, $i=2,\ldots,K$, in \eqref{eq:main2}
is to use such variables as $\ind(\BFx_t\geq \mathring{x})$, $\ind(\check{\BFz}_t\geq \mathring{z})$, and their interaction terms---for example, $ \ind(\BFx_t\geq \mathring{x})\BFx_t$ and $\ind(\check{\BFz}_t\geq \mathring{z})\check{\BFz}_t$ for some threshold values $\mathring{x}$ and $\mathring{z}$. See the simulation example in Section~\ref{sec:simulation} for an illustration.
Other nonlinear transformations of $\BFx_t$ and $\check{\BFz}_t$ may also work as IVs,
if
\begin{equation}\label{eq:conditional}  \mathbb{E}[\varepsilon_t|\BFx_t,\check{\BFz}_t]=0.
\end{equation}
This condition and the use of splines as IVs are widely employed in the conditional moment restriction literature \citep{ChenPouzo12}.
This implies Condition~\ref{cond:bounded} in Assumption \ref{assump:IV} and, thus, it is slightly stronger.
For optimal selection of IVs under the conditional moment restriction \eqref{eq:conditional},
we refer to \citet{DonaldImbensNewey09} and \citet{Belloni12}.
\end{remark}

\subsection{Regret Analysis}
In this subsection, we describe the long-run regret of the IV-Greedy algorithm. We then discuss a new technique---zig-zag induction---that we develop for analyzing algorithms associated with online learning environments.

\begin{theorem}
\label{theo:IV-Greedy-regret}
Let $\ttR = C_{\ttR} \log(T)$ and $\ttU = (C_{\ttR} + C_{\ttU}) \log(T)$ for some sufficiently large constants $ C_{\ttR}$ and $C_{\ttU}$.  Then, under Assumption~\ref{assump:IV}, the regret of the IV-Greedy algorithm is $\mathcal{O}(\log(T))$.
\end{theorem}

Note that
in the presence of endogeneity, an algorithm that does not produce consistent estimates of the coefficient vector $\BFalpha$ will generally suffer a linear regret \citep{NambiarSimchiWang19}.
Therefore, it is necessary to establish consistency of the coefficient estimate.
This requires analyzing the probability of the estimate $\widehat\BFalpha_t$ falling in the vicinity of $\BFalpha$
over Phase~3 of the algorithm.

From the 2SLS formula, it can be shown that the coefficient estimate $\widehat{\BFalpha}_t$ in Phase~3 depends on the covariance estimate $\widehat{\BFSigma}_{vz,{\ttR:t}}$. The calculation of $\widehat{\BFSigma}_{vz,{\ttR:t}}$, however, is not standard because it depends on the past estimates $\{\widehat{\BFalpha}_s:\ttR\leq  s\leq t-1\}$---that is,
\begin{equation}\label{eq:Sigma_hat_vz}
\widehat{\BFSigma}_{vz,\ttR:t} =\frac{1}{t-\ttR}\sum_{s=\ttR+1}^t \begin{pmatrix}
\ind(a_s=1)\BFv_s\\
\vdots \\
\ind(a_s=K)\BFv_s
\end{pmatrix} \BFz_s^\intercal,
\end{equation}
where $a_s$ is selected by the greedy algorithm and depends on $\widehat{\BFalpha}_s$.

Note that if the optimal arm $a^*_s$ were known and selected in each period, we would simply calculate
\begin{equation}\label{eq:sigma_vz*}
\widehat{\BFSigma}_{vz,\ttR:t}^*\coloneqq \frac{1}{t-\ttR}\sum\limits_{s=\ttR+1}^t \begin{pmatrix}
\ind(a_s^*=1)\BFv_s\\
\vdots\\
\ind(a_s^*=K)\BFv_s\\
\end{pmatrix} \BFz_s^\intercal,
\end{equation}
which is standard because it involves the average of samples from i.i.d. random variables.

However, because the optimal arms are unknown, the analysis of $\widehat{\BFSigma}_{vz,\ttR:t}$ poses a challenge. First, since $\widehat{\BFSigma}_{vz,\ttR:t}$ depends on the choice of arm $a_s$ for all $s$ from $\ttR+1$ to $t$, any past coefficient estimate $\widehat{\BFalpha}_s$, through its effect on the selected arm, has an effect on $\widehat{\BFSigma}_{vz,{\ttR:t}}$. Second, these coefficient estimates are serially correlated because they are all calculated from the same data source.
Lastly, not only does $\widehat{\BFSigma}_{vz,{\ttR:t}}$ depend on $\{\widehat{\BFalpha}_s:\ttR\leq  s\leq t-1\}$, $\widehat{\BFalpha}_t$ also depends on $\widehat{\BFSigma}_{vz,{\ttR:t}}$.
This two-way dependence implies that the coefficient estimates and the covariance estimates are entangled.

The two-way dependence, a common feature of online learning environments, makes it difficult for the coefficient estimates to converge to the true value in the long run. A poor coefficient estimate can lead to a wrong action, which contaminates the generated data  (since the data associated with the right arm is no longer observed). The wrong data then worsens the future coefficient estimates and, thus, the future data-generating process. In summary, errors made in the past do not disappear: they make future errors more likely.

To deal with this problem, we untangle the dependence via a ``zig-zag'' induction. As an overview of its structure, assume that, in period $t-1$ in Phase~3, the coefficient estimate $\widehat{\BFalpha}_s$ is sufficiently close to the true value for \emph{all} $s$ from $\ttU+1$ to $t-1$ with a sufficiently high probability.
A induction step is then used to show that this remains true for period $t$. To carry out the induction step, two sub-steps are needed.

The first sub-step shows that the covariance estimate $\widehat{\BFSigma}_{vz,{\ttR:t}}$ is sufficiently accurate in period $t$. This is the zig-step. In the second sub-step, we use this result from the zig-step to show that the induction hypothesis holds at time $t$. The difficult part of the induction is the zig-step. As mentioned earlier, the covariance estimate in the zig-step depends on all the coefficient estimates in the past, which are serially correlated.

\begin{figure}[ht]
    \centering
    \caption{The Zig-Zag Approach.}
    \label{fig:zigzag}
    \includegraphics[width=0.6\textwidth]{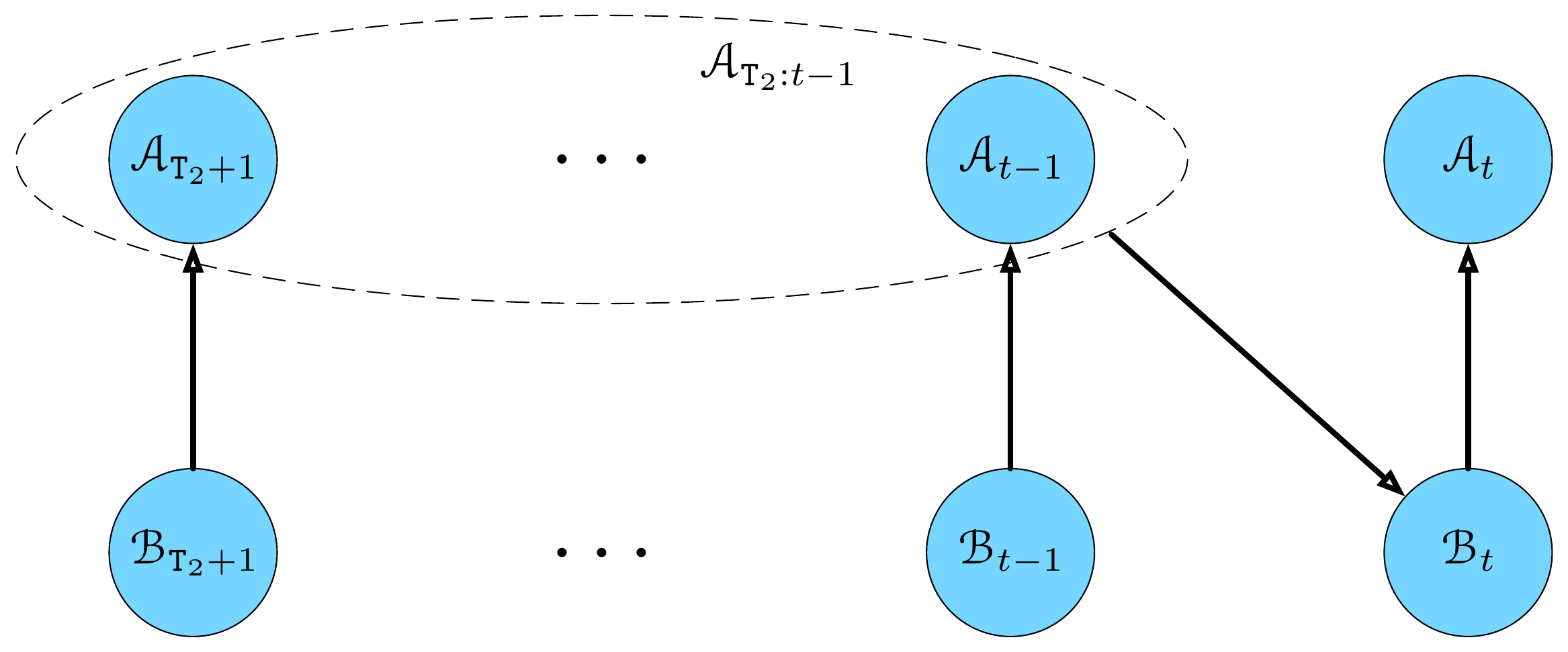}
\end{figure}

To deal with the challenge in the zig-step, let
\[\scrA_t\coloneqq \Biggl\{\|\widehat{\BFalpha}_t-\BFalpha\|\leq C_A \sqrt{\frac{\log(T)}{t-\ttR}}\Biggr\}
\qq{and}
\scrB_t\coloneqq \bigl\{\|\widehat{\BFSigma}_{vz,\ttR:t}-\BFSigma_{vz}^*\|\leq \eta\bigr\},
\]
for some constant $C_A>0$,
where $\|\cdot\|$ denotes the Euclidean norm for vectors and the spectral norm for matrices.
Further, let $\scrA_{\ttU:t}:=\cap_{s=\ttU+1}^{t} \scrA_{s}$, where we note that the intersection begins at the beginning of Phase~2. This is the event in which all the past coefficient estimates from Phase~2 are sufficiently accurate. With these definitions, the induction shows that there exists a constant $C>0$, such that
\begin{equation*}
\Pr( \scrA_{\ttU:t})\geq 1-C(t-\ttU)T^{-2}, \quad t=\ttU+1,\ldots,T.
\end{equation*}
This is achieved via a two-step induction using $\scrB_t$ as a bridge---first from $\scrA_{\ttU:t-1}$ to $\scrB_t$ and then from $\scrB_t$ to $\scrA_t$
(see Figure~\ref{fig:zigzag}).

Note that we perform the induction on $\scrA_{\ttU:t}$ instead of $\scrA_t$. This is because the zig-zag induction step requires the analysis of the covariance estimate. As we emphasize above, the covariance estimate depends on the past coefficient estimates. Therefore, showing that the covariance estimate is sufficiently accurate requires event $\scrA_{\ttU:t}$ to occur---that is, all the past coefficient estimates from Phase~2 on to be sufficiently accurate.

A key observation in our induction step is that, when the assumption holds in period $t-1$ of Phase~3, we can show (with details in the proof) that
there exists a constant $C'>0$, such that
\begin{equation}\label{eq:key_obs}
    \{a_s\neq a_s^*\}\subseteq  \cup_{1\leq j\neq a^*_s\leq K }\left\{|\BFv_s^\intercal \BFalpha_j -\BFv_s^\intercal \BFalpha_{a^*_s} |\leq C' \eta \right\},
\end{equation}
for all $s=\ttU+1,\ldots,t-1$.
Therefore, this relation  states that a wrong arm is selected only when the difference in the actual expected reward between this arm and some other arm is small.
(We can also show, with a similar argument, that \eqref{eq:key_obs} holds
all $s=\ttR+1,\ldots,\ttU$, i.e.,
Phase~2 of the algorithm, provided that
the coefficient estimate at the end of Phase~1 is sufficiently accurate.)

The intuition for this observation is that, when the coefficient estimate is sufficiently accurate, the  estimated expected reward
of choosing each arm (calculated from the coefficient estimate) is also close to the actual expected reward;
and thus the arm selected by the algorithm is optimal if the actual expected reward from this arm is sufficiently different than that from any other arm.
In other words, when a wrong arm is selected, it must be that the difference in the actual expected reward between this arm and some other arm is small.

A consequence of this observation is that, instead of estimating the probability that a wrong arm is selected, which depends on the past coefficient estimates, it suffices to estimate the probability that the covariate vector $\BFv_s$ falls within a bound.
Specifically, we can show that
\begin{align}
    \bigl\|\widehat{\BFSigma}_{vz,\ttR:t}-\widehat{\BFSigma}^*_{vz,\ttR:t}\bigr\| \leq{}&
    \tilde{C}\frac{1}{t-\ttR}\sum_{s=\ttR+1}^t \ind(a_s\neq a_s^*)  \nonumber \\
    \leq{}& \tilde{C} \frac{1}{t-\ttR}\sum_{s=\ttR+1}^t \sum_{1\leq j\neq a^*_s \leq K }\ind\left(|\BFv_s^\intercal \BFalpha_j - \BFv_s^\intercal \BFalpha_{a^*_s}|\leq C'\eta \right), \label{eq:iid_sum}
\end{align}
for some constant $\tilde{C}>0$.
Because the covariate vectors $\BFv_s$ are i.i.d. over different periods, so are $\ind\left(|\BFv_s^\intercal \BFalpha_j - \BFv_s^\intercal \BFalpha_i|\leq C'\eta \right)$.
In other words, the right-hand-side of the  inequality \eqref{eq:iid_sum} is a sample average of i.i.d. random variables; and therefore, bounds on its tail probability can be established using standard concentration inequalities. This allows us to show that, with a sufficiently high probability,  $\widehat{\BFSigma}_{vz,{\ttR:t}}$ is close to $\widehat{\BFSigma}_{vz,{\ttR:t}}^*$.

The induction shows that, with a sufficiently high probability, the coefficient estimates remains sufficiently accurate over the entire Phase~3 of the algorithm.
Standard calculation \citep{BastaniBayatiKhosravi21} can then be applied to show that the regret in Phase~3 is $\mathcal{O}(\log(T))$.
Since the durations of Phases~1 and 2 are both $\mathcal{O}(\log(T))$, this shows that the total regret is $\mathcal{O}(\log(T))$.
\cite{GoldenshlugerZeevi13} prove that
in a standard linear contextual bandit problem---with no endogeneity involved---the best possible lower bound
for any algorithm is $C\log(T)$ for some constant $C$.
Since our setup allows endogeneity in the covariates and includes standard linear contextual bandits as special cases,
$C\log(T)$ is also
a lower bound on regret for our setup.
Hence, Theorem~\ref{theo:IV-Greedy-regret} shows that IV-Greedy achieves the asymptotically minimal regret.

\subsection{Statistical Inference}

The contextual bandit literature mostly focuses on regret analysis, whereas results on statistical inference are relatively scarce.
We provide  such results for the coefficient estimates.

\begin{theorem}
\label{theorem:asymp}
Let $\ttR = C_{\ttR} \log(T)$ and $\ttU = (C_{\ttR}+ C_{\ttU}) \log(T)$ for some sufficiently large constants $ C_{\ttR}$  and  $C_{\ttU}$.
Then, under Assumption~\ref{assump:IV}, the estimator $\widehat{\BFalpha}_T$ of Algorithm~\ref{algo:IV-bandits} satisfies
\begin{gather}\label{eq:CLT}
\sqrt{T-\ttR} (\widehat{\BFalpha}_{T} - \BFalpha )
\rightsquigarrow \calN\Bigl(\BFzero,\sigma^2\bigl(\BFSigma_{vz}^*\BFSigma_{zz}^{-1}(\BFSigma_{vz}^*)^\intercal\bigr)^{-1}\Bigr),
\end{gather}
as $T\to\infty$, where $\BFSigma_{vz}^* = \mathbb{E}[\BFv_t^* \BFz_t^\intercal]$, $\BFSigma_{zz}=\mathbb{E}[\BFz_t \BFz_t^\intercal]$, and $\calN(\BFzero, \BFSigma)$ denotes the multivariate normal distribution with zero mean vector and covariance matrix $\BFSigma$.
\end{theorem}

To construct a confidence interval for $\BFalpha$, we need a consistent estimator of the covariance matrix $\sigma^2\bigl(\BFSigma_{vz}^*\BFSigma_{zz}^{-1}(\BFSigma_{vz}^*)^\intercal\bigr)^{-1}$ in \eqref{eq:CLT}.
To this end, note that
$\widehat{\sigma}^2_T$, $\widehat{\BFSigma}_{vz,\ttR:T}$, and $\widehat{\BFSigma}_{zz,\ttR:T}$ are consistent estimators of $\sigma^2$, $\BFSigma_{vz}^*$, and $\BFSigma_{zz}$, respectively.
Moreover, by the definition of $\widehat{\BFOmega}_T$ in Algorithm~\ref{algo:IV-bandits},
\[
(T-\ttR) \widehat{\BFOmega}_T = \bigl(\widehat{\BFSigma}_{vz,\ttR:T}\widehat{\BFSigma}_{zz,\ttR:T}^{-1}(\widehat{\BFSigma}_{vz,\ttR:T})^\intercal\bigr)^{-1}.
\]
Thus, we use $(T-\ttR)\widehat{\sigma}^2_T \widehat{\BFOmega}_T$ to consistently estimate $\sigma^2\bigl(\BFSigma_{vz}^*\BFSigma_{zz}^{-1}(\BFSigma_{vz}^*)^\intercal\bigr)^{-1}$.
Then,
standard calculation  leads to the following result,
which facilitates inference on $\BFalpha$.

\begin{corollary}\label{coro:inf}
Let $\ttR = C_{\ttR} \log(T)$ and $\ttU = (C_{\ttR}+ C_{\ttU}) \log(T)$ for some sufficiently large constants $ C_{\ttR}$  and  $C_{\ttU}$.
Then, under Assumption~\ref{assump:IV}, the estimator $\widehat{\BFalpha}_T$ of Algorithm~\ref{algo:IV-bandits} satisfies
\begin{gather*}
(T-\ttR)^\frac{1}{2}\bigl(\widehat{\sigma}_T^2\widehat{\BFOmega}_T\bigr)^{-\frac{1}{2}}(\widehat\BFalpha_T-\BFalpha) \rightsquigarrow \mathcal{N}(0,\BFI_{Kp}), \\
    (T-\ttR)(\widehat\BFalpha_T-\BFalpha)^\intercal \bigl(\widehat{\sigma}_T^2\widehat{\BFOmega}_T\bigr)^{-1} (\widehat\BFalpha_T-\BFalpha) \rightsquigarrow \chi^2_{Kp}.
\end{gather*}
as $T\to\infty$, where
$\widehat{\sigma}_T$ and $\widehat{\BFOmega}_T$ are defined in Algorithm~\ref{algo:IV-bandits},
$\BFI_{Kp}$ is the identity matrix of size $(Kp)\times (Kp)$,
and $\chi^2_{Kp}$ denotes the chi-square distribution with $Kp$ degrees of freedom.
\end{corollary}

\section{Simulation Experiments}\label{sec:simulation}

In this section, we present a simulation study to demonstrate the performance of IV-Greedy and compare it with the following alternatives.
These alternatives---similar to IV-Greedy---all have a randomization phase of duration $\ttR$, in which each arm is selected uniformly at random.
\begin{enumerate}[label=(\roman*)]
    \item Naive-IV-Greedy.  In each period $t=\ttR+1,\ldots,T$, this algorithm performs the arm-specific 2SLS method, as detailed in Equation~\eqref{eq:Arm-2SLS}, to obtain an estimate of $\BFalpha$ and selects an arm following a greedy manner.
    \item OLS-UCB. In each period $t=\ttR+1,\ldots,T$, this algorithm performs the arm-specific OLS method, as detailed in Equation~\eqref{eq:OLS}, to obtain an estimate of $\BFalpha$ and selects an arm following the upper-confidence-bound (UCB) criterion. This algorithm is the standard benchmark in context bandits literature for linear reward functions; see \cite{ZhongHongLiu21} for reference.
    \item Randomize-then-commit (RTC). After the first $\ttR$ periods, this algorithm performs arm-specific 2SLS to obtain an estimate of $\BFalpha$ and commit to it. In each period $t=\ttR+1,\ldots,T$, the algorithm continues to use this estimate without updates to select an arm in a greedy manner. The details of this algorithm are discussed in Section~\ref{sec:joint-2sls}.
\end{enumerate}
For all the four algorithms, we set $\ttR=50$ and the total run length $T=20000$.
In addition, we set $\ttU=100$ for IV-Greedy.

We evaluate these algorithms based on regret and the number of wrong arms pulled---both of which are common metrics for online learning algorithms. Additionally, we perform statistical inference on the coefficients in the reward function upon the termination of the algorithms.
To this end, we consider a two-armed linear contextual bandit problem with three-dimensional covariates, i.e., $K=2$.
Let the covariate vector $\BFv_t = (1, x_t, d_t)^\intercal$, where  $x_t\in\Real$ is an exogenous variable and $d_t\in\Real$ is an endogenous variable.
The endogeneity is introduced in the following manner.
Suppose the random reward at time $t$ is
\[
R_t = \sum_{i=1}^2 \mu_i(\BFv_t)\ind(a_t=i) +\varepsilon_t,
\]
where $\mu_i(\BFv_t) = \beta_{i,0} + \beta_{i, 1}x_t + \gamma_{i} d_t$,
for some parameters   $\beta_{i,0}$, $\beta_{i,1}$, and $\gamma_i$.
Suppose also
\[d_t = \sqrt{x_t} + \rho_{\check{z}} \check{z}_t + \rho_\eta \eta_t \qq{and} \varepsilon_t = \tilde{\varepsilon}_t +  2\eta_t,\]
where $\rho_{\check{z}}$ and $\rho_\eta$ are some positive constants,
and
$x_t$, $\check{z}_t$, $\eta_t$, and $\tilde{\varepsilon}_t$ are independent random variables with the following distributions:
\begin{gather*}
    x_t \sim{}  \mathsf{TruncNorm}(\mu_x, \sigma_x^2, l_x, u_x), \quad
    \check{z}_t \sim{} \mathsf{TruncNorm}(\mu_{\check{z}}, \sigma_{\check{z}}^2, l_{\check{z}}, u_{\check{z}}), \\
    \eta_t \sim{} \mathsf{TruncNorm}(\mu_\eta, \sigma_\eta^2, l_\eta, u_\eta), \quad
    \tilde{\varepsilon}_t \sim \mathcal{N}(0, \sigma^2_{\tilde{\varepsilon}}).
\end{gather*}
Here, we use $\mathsf{TruncNorm}(\mu, \sigma^2, l, u)$ to denote the truncated normal distribution that is derived from bounding the $\mathcal{N}(\mu,\sigma^2)$ distribution  on the interval $(l, u)$.
The parameters are specified as follows:
\begin{gather*}
(\beta_{1,0}, \beta_{1,1}, \gamma_1) = (1, 4, 4), \quad (\beta_{2,0}, \beta_{2,1}, \gamma_2) = (8, 2, 2), \\
\rho_{\check{z}} = 0.5, \quad \rho_\eta = 1.5, \quad \sigma^2_{\tilde{\varepsilon}} = 0.25, \\
(\mu_x, \sigma_x^2, l_x, u_x) = (0, 1, 0, 10), \quad
(\mu_{\check{z}}, \sigma_{\check{z}}^2, l_{\check{z}}, u_{\check{z}}) = (0, 4, 0, 10), \\
(\mu_\eta, \sigma_\eta^2, l_\eta, u_\eta) = (0, 0.25, -5, 5).
\end{gather*}

Note that $\varepsilon_t$ and $d_t$ are correlated through $\eta_t$ and, thus, $d_t$ becomes endogenous.
Moreover,
$\check{z}_t$ is correlated with $d_t$ but is uncorrelated with $\varepsilon_t$,
so $\check{z}_t$ serves as an IV for $d_t$.
However, for IV-Greedy to work, the number of IVs must be at least twice as many as the number of endogenous variables in order to account for the endogeneity spillover issue; see Remark~\ref{comment:decomp}.
As discussed in Remark~\ref{comment:multipe-IVs}, we create additional IVs using
$\ind(x_t\geq \mathring{x})$, $\ind(\check{z}_t\geq \mathring{z})$, and the interaction terms between them and $x_t$ or $\check{z}_t$, for some threshold values $\mathring{x}$ and $\mathring{z}$.
Specifically, we use the following list of IVs:
\[
\bigl\{1, \; x_t,\; \check{z}_t, \;\ind(x_t\geq 1), \;\ind(x_t\geq 1)\check{z}_t,\; \ind(x_t\geq 1.5),\; \ind(x_t\geq 1.5)\check{z}_t, \;\ind(\check{z}_t\geq 2), \;\ind(\check{z}_t\geq 2)\check{z}_t\bigr\}.
\]

\begin{figure}[ht]
    \begin{center}
    \caption{Regret}
    \label{fig:regret}
    \includegraphics[height=0.35\textwidth]{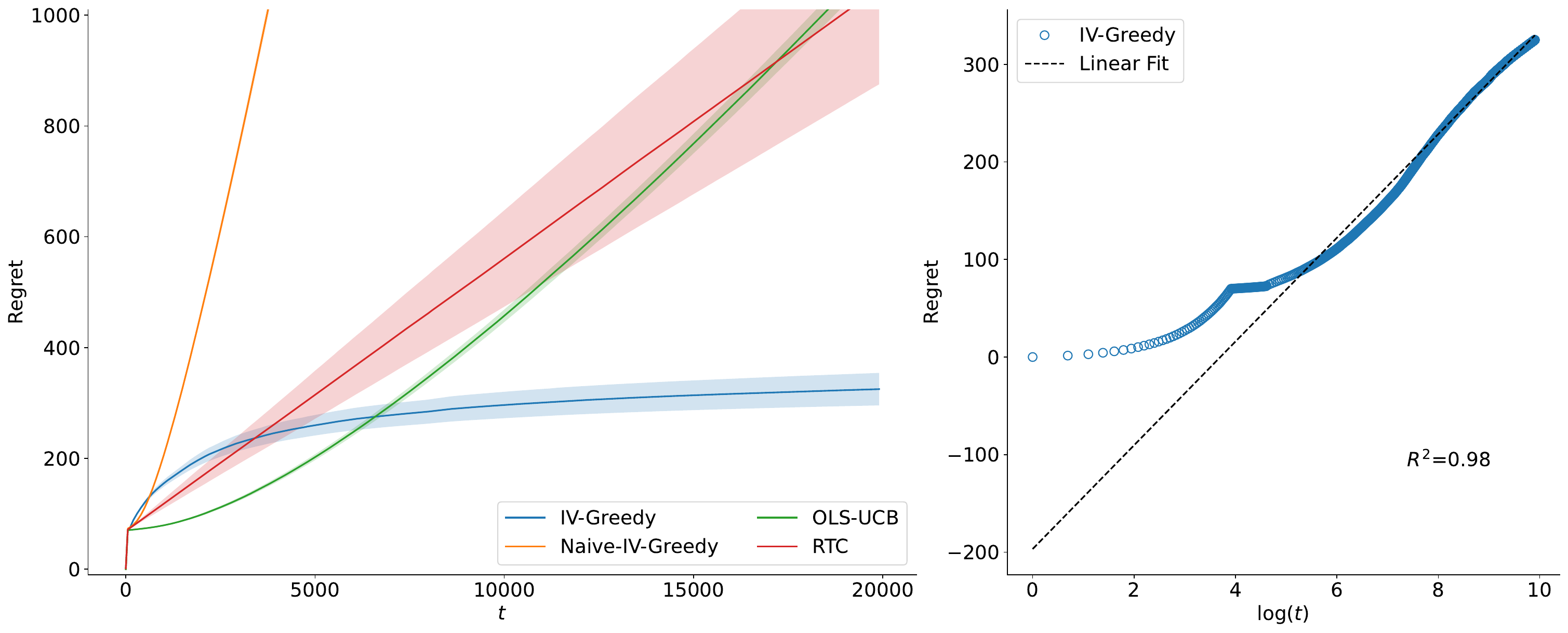}
    \end{center}
    \footnotesize{\emph{Note.} The shaded areas represent 95\% confidence bands calculated from 1000 replications.}
\end{figure}

\begin{figure}[ht]
    \begin{center}
    \caption{Biases in Coefficient Estimates.}
    \label{fig:bias}
    \includegraphics[width=0.9\textwidth]{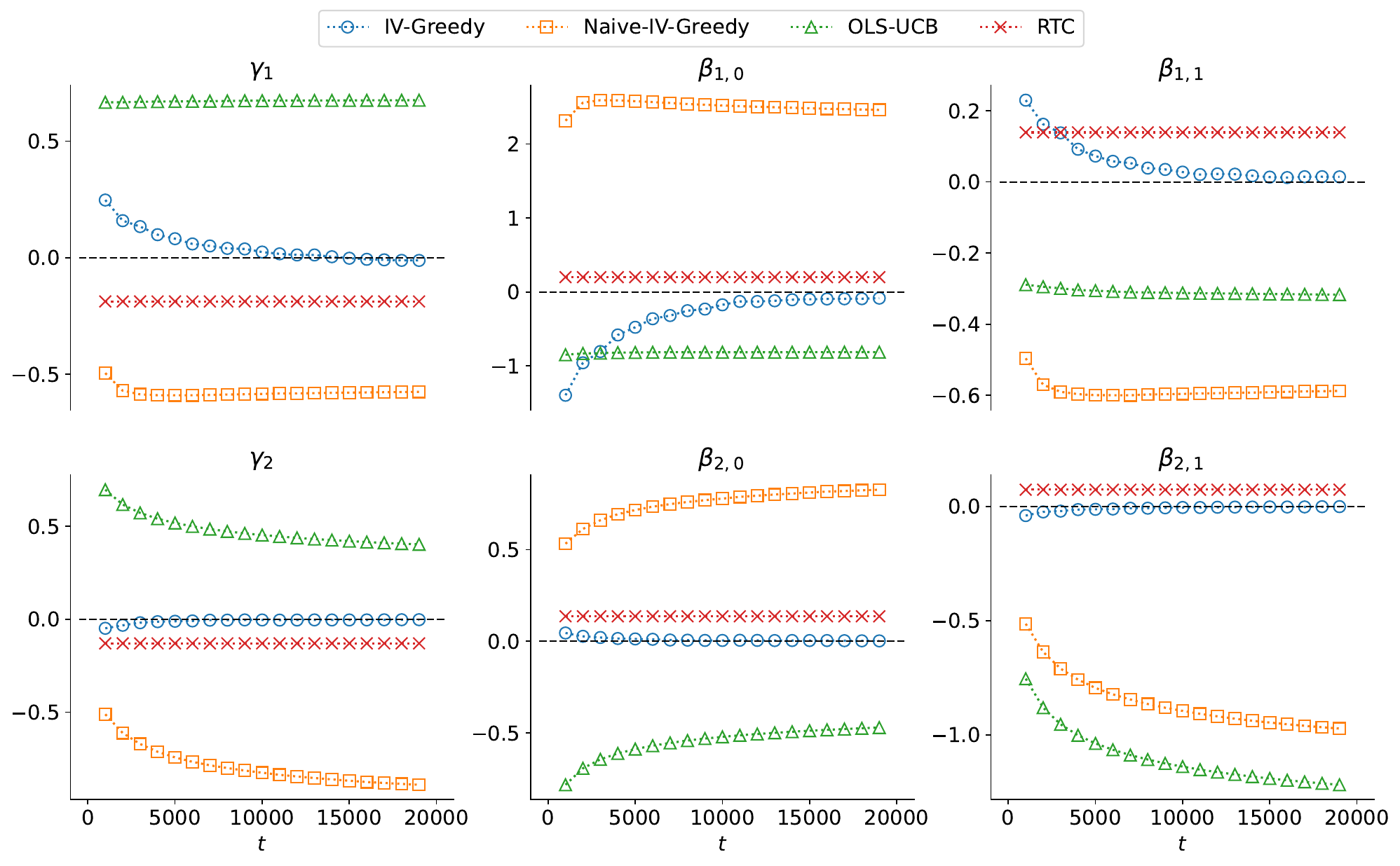}
    \end{center}
\footnotesize{\emph{Note.} For each parameter $\alpha$, the bias $\widehat{\alpha}_t - \alpha$ is calculated for 1000 replications to obtain an average.}
\end{figure}

Further, we run the four algorithms (IV-Greedy, Naive-IV-Greedy, OLS-UCB, and RTC) 1000 times each and present the results in Figures~\ref{fig:regret} and \ref{fig:bias} as well as Table~\ref{tab:stats}.

Figure~\ref{fig:regret} presents the regret of each algorithm.
First, in contrast to the others, IV-Greedy  achieves a regret with a diminishing growth rate after a sufficient number of iterations (e.g., $T=5000$).
The confidence bands further show that the performances of these two algorithms are stable in the learning process.
Second, to verify our theoretical result in Theorem \ref{theo:IV-Greedy-regret}, we conduct a linear regression analysis of the regrets of IV-Greedy on $\log(t)$.
The value of $R^2$ is $0.99$, thereby
providing strong empirical evidence for our theory.
Third, although the theoretical analysis requires that Phases~1 and 2 of IV-Greedy  should be ``sufficiently long,''
the simulation results indicate that they can be set to be moderate values ($\ttR=50$ and $\ttU=100$) in practice.

Figure~\ref{fig:bias} presents the biases of the coefficient estimates.
For the estimates produced by IV-Greedy, the bias vanishes as the number of iterations increases, while the other algorithms are generally biased.
In particular, the fact that Naive-IV-Greedy produces inconsistent estimates of the coefficients demonstrates the existence of endogeneity spillover in this online learning example (the endogeneity in the covariates is addressed by the use of IVs in the arm-specific 2SLS, but the endogeneity that spills over to the  actions remains).

Secondly, it is well known in the online learning literature that the use of the UCB criterion encourages the ``exploration'' (in the conventional sense) of seemingly inferior arms.
Assuming the absence of endogeneity, this would allow OLS-UCB to provide consistent estimates of the coefficients at the cost of a sub-linear regret.
The observation that OLS-UCB does not yield consistent estimates aligns with our discussion at the end of Section~\ref{sec:double-endog}.
Namely, to address the endogeneity issue in an online learning environment, one must undertake what is termed ``ex-ante exploration''.
This process involves perturbing the data-generating process, and IVs serve this purpose effectively.

Lastly, RTC first performs arm-specific 2SLS to data from randomization and then commits to the coefficient estimates.
This results in a dilemma: on the one hand, if the randomization period is long, then the regret is large, because the randomization produces a linear regret; on the other hand, if the randomization period is short, then the coefficient estimates are inaccurate,\footnote{In general, the 2SLS estimates have a second-order bias if the sample size is small or the IVs are weak \citep{AndrewsStockSun19}.}
which also leads to a large regret in the long run.
Thus, the regret of RTC is, at best, a fractal polynomial of $T$, which is significantly worse than $\log(T)$.

\begin{table}[ht]
\begin{center}
\caption{Statistics at Time $T=20000$.}
    \label{tab:stats}
\footnotesize{
\begin{tabular}{lccccccccc}
\toprule
          & \multicolumn{3}{c}{$\gamma_1$}   & \multicolumn{3}{c}{$\beta_{1,0}$} & \multicolumn{3}{c}{$\beta_{1,1}$} \\
\cmidrule(l){2-4}
\cmidrule(l){5-7}
\cmidrule(l){8-10}
          & Bias  & SD      & Coverage & Bias   & SD    & Coverage & Bias   & SD     & Coverage \\
IV-Greedy & -0.009 & 0.677 &  0.948    & -0.081 & 0.736  & 0.957    & 0.011  & 0.125 &  0.956    \\
Naive-IV-Greedy    &  -0.575     &   0.095       &    0.003      &   2.456     &    0.118   &      0.003       &   -0.586    &  0.063      &    0.003     \\
OLS-UCB   &    0.676   &   0.083    &    0.000      &    -0.814    &    0.208        &    0.000      &    -0.317    &  0.078       &   0.000      \\
RTC    &  -0.188     &    2.079     &    0.920     &   0.200     &    2.373      &   0.926       &  0.139      &    2.023     &    0.938     \\
\midrule
& \multicolumn{3}{c}{$\gamma_2$}   & \multicolumn{3}{c}{$\beta_{2,0}$} & \multicolumn{3}{c}{$\beta_{2,1}$} \\
\cmidrule(l){2-4}
\cmidrule(l){5-7}
\cmidrule(l){8-10}
          & Bias  & SD    & Coverage & Bias   & SD    & Coverage & Bias   & SD    & Coverage \\
IV-Greedy    &   -0.001    &  0.032      &    0.956      &   0.001     &    0.042    &       0.956     &  -0.001      &   0.038     &     0.946    \\
Naive-IV-Greedy    &   -0.894    &    0.118       &  0.003        &   0.829     &    0.086   &     0.003     &     -0.978  &    0.091    &    0.003     \\
OLS-UCB    &   0.399    &  0.069       &     0.000     &    -0.467     &    0.050   &    0.000       &  -1.224      &   0.078     &   0.000      \\
RTC    &   -0.129    &   0.851     &     0.908     &   0.136     &  1.267      &       0.921   &     0.073   &  0.643 &     0.947    \\
\bottomrule
\end{tabular}
}
\end{center}
\footnotesize{\emph{Note.} The coverage is calculated for the 95\% confidence interval based on 1000 replications.}

\end{table}

In Table~\ref{tab:stats}, we report main statistics for the coefficient estimates after 20000 iterations of the five algorithms.
For all the coefficients, IV-Greedy results in almost zero bias and the corresponding confidence intervals achieve proper coverage of the true values (the coverage is close to 95\% for 95\% confidence intervals).
Note that Naive-IV-Greedy produces both substantially biased estimates and confidence intervals with zero coverage,
which, again, demonstrates the endogeneity spillover issue.
Note also that the quality of the inference results produced by RTC is fair, but the SD is much larger than IV-Greedy.

\section{Conclusions}\label{sec:conclusions}
We study the dynamic selection problems that arise in algorithmic decision-making when the data has endogeneity problems. In a contextual bandit model, endogeneity in data influences the arms pulled, causing a non-random sampling of data that results in self-fulfilling bias.
We correct for the bias by incorporating IVs to existing online learning algorithms. The new algorithms lead to the true parameter values and meanwhile attain low (logarithmic-like) regret levels.
We also prove a central limit theorem for statistical inference of the parameters of interest.
To establish these properties, we develop a general technique that untangles the interdependence between data and actions.

The use of IVs corrects the bias because they perturb the data-generating process.
Therefore, our approach features ``ex-ante exploration'', which contrasts with the ``ex-post exploration''---that is, randomizing the choices given to the users once they arrive. The ex-post exploration, typically under the name of exploration, is the focus of most research.
However, in the presence of endogenous data, our analysis highlights the importance of ex-ante exploration. Understanding how best to carry out ex-ante exploration and how these two types of exploration interact offer fruitful venues for future research.

\bibliographystyle{aea}
\bibliography{ref}

\clearpage
\begin{appendix}
\numberwithin{equation}{section}

\begin{center}
\Large{For Online Publication Only.}
\end{center}

\section{Proof of Proposition~\ref{prop:multiplicity}}

By direct calculations and mathematical induction, we can show that for any $k\geq 1$ and  $b\in[0,1]$,
\[
\Cov[v_t,v_t^k|v_t>b] = \frac{(1-b)^2}{2(k+1)(k+2)} \sum_{i=0}^{k-1}(k-i)(i+1)b^i.
\]
Consequently,
\begin{equation}\label{eq:SF-bias}
\frac{\Cov[v_t, \varepsilon_t|v_t > b]}{\Var[v_t|v_t > b]} =
\frac{\sum_{k=1}^n \beta_k \Cov[v_t, v_t^k|v_t > b]}{\Cov[v_t, v_t|v_t > b]}
= \sum_{k=1}^n \frac{6 \beta_k}{(k+1)(k+2)}\sum_{i=0}^{k-1}(k-i)(i+1)b^i .
\end{equation}

By setting $b=0$ in equation~\eqref{eq:SF-bias}, we derive the limit of the OLS estimate of $\alpha$ in the following manner:
\[
\widehat{\alpha}_{\mathrm{OLS}} \coloneqq \alpha+ \frac{\Cov[v_t, \varepsilon_t]}{\Var[v_t]} = \alpha + \frac{\Cov[v_t, \varepsilon_t|v_t > 0]}{\Var[v_t|v_t > 0]} =  \alpha + \sum_{k=1}^n \frac{6k}{(k+1)(k+2)}\beta_k.
\]
Moreover, applying equation~\eqref{eq:SF-bias} to equation~\eqref{eq:bias} with $b=c/\widehat{\alpha}$ yields that
\begin{equation}\label{eq:self-1}
    \widehat{\alpha}^n -  \biggl( \alpha + \sum_{k=1}^n \frac{6 k \beta_k}{(k+1)(k+2)} \biggr)\widehat{\alpha}^{n-1} -  \sum_{i=1}^{n-1} \biggl( \sum_{k=i+1}^n \frac{6 (k-i) \beta_k}{(k+1)(k+2)}\biggr) (i+1) c^i \widehat{\alpha}^{n-1-i} =0,
\end{equation}
which is a $n$-th degree polynomial equation in terms of $\widehat{\alpha}$.

Assume that equation~\eqref{eq:self-1} has $n$ real roots: $\{r_k: k=1,\ldots,n\}$.
Then, $\prod_{k=1}^n (\widehat{\alpha}- r_k) = 0$, that is,
\begin{equation}\label{eq:self-2}
\widehat{\alpha}^{n} +
\sum_{i=1}^n \biggl( \sum_{j_1,\ldots, j_i} \prod_{k=1}^i r_{j_k} \biggr)(-1)^i \widehat{\alpha}^{n-i}
=0,
\end{equation}
where   $\{j_1,\ldots,j_i\}$ are mutually different numbers in $\{1,\ldots,n\}$ for each $i=1,\ldots,n$.

By matching the coefficients of $\widehat{\alpha}^{n-i}$ in equation~\eqref{eq:self-1} and equation~\eqref{eq:self-2} for each $i=1,\ldots,n$, we have
\begin{equation}\label{eq:coeff-roots}
\left\{
\begin{aligned}
-\biggl(\frac{6 \beta_n }{(n+1)(n+2)} \biggr)n c^{n-1}
={}& (-1)^n \prod_{j=1}^n r_j
\qquad (\mbox{matching for } \widehat{\alpha}^0), \\
- \biggl(\frac{6\beta_{n-1}}{n(n+1)} + \frac{12 \beta_n}{(n+1)(n+2)} \biggr) (n-1) c^{n-2}
={}& (-1)^{n-1}  \sum_{j_1,\ldots, j_{n-1}} \prod_{k=1}^{n-1} r_{j_k}
\qquad (\mbox{matching for } \widehat{\alpha}^1), \\
& \vdots \\
- \biggl( \sum_{k=2}^n \frac{6 (k-1) \beta_k}{(k+1)(k+2)}\biggr) 2 c
={}& (-1)^2 \sum_{j_1\neq j_2} r_{j_1}r_{j_2}
\qquad (\mbox{matching for } \widehat{\alpha}^{n-2}), \\
- \biggl(\alpha + \sum_{k=1}^n \frac{6 k \beta_k}{(k+1)(k+2)}\biggr) ={}& (-1)\sum_{j=1}^n r_j
\qquad (\mbox{matching for } \widehat{\alpha}^{n-1}).
\end{aligned}
\right.
\end{equation}
This is a system of linear equations in $\{\beta_k:k=1,\ldots,n\}$ and can easily be solved.
In particular, note that the first equation in \eqref{eq:coeff-roots} involves only $\beta_n$, the second involves only $(\beta_{n-1},\beta_n)$, and so forth.
Thus, we begin with the first equation,  calculating the value of $\beta_n$, then plug its value into the second equation to solve for $\beta_{n-1}$, and continue with the other equations in the same fashion to solve for $\beta_{n-2},\ldots,\beta_2, \beta_1$ one at a time.

Further, we may set $\beta_0$ such that  $\E[\varepsilon_t] = \sum_{k=0}^n \beta_k v_t^k = 0$, that is,
\begin{equation}\label{eq:coeff-roots2}
\beta_0 = - \sum_{k=1}^n \beta_k \E[v_t^k] = - \sum_{k=1}^n \frac{1}{k+1}\beta_k.
\end{equation}

Therefore,
for any arbitrary sequence $\{r_k: k=1,\ldots,n\}$ such that $c/r_k\in(0,1)$ for each $k$,
there exist a sequence $\{\beta_k:k=0,1,\ldots,n\}$---given by equations \eqref{eq:coeff-roots} and \eqref{eq:coeff-roots2}---such that $\{r_k: k=1,\ldots,n\}$ are $n$ real fixed points of equation~\eqref{eq:bias}.
Note that our method of constructing multiple self-fulling biases can be generalized. For an arbitrary number of self-fulling biases with arbitrary values, we can again let the covariates be uniformly distributed and construct the noise as a higher-order polynomial of the covariates.
The coefficients of the polynomial can again be solved via linear equations.

\section{Concentration Inequalities}

\begin{definition}
[Singular Values]
Let $\BFM\in\Real^{k\times \ell}$ be a real matrix.
If $k\geq \ell$, then
a non-negative real number $\SFs$ is said to be a \emph{singular value} of $\BFM$ if $\SFs^2$ is an eigenvalue of $\BFM^\intercal \BFM\in \Real^{\ell\times\ell}$.
If $k<\ell$, then
a non-negative real number $\SFs$ is said to be a singular value of $\BFM$ if $\SFs^2$ is an eigenvalue of $\BFM \BFM^\intercal \in\Real^{k\times k}$.
Hence, $\BFM\in\Real^{k\times \ell}$ has $\min\{k,\ell\}$ singular values in total.
Let $\phi_{\max}(\BFM)$ and $\phi_{\min}(\BFM)$ denote the  largest and smallest singular values of $\BFM$,  respectively.
\end{definition}

\begin{definition}
[Matrix Norms]
The \emph{spectral norm} of a matrix $\BFM\in\Real^{k\times \ell}$  is defined as
$
\|\BFM\| \coloneqq
\sup_{\BFu\in\Real^\ell: \|\BFu\|=1}\|\BFM\BFu\|
$,
and its \emph{max norm} is defined as
$\|\BFM\|_{\max} \coloneqq \max_{1\leq i\leq k,1\leq j\leq \ell}|M_{ij}|$,
where $M_{ij}$ denotes the $(i,j)$-th entry of $\BFM$.
\end{definition}
\begin{definition}
[Sub-Gaussian Random Variables]
A random variable $X$ is said to be \emph{sub-Gaussian} with variance proxy $\varsigma^2$
if $\E\bigl[e^{\lambda(X-\mathbb{E}[X])} \bigr] \leq e^{\lambda^2\varsigma^2/2}$
for all $\lambda\in\Real$.
\end{definition}

\begin{definition}[Sub-Exponential Random Variables]\label{def:sub-expon}
A random variable $X$ is said to be
\emph{sub-exponential} if its \emph{sub-exponential norm},  defined as
$
\|X\|_{\psi_1} \coloneqq \inf\left\{w>0: \E\bigl[e^{|X|/w}\bigr]\leq 2\right\}$, is finite.
\end{definition}

\begin{lemma}\label{lemma:var-char-SV}
For any matrix $\BFM\in\Real^{k\times \ell}$,
\[
\phi_{\max}(\BFM) = \sup_{\BFu\in\Real^\ell: \|\BFu\|=1}\|\BFM\BFu\|
\qq{and}
\phi_{\min}(\BFM) = \inf_{\BFu\in\Real^\ell: \|\BFu\|=1}\|\BFM\BFu\|,
\]
where $\|\BFu\|$ denotes the Euclidean norm of $\BFu$.
\end{lemma}
\begin{proof}[Proof.]
See page 78 of \cite{GolubLoan13}.
\end{proof}

\begin{lemma}\label{lemma:singular-value}
For any matrices $\BFM\in\Real^{k\times \ell}$ and  $\BFN\in\Real^{\ell\times m}$, we have
\begin{equation}\label{eq:SV-1}
\phi_{\max}(\BFM\BFN) \leq \phi_{\max}(\BFM) \phi_{\max}(\BFN)\qq{and} \phi_{\min}(\BFM\BFN) \geq \phi_{\min}(\BFM) \phi_{\min}(\BFN).
\end{equation}
If $\BFM$ is an invertible square matrix, then
\begin{equation}\label{eq:SV-2}
\phi_{\max}(\BFM^{-1}) = (\phi_{\min}(\BFM))^{-1}
\qq{and} \phi_{\min}(\BFM^{-1}) = (\phi_{\max}(\BFM))^{-1}.
\end{equation}
\end{lemma}
\begin{proof}[Proof.]
The inequalities in \eqref{eq:SV-1} follow from Theorem~3 of \cite{WangXi97}, and
the identities in \eqref{eq:SV-2} follow from the definition of singular values.
\end{proof}

\begin{lemma}\label{lemma:matrix-norm}
For any matrix $\BFM\in\Real^{k\times \ell}$, $\|\BFM\| \leq \sqrt{k\ell} \|\BFM\|_{\max}$.
\end{lemma}
\begin{proof}[Proof.]
See page 72 of \cite{GolubLoan13}.
\end{proof}

\begin{lemma}\label{lemma:Hoeffding_matrix}
Let $\{\BFM_i\}_{i=1}^N$ be a sequence of zero-mean independent $k$-by-$\ell$ random matrices.
For each $i=1,\ldots,N$,
assume that $\|\BFM_i\| \leq b$ almost surely for some constant $b>0$.
Then, for all $\tau\geq 0$,
\begin{gather*}
\Pr(\biggl\|\frac{1}{N}\sum_{i=1}^N \BFM_i \biggr\| \geq \tau ) \leq 2(k+\ell) \exp(\frac{- N \tau^2  }{2 b^2}),
\end{gather*}
Moreover, if $k=\ell$, then the leading constant $2(k+\ell)$ on the right-hand-side of the above inequalities can be reduced to $2k$.
\end{lemma}
\begin{proof}[Proof.]
See Hoeffding's inequality for bounded random matrices
\citep[pages~174--175]{Wainwright19}.
\end{proof}

\begin{corollary}\label{coro:Hoeffding_matrix}
Let $\{\BFM_i\}_{i=1}^N$ be a sequence of independent $k$-by-$\ell$ random matrices.
For each $i=1,\ldots,N$,
assume that $\|\BFM_i\| \leq b$ almost surely for some constant $b>0$.
Then, for all $\tau\geq 0$,
\begin{gather}
\Pr(\phi_{\max}\biggl(\frac{1}{N}\sum_{i=1}^N \BFM_i \biggr) \geq \phi_{\max}(\mathbb{E}[\BFM_1])+\tau ) \leq 2(k+\ell) \exp(\frac{- N \tau^2  }{8 b^2}),  \label{eq:max-SV-bound} \\
\Pr(\phi_{\min}\biggl(\frac{1}{N}\sum_{i=1}^N \BFM_i\biggr) \leq \phi_{\min}(\mathbb{E}[\BFM_1])-\tau ) \leq 2(k+\ell) \exp(\frac{- N \tau^2  }{8 b^2}). \label{eq:min-SV-bound}
\end{gather}
Moreover, if $k=\ell$, then the leading constant $2(k+\ell)$ on the right-hand-side of the above inequalities can be reduced to $2k$.
\end{corollary}

\begin{proof}[Proof.]
Let $\widetilde{\BFM}_i = \BFM_i - \mathbb{E}[\BFM_i]$. Then, $\|\widetilde{\BFM}_i\|\leq \|{\BFM}_i\|+\|\mathbb{E}[\BFM_i] \|\leq 2b$.
By Lemmas~\ref{lemma:var-char-SV} and \ref{lemma:Hoeffding_matrix},
\begin{equation}\label{eq:phi-max-M}
    \phi_{\max}\biggl(\frac{1}{N}\sum_{i=1}^N \widetilde{\BFM}_i\biggr) = \biggl\| \frac{1}{N}\sum_{i=1}^N \widetilde{\BFM}_i \biggr\| \leq \tau
\end{equation}
holds with probability at least $1- 2(k+\ell)\exp(-\frac{N\tau^2}{8b^2})$.

Given the event  \eqref{eq:phi-max-M} holds,  it follows from Lemma~\ref{lemma:var-char-SV} that
\begin{align*}
    \phi_{\max}\biggl(\frac{1}{N}\sum_{i=1}^N \BFM_i\biggr)  ={}& \sup_{\BFu\in\Real^\ell:\|\BFu\|=1}\biggl\|\frac{1}{N}\sum_{i=1}^N {\BFM}_i \BFu \biggr\|  \\
    \leq{}& \sup_{\BFu\in\Real^\ell:\|\BFu\|=1} \|\mathbb{E}[\BFM_1]\BFu\|  + \sup_{\BFu\in\Real^\ell:\|\BFu\|=1} \biggl\|\frac{1}{N}\sum_{i=1}^N \widetilde{\BFM}_i \BFu \biggr\|   \\
     \leq{} &  \phi_{\max}(\mathbb{E}[\BFM_1])+\tau
\end{align*}
holds with probability at least $1- 2(k+\ell)\exp(-\frac{N\tau^2}{8b^2})$, proving \eqref{eq:max-SV-bound}.
Likewise, given the event  \eqref{eq:phi-max-M} holds,
\begin{align*}
    \phi_{\min}\biggl(\frac{1}{N}\sum_{i=1}^N \BFM_i\biggr)  ={}& \inf_{\BFu\in\Real^\ell:\|\BFu\|=1}\biggl\|\frac{1}{N}\sum_{i=1}^N {\BFM}_i \BFu \biggr\|  \\
    \geq{}& \inf_{\BFu\in\Real^\ell:\|\BFu\|=1} \|\mathbb{E}[\BFM_1]\BFu\|  + \sup_{\BFu\in\Real^\ell:\|\BFu\|=1} \biggl\|\frac{1}{N}\sum_{i=1}^N \widetilde{\BFM}_i \BFu \biggr\|   \\
     \geq{} &  \phi_{\min}(\mathbb{E}[\BFM_1]) - \tau,
\end{align*}
which proves \eqref{eq:min-SV-bound}.
\end{proof}

\begin{lemma}\label{lemma:Hoeffding_SG}
Let $\{X_i\}_{i=1}^N$ be a sequence of i.i.d. zero-mean sub-Gaussian random variables with variance proxy $\varsigma^2$. Then, for all $\tau\geq 0$,
\[
\Pr(\biggl| \frac{1}{N}\sum_{i=1}^N X_i \biggr| \geq \tau ) \leq 2\exp(-\frac{N\tau^2  }{2\varsigma^2}).
\]
\end{lemma}

\begin{proof}[Proof.]
See Hoeffding's inequality for sub-Gaussian random variables \citep[page 24]{Wainwright19}.
\end{proof}

\begin{corollary}\label{coro:Hoeffding_vector}
Let $\BFg\in\Real^k$ be a vector of zero-mean sub-Gaussian random variables with variance proxy $\varsigma^2$. Let $\{\BFg_i\}_{i=1}^N$ be a sequence of i.i.d. random vectors having the same distribution as $\BFg$.
Then, for all $\tau\geq 0$,
\[
\Pr(\biggl\| \frac{1}{N}\sum_{i=1}^N \BFg_i  \biggr\| \geq \tau) \leq  2k\exp(-\frac{N\tau^2}{2 k^2\varsigma^2}).
\]
\end{corollary}

\begin{proof}[Proof.]
Let $g_{ij}$ denote the $j$-th entry of $\BFg_i$, for all $i=1,\ldots,N$ and $j=1,\ldots,k$.
Then, for each $j$, $\{g_{ij}\}_{i=1}^N$ is a sequence of i.i.d. zero-mean sub-Gaussian random variables with variance proxy $\varsigma^2$.
Note that if  $|\frac{1}{N}\sum_{i=1}^N g_{ij} | \leq \frac{\tau}{k}$ for all $j=1,\ldots,k$, then
$\| \frac{1}{N}\sum_{i=1}^N \BFg_i \| \leq \tau $.
Hence,
\begin{align*}
\Pr(\biggl\| \frac{1}{N}\sum_{i=1}^N \BFg_i  \biggr\| \geq \tau) \leq{}&
\Pr(\biggl|\frac{1}{N}\sum_{i=1}^N g_{ij} \biggr| \geq \frac{\tau}{k}\mbox{ for some }j=1,\ldots,k)  \\
\leq{}& \sum_{j=1}^k \Pr(\biggl|\frac{1}{N}\sum_{i=1}^N g_{ij} \biggr| \geq \frac{\tau}{k})
\leq  2k\exp(-\frac{N\tau^2}{2 k^2\varsigma^2}),
\end{align*}
where the last inequality follows from Lemma~\ref{lemma:Hoeffding_SG}.
\end{proof}

\begin{lemma}\label{lemma:Bernstein}
Let $\{X_i\}_{i=1}^N$ be a sequence of i.i.d. zero-mean sub-exponential random variables. Then, there exists a constant $\Cbern>0$ such that for all $\tau\geq 0$,
\[
\Pr\biggl(\biggl| \frac{1}{N}\sum_{i=1}^N X_i\biggr| \geq \tau \biggr) \leq 2\exp(- N \Cbern \cdot \min\biggl\{\frac{ \tau^2 }{\|X_1\|_{\psi_1}^2},\frac{\tau}{\|X_1\|_{\psi_1}}\biggr\} ).
\]
\end{lemma}

\begin{proof}[Proof.]
See Bernstein's inequality for sub-exponential random variables \citep[Theorem~2.8.2]{Vershynin18}.
\end{proof}

\section{Coefficient Stabilization Phase: $t=1,\ldots, \ttR$}
Throughout the rest of the Supplemental Material,
we let $\ttR= C_{\ttR}\log(T)$ and $\ttU = (C_{\ttR}+C_{\ttU})\log(T)$ for some constant $C_{\ttU}$ for some constants $C_{\ttR}$ and $C_{\ttU}$;
we also suppose that Assumption~\ref{assump:IV} holds and the data are generated by Algorithm~\ref{algo:IV-bandits}.

We first define a set of notation before the analysis.
For any $t$, let
\begin{gather*}
\widetilde{\BFv}_{t}\coloneqq
\begin{pmatrix}
\ind(a_t=1)\BFv_t  \\
\vdots\\
\ind(a_t=K)\BFv_t
\end{pmatrix}\in\Real^{Kp},
\quad
\widetilde{\BFz}_{t}\coloneqq
\begin{pmatrix}
\ind(a_t=1)\BFz_t  \\
\vdots \\
\ind(a_t=K)\BFz_t
\end{pmatrix}\in\Real^{Kq}, \\
\widetilde{\BFV}_{t}\coloneqq \begin{pmatrix}
\widetilde{\BFv}_{1}^\intercal \\
\vdots\\
\widetilde{\BFv}_{t}^\intercal
\end{pmatrix}\in\Real^{t\times (Kp)},
\quad
\widetilde{\BFZ}_{t}\coloneqq \begin{pmatrix}
\widetilde{\BFz}_{1}^\intercal \\
\vdots\\
\widetilde{\BFz}_{t}^\intercal
\end{pmatrix}\in\Real^{t\times (Kq)},
\quad
\BFvarepsilon_{t}\coloneqq \begin{pmatrix}
\varepsilon_{1} \\
\vdots\\
\varepsilon_{t}
\end{pmatrix}\in\Real^{t}, \\
\widetilde{\BFSigma}_{vz,t} \coloneqq \frac{1}{t}\sum_{s=1}^t \widetilde{\BFv}_s \widetilde{\BFz}_s^\intercal =  \frac{1}{t}\widetilde{\BFV}_{t}^\intercal\widetilde{\BFZ}_{t} \in \Real^{(Kp)\times (Kq)},\\
\widetilde{\BFSigma}_{zz,t}\coloneqq \frac{1}{t}\sum_{s=1}^t \widetilde{\BFz}_s \widetilde{\BFz}_s^\intercal =
\frac{1}{t}\widetilde{\BFZ}_{t}^\intercal \widetilde{\BFZ}_{t}\in \Real^{(Kq)\times (Kq)}, \\
\widetilde{\BFG}_{t}\coloneqq \frac{1}{t}\sum_{s=1}^t \widetilde{\BFz}_s \varepsilon_s = \frac{1}{t}\widetilde{\BFZ}_{t}^\intercal \BFvarepsilon_{t}\in\Real^{2q}, \\
\BFSigma_{vz}\coloneqq \mathbb{E}[\BFv_t  \BFz_t^\intercal]     \in\Real^{p\times q},
\quad
\BFSigma_{zz}\coloneqq \mathbb{E}[\BFz_t \BFz_t^\intercal]\in\Real^{q\times q}, \\
\breve{\BFSigma}_{vz}\coloneqq \mathbb{E}[\widetilde{\BFv}_t  \widetilde{\BFz}_t^\intercal]     \in\Real^{(Kp)\times (Kq)},
\quad
\breve{\BFSigma}_{zz}\coloneqq \mathbb{E}[\widetilde{\BFz}_t \widetilde{\BFz}_t^\intercal]\in\Real^{(Kq)\times (Kq)}.
\end{gather*}

\begin{lemma}\label{lemma:full-rank}
Both $\breve{\BFSigma}_{zz}$ and $\breve{\BFSigma}_{vz}$ have full rank.
\end{lemma}

\begin{proof}[Proof.]
By
condition~(iii) of Assumption \ref{assump:IV}, $\BFSigma_{vz}^*=\mathbb{E}[\BFv_t^* \BFz_t^\intercal]$ is full rank and $\BFSigma_{zz}=\mathbb{E}[\BFz_t \BFz_t^\intercal]$ is positive definite. By definition, for $t=1,2,\ldots,\ttR$, $a_t$ is randomly selected
such that $\Pr(a_t=i)  = \frac{1}{K}$ for all $i=1,\ldots,K$.
Thus,
\begin{equation*}
    \breve{\BFSigma}_{vz}=\frac{1}{K}\begin{pmatrix}
    \BFSigma_{zz} & \cdots & \BFzero \\
    \vdots & \ddots & \vdots \\
     \BFzero & \cdots&   \BFSigma_{zz}
    \end{pmatrix},
\end{equation*}
and thus $\breve{\BFSigma}_{vz}$ is full rank and positive definite.

Because  $\sum_{i=1}^K\ind(a_t^*=i)=1$ and $\BFSigma_{vz}^*$ is a full-rank matrix of size $(Kp)\times q$,
$\BFSigma_{vz}=\begin{pmatrix}
    \BFI_p & \cdots &  \BFI_p
    \end{pmatrix} \BFSigma_{vz}^*$
is a full-rank matrix of size $p\times q$, where $\BFI_p$ is the identity matrix of size $p\times p$.
As a result,
\begin{align*}
    \breve{\BFSigma}_{vz} &= \mathbb{E}[\widetilde{\BFv}_t \widetilde{\BFz}_t^\intercal]
     = \begin{pmatrix}
    \mathbb{E}[\ind(a_t=1)\BFv_t \BFz_t^\intercal] & \cdots & \BFzero \\
    \vdots & \ddots & \vdots \\
    \BFzero &  \cdots & \mathbb{E}[\ind(a_t=2)\BFv_t \BFz_t^\intercal]
    \end{pmatrix}
     = \frac{1}{K}\begin{pmatrix}
    \BFSigma_{vz} & \cdots &  \BFzero \\
    \vdots & \ddots & \vdots \\
    \BFzero & \cdots &  \BFSigma_{vz}
    \end{pmatrix}.
\end{align*}
Thus, $\breve{\BFSigma}_{vz}$ is a full-rank matrix of size $(Kp)\times (Kq)$.
\end{proof}

\begin{lemma}\label{lemma:RanP}
Define
\begin{equation}\label{eq:kappa-breve}
\breve{\kappa}\coloneqq  \frac{2\phi_{\max}(\breve{\BFSigma}_{vz})\phi_{\max}(\breve{\BFSigma}_{zz})}{ (\phi_{\min}(\breve{\BFSigma}_{vz}))^2\phi_{\min}(\breve{\BFSigma}_{zz}) }
\end{equation}
and
\begin{equation}\label{eq:def-A_T1}
\scrA_{\ttR}\coloneqq  \Bigl\{\bigl\|\widehat{\BFalpha}_{\ttR} - \BFalpha \bigr\| \leq  2 K q\bar{z}\varsigma \breve{\kappa}  C_{\ttR}^{-\frac{1}{2}} \Bigr\}.
\end{equation}
Suppose
\begin{equation}\label{eq:const-C_T1}
C_{\ttR} \geq \left(\frac{40K \max\{\sqrt{pq}\bar{v}\bar{z},\; q\bar{z}^2\}}{\min\{\phi_{\min}(\breve{\BFSigma}_{vz}),\phi_{\min}(\breve{\BFSigma}_{zz})\}}\right)^2.
\end{equation}
Then, $\Pr(\scrA_{\ttR})\geq 1- 2K (2p + 5 q) T^{-2} $  for all $T\geq \ttR$.
\end{lemma}

\begin{proof}[Proof.]
It follows from Lemma~\ref{lemma:full-rank} that
$0<\phi_{\min}(\breve{\BFSigma}_{zz})\leq \phi_{\max}(\breve{\BFSigma}_{zz})<\infty$ and  $0<\phi_{\min}(\breve{\BFSigma}_{vz})\leq \phi_{\max}(\breve{\BFSigma}_{vz})<\infty$.
Hence, $\breve{\kappa}$ is a positive finite constant.

Because in the coefficient stabilization phase
$a_t$ is randomly selected for all $t=1,\ldots,\ttR$,
we may rewrite the arm-specific 2SLS for the $K$ arms together equivalently as the joint-2SLS using the notation $\widetilde{\BFV}_{\ttR}$  and $\widetilde{\BFZ}_{\ttR}$.
Hence,
\begin{align*}
    \widehat{\BFalpha}_{\ttR} - \BFalpha = {}&   \left(\widetilde{\BFV}_{\ttR}^\intercal \calP[\widetilde{\BFZ}_{\ttR}] \widetilde{\BFV}_{\ttR}\right)^{-1}\widetilde{\BFV}_{\ttR}^\intercal \calP[\widetilde{\BFZ}_{\ttR}]\BFvarepsilon_{\ttR} \nonumber \\
    ={}& \Bigl(\widetilde{\BFV}_{\ttR}^\intercal \widetilde{\BFZ}_{\ttR} (\widetilde{\BFZ}_{\ttR}^\intercal \widetilde{\BFZ}_{\ttR})^{-1}
    \widetilde{\BFZ}_{\ttR}^\intercal  \widetilde{\BFV}_{\ttR}\Bigr)^{-1}
    \widetilde{\BFV}_{\ttR}^\intercal \widetilde{\BFZ}_{\ttR} (\widetilde{\BFZ}_{\ttR}^\intercal \widetilde{\BFZ}_{\ttR})^{-1}\widetilde{\BFZ}_{\ttR}^\intercal\BFvarepsilon_{\ttR} \nonumber
    \\
    ={}& \left(\widetilde{\BFSigma}_{vz,\ttR} \widetilde{\BFSigma}_{zz,\ttR}^{-1}\widetilde{\BFSigma}_{vz,\ttR}^\intercal\right)^{-1}\widetilde{\BFSigma}_{vz,\ttR} \widetilde{\BFSigma}_{zz,\ttR}^{-1} \widetilde{\BFG}_{\ttR},
\end{align*}
where the projection operator $\calP$ maps a matrix $\BFZ$ to $\BFZ(\BFZ^\intercal \BFZ)^{-1} \BFZ^\intercal$.
Therefore,
\begin{equation} \label{eq:norm-alpha_est_T1}
\|\widehat{\BFalpha}_{\ttR} - \BFalpha\| \leq \Bigl\| \left(\widetilde{\BFSigma}_{vz,\ttR} \widetilde{\BFSigma}_{zz,\ttR}^{-1}\widetilde{\BFSigma}_{vz,\ttR}^\intercal\right)^{-1}\widetilde{\BFSigma}_{vz,\ttR} \widetilde{\BFSigma}_{zz,\ttR}^{-1}\Bigr\| \cdot \bigl\| \widetilde{\BFG}_{\ttR} \bigr\|
\end{equation}
In what follows, we analyze the two norms on the right-hand-side of \eqref{eq:norm-alpha_est_T1} separately.

Because the samples are i.i.d.,
\begin{gather*}
\mathbb{E}[\widetilde{\BFSigma}_{vz,\ttR}]=\mathbb{E}[\widetilde\BFv_s  \widetilde{\BFz}_s^\intercal]= \breve{\BFSigma}_{vz} \qq{and}
\mathbb{E}[\widetilde{\BFSigma}_{zz,\ttR}] = \mathbb{E}[\widetilde{\BFz}_s \widetilde{\BFz}_s^\intercal ] = \breve{\BFSigma}_{zz}.
\end{gather*}
Moreover, because  $\widetilde{\BFv}_s \widetilde{\BFz}_s^\intercal $ is of size $(Kp)$-by-$(Kq)$, we know by Lemma~\ref{lemma:matrix-norm} that
\begin{align*}
\|\widetilde{\BFv}_s \widetilde{\BFz}_s^\intercal \| \leq  \sqrt{(Kp)\times (Kq)} \|\widetilde{\BFv}_s \widetilde{\BFz}_s^\intercal\|_{\max} \leq
K\sqrt{pq} \bar{v}\bar{z}.
\end{align*}
Likewise, $\|\widetilde{\BFz}_s \widetilde{\BFz}_s^\intercal\| \leq Kq \bar{z}^2$.
Hence, if  we define the following events:
\begin{align*}
\scrS_{vz, \min} \coloneqq{}& \Bigl\{\phi_{\min}(\widetilde{\BFSigma}_{vz,\ttR})\leq \phi_{\min}(\breve{\BFSigma}_{vz})-\tau\Bigr\},  \\
\scrS_{vz, \max} \coloneqq{}&
\Bigl\{\phi_{\max}(\widetilde{\BFSigma}_{vz,\ttR})\geq \phi_{\max}(\breve{\BFSigma}_{vz})+\tau\Bigr\}, \\
\scrS_{zz, \min} \coloneqq{}&
\Bigl\{\phi_{\min}(\widetilde{\BFSigma}_{zz,\ttR})\leq \phi_{\min}(\breve{\BFSigma}_{zz})-\tau\Bigr\}, \\
\scrS_{zz, \max} \coloneqq{}&
\Bigl\{\phi_{\max}(\widetilde{\BFSigma}_{zz,\ttR})\geq  \phi_{\max}(\breve{\BFSigma}_{zz})+\tau\Bigr\},
\end{align*}
then we can apply Corollary~\ref{coro:Hoeffding_matrix} to conclude that for all $\tau\geq 0$,
\begin{gather*}
    \Pr(\scrS_{vz, \min}) \leq 2K(p+q)\exp(-\frac{  \tau^2 \ttR}{ 8K^2 pq\bar{v}^2 \bar{z}^2 }),\\
    \Pr(\scrS_{vz, \max}) \leq 2K(p+q)\exp(-\frac{  \tau^2 \ttR}{ 8K^2 pq\bar{v}^2 \bar{z}^2}),\\
    \Pr(\scrS_{zz, \min}) \leq 2K q\exp(-\frac{  \tau^2 \ttR}{  8K^2 q^2\bar{z}^4 }),\\
    \Pr(\scrS_{zz, \max}) \leq 2K q\exp(-\frac{  \tau^2 \ttR}{  8K^2 q^2\bar{z}^4}).
\end{gather*}

Setting $\tau = \zeta_1\coloneqq  4K \max\{\sqrt{pq}\bar{v}\bar{z},\; q\bar{z}^2\}C_{\ttR}^{-\frac{1}{2} } $ leads to
\begin{gather*}
    \exp(-\frac{ \zeta_1^2 \ttR}{8K^2 pq\bar{v}^2\bar{z}^2})\leq T^{-2} \qq{and}
    \exp(-\frac{ \zeta_1^2 \ttR}{8K^2 q^2\bar{z}^4})\leq T^{-2}.
\end{gather*}
Moreover, by the condition \eqref{eq:const-C_T1}, we have
\begin{equation}\label{eq:upper-bound-eta}
\zeta_1 = 4K \max\{\sqrt{pq}\bar{v}\bar{z}, q\bar{z}^2\}C_{\ttR}^{-\frac{1}{2} } \leq  \frac{1}{10}\min\left\{\phi_{\min}(\breve{\BFSigma}_{vz}),\phi_{\min}(\breve{\BFSigma}_{zz})\right\}.
\end{equation}
Therefore, upon  defining the joint event
$\scrS_{\phi,\ttR}\coloneqq  \scrS_{vz, \min}^{\SFc}\cap \scrS_{vz, \max}^{\SFc} \cap \scrS_{zz, \min}^{\SFc} \cap \scrS_{zz, \max}^{\SFc}$,
it follows from the Bonferroni inequality that
\begin{align}
\Pr(\scrS_{\phi,\ttR}) \geq{}&
1 - \bigl[\Pr(\scrS_{vz, \min}) + \Pr(\scrS_{vz, \max}) +
\Pr(\scrS_{zz, \min}) + \Pr(\scrS_{zz, \max})\bigr] \nonumber \\
\geq {}&
1-4K(p+2q) T^{-2}.  \label{eq:prob-A_phi_T1}
\end{align}

Note that conditional on the event $\scrS_{\phi,\ttR}$ with $\tau=\zeta_1$, we have
\begin{equation}\label{eq:phi-pm-eta}
\begin{gathered}
\phi_{\min}(\widetilde{\BFSigma}_{vz,\ttR})> \phi_{\min}(\breve{\BFSigma}_{vz})-\zeta_1>0, \quad
\phi_{\max}(\widetilde{\BFSigma}_{vz,\ttR}) < \phi_{\max}(\breve{\BFSigma}_{vz})+\zeta_1, \\
\phi_{\min}(\widetilde{\BFSigma}_{zz,\ttR}) >  \phi_{\min}(\breve{\BFSigma}_{zz})-\zeta_1>0, \quad
\phi_{\max}(\widetilde{\BFSigma}_{zz,\ttR})<  \phi_{\max}(\breve{\BFSigma}_{zz})+\zeta_1.
\end{gathered}
\end{equation}
This implies that
\begin{align}
& \Bigl\|\Bigl(\widetilde{\BFSigma}_{vz,\ttR} \widetilde{\BFSigma}_{zz,\ttR}^{-1}\widetilde{\BFSigma}_{vz,\ttR}^\intercal\Bigr)^{-1}\widetilde{\BFSigma}_{vz,\ttR} \widetilde{\BFSigma}_{zz,\ttR}^{-1} \Bigr\| \nonumber \\
={}&
    \phi_{\max}\Bigl(\left(\widetilde{\BFSigma}_{vz,\ttR} \widetilde{\BFSigma}_{zz,\ttR}^{-1}\widetilde{\BFSigma}_{vz,\ttR}^\intercal\right)^{-1}\widetilde{\BFSigma}_{vz,\ttR} \widetilde{\BFSigma}_{zz,\ttR}^{-1}\Bigr) \nonumber \\
    \leq{}&  \frac{\phi_{\max}(\widetilde{\BFSigma}_{vz,\ttR})\phi_{\max}(\widetilde{\BFSigma}_{zz,\ttR})}{(\phi_{\min}(\widetilde{\BFSigma}_{vz,\ttR}))^2\phi_{\min}(\widetilde{\BFSigma}_{zz,\ttR})} \label{eq:step1} \\
    \leq{}&   \frac{(\phi_{\max}(\breve{\BFSigma}_{vz})+\zeta_1)(\phi_{\max}(\breve{\BFSigma}_{zz})+\zeta_1)}{(\phi_{\min}(\breve{\BFSigma}_{vz})-\zeta_1)^2(\phi_{\min}(\breve{\BFSigma}_{zz})-\zeta_1)} \label{eq:step2}\\
    \leq {}& \left(\frac{11}{10}\right)^2 \left(\frac{10}{9}\right)^3 \frac{\phi_{\max}(\breve{\BFSigma}_{vz})\phi_{\max}(\breve{\BFSigma}_{zz})}{(\phi_{\min}(\breve{\BFSigma}_{vz}))^2\phi_{\min}(\breve{\BFSigma}_{zz}) } \label{eq:step3} \\
    \leq{}& \breve{\kappa}, \label{eq:step4}
\end{align}
where \eqref{eq:step1} follows from Lemma~\ref{lemma:singular-value},
\eqref{eq:step2} from \eqref{eq:phi-pm-eta},  \eqref{eq:step3} from \eqref{eq:upper-bound-eta},
and \eqref{eq:step4} from \eqref{eq:kappa-breve}.
Therefore,
\begin{equation}\label{eq:bound-on-part1}
\scrS_{\phi,\ttR} \subseteq
\Bigl\{\Bigl\| \left(\widetilde{\BFSigma}_{vz,\ttR} \widetilde{\BFSigma}_{zz,\ttR}^{-1}\widetilde{\BFSigma}_{vz,\ttR}^\intercal\right)^{-1}\widetilde{\BFSigma}_{vz,\ttR} \widetilde{\BFSigma}_{zz,\ttR}^{-1}\Bigr\| \leq \breve{\kappa}\Bigr\}.
\end{equation}

Note that $\mathbb{E}[\widetilde{\BFG}_{\ttR}]=\mathbb{E}[\widetilde{\BFz}_{s}\varepsilon_s]=\BFzero$.
Moreover, note  that $\widetilde{\BFz}_s \varepsilon_s \in\Real^{Kq}$ is a vector of zero-mean sub-Gaussian random variables with variance proxy $\bar{z}^2\varsigma^2$.
Hence,
applying Corollary~\ref{coro:Hoeffding_vector} to  $\widetilde{\BFG}_{\ttR}$ yields that for all $\tau\geq 0$,
\begin{gather*}
    \Pr(\bigl\|\widetilde{\BFG}_{\ttR}\bigr\|\geq \tau)
    \leq 2K q\exp(-\frac{\tau^2 \ttR}{2K^2 q^2\bar{z}^2 \varsigma^2}).
\end{gather*}
Thus,
setting $\tau=\zeta_2\coloneqq 2K q\bar{z} \varsigma C_{\ttR}^{-\frac{1}{2}}$ yields
\begin{gather}\label{eq:prob-G_T1}
\Pr(\bigl\|\widetilde{\BFG}_{\ttR}\bigr\|>\zeta_2)\leq
2K q\exp(- 2 \ttR C_{\ttR}^{-1})
= 2K q \exp(-2 \log(T))
= 2K q T^{-2}.
\end{gather}

It follows from \eqref{eq:norm-alpha_est_T1} and \eqref{eq:bound-on-part1} that
\[
\scrS_{\phi,\ttR}\cap \bigl\{\bigl\|\widetilde{\BFG}_{\ttR} \bigr\|\leq \zeta_2\bigr\} \subseteq
\bigl\{
\|\widehat{\BFalpha}_{\ttR} - \BFalpha \| \leq  \zeta_2\breve{\kappa}
\bigr\} = \scrA_{\ttR}.
\]
Hence,
\begin{align*}
    \Pr(\scrA_{\ttR})
    \geq{}& \Pr(\scrS_{\phi,\ttR}\cap \bigl\{\|\widetilde{\BFG}_{\ttR}\|\leq \zeta_2\bigr\}) \geq
    1- 2K(2p + 5 q) T^{-2},
\end{align*}
where the second inequality follows from \eqref{eq:prob-A_phi_T1} and \eqref{eq:prob-G_T1}.
\end{proof}

\section{Covariance Stabilization Phase and Beyond: $t>\ttR$}\label{sec:cov-stab}
For any $t$, let
\begin{gather*}
\widetilde{\BFv}_{t}\coloneqq
\begin{pmatrix}
\ind(a_t=1)\BFv_t  \\
\vdots \\
\ind(a_t=K)\BFv_t
\end{pmatrix}\in\Real^{Kp},
\quad
\BFv_{t}^*\coloneqq
\begin{pmatrix}
\ind(a_t^*=1)\BFv_t  \\
\vdots \\
\ind(a_t^*=K)\BFv_t
\end{pmatrix}\in\Real^{Kp}.
\end{gather*}

For any $t_1<t_2$, let
\begin{gather*}
\widetilde{\BFV}_{t_1:t_2}\coloneqq \begin{pmatrix}
\widetilde{\BFv}_{t_1+1}^\intercal \\
\vdots\\
\widetilde{\BFv}_{t_2}^\intercal
\end{pmatrix}\in\Real^{(t_2-t_1)\times (Kp)},
\quad
\widetilde{\BFV}_{t_1:t_2}^*\coloneqq \begin{pmatrix}
(\BFv^*_{t_1+1})^\intercal \\
\vdots\\
(\BFv^*_{t_2})^\intercal
\end{pmatrix}\in\Real^{(t_2-t_1)\times (Kp)},
\\
\BFZ_{t_1:t_2}\coloneqq \begin{pmatrix}
\BFz_{t_1+1}^\intercal \\
\vdots\\
\BFz_{t_2}^\intercal
\end{pmatrix}\in\Real^{(t_2-t_1)\times q},\quad  \BFR_{t_1:t_2}\coloneqq \begin{pmatrix}
R_{t_1+1} \\
\vdots\\
R_{t_2}
\end{pmatrix}\in\Real^{t_2-t_1},
\quad
\BFvarepsilon_{t_1:t_2}\coloneqq \begin{pmatrix}
\varepsilon_{t_1+1} \\
\vdots\\
\varepsilon_{t_2}
\end{pmatrix}\in\Real^{t_2-t_1},\\
\widehat{\BFSigma}_{vz,t_1:t_2} \coloneqq \frac{1}{t_2-t_1}\sum_{s=t_1+1}^{t_2} \widetilde\BFv_s  \BFz_s^\intercal
= \frac{1}{t_2-t_1} \widetilde{\BFV}_{t_1:t_2}^\intercal \BFZ_{t_1:t_2}
\in\Real^{(Kp)\times q}, \\
\widehat{\BFSigma}_{vz,t_1:t_2}^*:=\frac{1}{t_2-t_1}\sum_{s=t_1+1}^{t_2} \BFv_t^* \BFz_t^\intercal = \frac{1}{t_2-t_2}(\BFV_{t_1:t_2}^*)^\intercal \BFZ_{t_1:t_2} \in \Real^{(Kp)\times q},
\\
\widehat{\BFSigma}_{zz,t_1:t_2} \coloneqq \frac{1}{t_2-t_1} \sum_{s=t_1+1}^{t_2} \BFz_s \BFz_s^\intercal
=\frac{1}{t_2-t_1} \BFZ_{t_1:t_2}^\intercal \BFZ_{t_1:t_2} \in\Real^{q\times q} ,\\
\widehat{\BFG}_{t_1:t_2}\coloneqq \frac{1}{t_2-t_1}\sum_{s=t_1+1}^{t_2} \BFz_s \varepsilon_s
= \frac{1}{t_2-t_1} \BFZ_{t_1:t_2}^\intercal \BFvarepsilon_{t_1:t_2}
\in\Real^{q}, \label{eq:def-G-hat} \\
\BFSigma_{vz}^*\coloneqq \mathbb{E}[\BFv^*_t  \BFz_t^\intercal]     \in\Real^{(Kp)\times q},
\quad
\BFSigma_{zz}\coloneqq \mathbb{E}[\BFz_t \BFz_t^\intercal]\in\Real^{q \times q}.
\end{gather*}

Note that for $t>\ttR$, the 2SLS estimator $\widehat{\BFalpha}_t$ in Algorithm~\ref{algo:IV-bandits} satisfies
\begin{align}
\widehat{\BFalpha}_{t} - \BFalpha={}& \bigl(\widetilde{\BFV}_{\ttR:t}^\intercal \calP[\BFZ_{\ttR:t}] \widetilde{\BFV}_{\ttR:t}\bigr)^{-1} \widetilde{\BFV}_{\ttR:t}^\intercal\calP[\BFZ_{\ttR:t}] \BFvarepsilon_{\ttR:t} \nonumber\\
={}& \Bigl(\widehat{\BFSigma}_{vz,{\ttR:t}}\widehat{\BFSigma}_{zz,{\ttR:t}}^{-1} \widehat{\BFSigma}_{vz,{\ttR:t}}^\intercal\Bigr)^{-1}\widehat{\BFSigma}_{vz,\ttR:t}\widehat{\BFSigma}_{zz,{\ttR:t}}^{-1} \widehat{\BFG}_{{\ttR:t}}.\label{eq:main-2sls}
\end{align}
Moreover, recall that in Algorithm~\ref{algo:IV-bandits} we use the following estimator for the variance of noise terms:
\begin{equation}\label{eq:variance-estimator}
\widehat{\sigma}_t \coloneqq (t-\ttR)^{-\frac{1}{2}} \bigl\|\BFR_{\ttR: t} -  \widetilde{\BFV}_{\ttR: t}
\widehat{\BFalpha}_t
\bigr\|\in\Real_+.
\end{equation}

In order to establish an probabilistic bound on $\|\widehat{\BFalpha}_t-\BFalpha\|$, we establish below bounds on a set of events that characterize how accurate the estimators such as  $\widehat{\BFSigma}_{vz,\ttR:t}$, $\widehat{\BFSigma}_{zz,\ttR:t}$, and $\widehat\sigma_t^2$ are relative to their corresponding true values.

Specifically, we fix arbitrary constants $C_{\scrA}, \eta, \Delta>0$,
and define for any $t>\ttR$ the following events:
\begin{gather*}
    \scrA_t\coloneqq \Biggl\{\|\widehat{\BFalpha}_t - \BFalpha \|\leq C_{\scrA}\sqrt{\frac{\log(T)}{t-\ttR}}\Biggr\}, \quad
    \scrB_t\coloneqq \bigl\{\|\widehat{\BFSigma}_{vz,\ttR:t}-\BFSigma_{vz}^*\|\leq \eta\bigr\}, \\
    \scrC_t\coloneqq \bigl\{\|\widehat{\BFSigma}_{zz,\ttR:t}-\BFSigma_{zz}\|\leq \eta\bigr\},  \quad
    \scrD_t\coloneqq \bigl\{|\widehat\sigma_t^2 -\sigma^2|\leq \Delta\bigr\}.
\end{gather*}
In Lemmas~\ref{lemma:conditional-A}--\ref{lemma:iter-bounds} below, we will establish probabilistic bounds of these events and their various intersections.

\begin{lemma}
\label{lemma:conditional-A}
Define
\begin{equation}\label{eq:kappa-bar}
\bar{\kappa} \coloneqq  \frac{2\phi_{\max}(\BFSigma_{vz}^*)\phi_{\max}(\BFSigma_{zz})}{(\phi_{\min}(\BFSigma_{vz}^*))^2 \phi_{\min}(\BFSigma_{zz})}.
\end{equation}
Suppose that  the constants $C_{\scrA}$ and $\eta$ in the definitions of $\scrA_t$, $\scrB_t$ and
$\scrC_t$ satisfy
\begin{gather}
C_{\scrA} \geq 2q \bar{z}\varsigma \bar{\kappa}, \label{eq:condition-C_A} \\
\eta\leq \frac{1}{10}\min\Bigl\{\phi_{\min}(\BFSigma_{vz}^*),\; \phi_{\min}(\BFSigma_{zz})\Bigr\}.    \label{eq:condition-eta}
\end{gather}
Then, for all $t=\ttR+1,\ldots,T$,
\begin{equation*}
 \Pr( \scrA_t^\SFc \cap \scrB_t\cap \scrC_t)\leq  2q T^{-2}.
\end{equation*}
\end{lemma}

\begin{proof}[Proof.]

It follows from Assumption~\ref{assump:IV} that
$0<\phi_{\min}(\BFSigma_{zz})\leq \phi_{\max}(\BFSigma_{zz})<\infty$ and $0<\phi_{\min}(\BFSigma_{vz}^*)\leq \phi_{\max}(\BFSigma_{vz}^*)<\infty$.
Hence, $\bar{\kappa}$ is a positive finite constant.

We apply an analysis that is similar to that used in the proof of Lemma~\ref{lemma:RanP}.
Specifically, it follows from \eqref{eq:main-2sls} that for all $t>\ttR$,
\begin{equation} \label{eq:norm-alpha_est_t}
\|\widehat{\BFalpha}_{t} - \BFalpha\| \leq \Bigl\|
\Bigl(\widehat{\BFSigma}_{vz,{\ttR:t}}\widehat{\BFSigma}_{zz,{\ttR:t}}^{-1} \widehat{\BFSigma}_{vz,{\ttR:t}}^\intercal\Bigr)^{-1}\widehat{\BFSigma}_{vz,\ttR:t}\widehat{\BFSigma}_{zz,{\ttR:t}}^{-1} \Bigr\| \cdot \bigl\| \widehat{\BFG}_{{\ttR:t}} \bigr\|.
\end{equation}
We analyze below the two norms on the right-hand-side of \eqref{eq:norm-alpha_est_t} separately.

Note that conditional on the event $\scrB_t\cap\scrC_t$ with $\eta <
\min\{\phi_{\min}(\BFSigma_{vz}^*), \phi_{\min}(\BFSigma_{zz})\} $, which is implied by condition \eqref{eq:condition-eta},
 we have
\begin{equation}\label{eq:phi-pm-eta'}
\begin{gathered}
\phi_{\min}(\widehat{\BFSigma}_{vz,\ttR:t})> \phi_{\min}(\BFSigma_{vz}^*)-\eta>0, \quad
\phi_{\max}(\widehat{\BFSigma}_{vz,\ttR:t}) < \phi_{\max}(\BFSigma_{vz}^*)+\eta, \\
\phi_{\min}(\widehat{\BFSigma}_{zz,\ttR:t}) >  \phi_{\min}(\BFSigma_{zz})-\eta>0, \quad
\phi_{\max}(\widehat{\BFSigma}_{zz,\ttR:t})<  \phi_{\max}(\BFSigma_{zz})+\eta.
\end{gathered}
\end{equation}
This implies that
\begin{align}
    & \left\| \left(\widehat{\BFSigma}_{vz,\ttR:t}\widehat{\BFSigma}_{zz,\ttR:t}^{-1} \widehat{\BFSigma}_{vz,\ttR:t}^\intercal\right)^{-1}\widehat{\BFSigma}_{vz,\ttR:t}\widehat{\BFSigma}_{zz,\ttR:t}^{-1}\right\| \nonumber \\
    ={}& \phi_{\max}\left(\left(\widehat{\BFSigma}_{vz,\ttR:t}\widehat{\BFSigma}_{zz,\ttR:t}^{-1} \widehat{\BFSigma}_{vz,\ttR:t}^\intercal\right)^{-1}\widehat{\BFSigma}_{vz,\ttR:t}\widehat{\BFSigma}_{zz,\ttR:t}^{-1}\right) \nonumber \\
    \leq {}&  \frac{\phi_{\max}(\widehat{\BFSigma}_{vz,\ttR:t})\phi_{\max}(\widehat{\BFSigma}_{zz,\ttR:t})}{(\phi_{\min}(\widehat{\BFSigma}_{vz,\ttR:t}))^2 \phi_{\min}(\widehat{\BFSigma}_{zz,\ttR:t})} \label{eq:step1'}  \\
    \leq{}& \frac{(\phi_{\max}({\BFSigma}_{vz}^*)+\eta)(\phi_{\max}({\BFSigma}_{zz})+\eta)}{(\phi_{\min}({\BFSigma}_{vz}^*)-\eta)^2 (\phi_{\min}({\BFSigma}_{zz})-\eta)} \label{eq:step2'} \\
    \leq{}& \biggl(\frac{11}{10}\biggr)^2\biggl(\frac{10}{9}\biggr)^3 \frac{\phi_{\max}(\BFSigma_{vz}^*)\phi_{\max}(\BFSigma_{zz})}{(\phi_{\min}(\BFSigma_{vz}^*))^2 \phi_{\min}(\BFSigma_{zz})} \label{eq:step3'} \\
    \leq{}&  \bar{\kappa}, \label{eq:step4'}
\end{align}
where \eqref{eq:step1'} follows from   Lemma~\ref{lemma:singular-value},
\eqref{eq:step2'} from \eqref{eq:phi-pm-eta'},  \eqref{eq:step3'} from \eqref{eq:condition-eta},
and \eqref{eq:step4'} from \eqref{eq:kappa-bar}.
Therefore,
\begin{equation}\label{eq:bound-on-part1'}
\scrB_t\cap \scrC_t \subseteq
\Bigl\{\Bigl\| \left(\widehat{\BFSigma}_{vz,\ttR:t}\widehat{\BFSigma}_{zz,\ttR:t}^{-1} \widehat{\BFSigma}_{vz,\ttR:t}^\intercal\right)^{-1}\widehat{\BFSigma}_{vz,\ttR:t}\widehat{\BFSigma}_{zz,\ttR:t}^{-1}\Bigr\| \leq \bar{\kappa}\Bigr\}.
\end{equation}

Next, note that $\mathbb{E}[\widehat{\BFG}_{\ttR:t}]=\mathbb{E}[\BFz_{s}\varepsilon_s]=\BFzero$.
Moreover, it is straightforward to see that $\BFz_s \varepsilon_s \in\Real^{q}$ is a vector of zero-mean sub-Gaussian random variables with variance proxy $\bar{z}^2\varsigma^2$.
Hence,
applying Corollary~\ref{coro:Hoeffding_vector} to  $\widehat{\BFG}_{\ttR:t}$ yields that for all $\tau\geq 0$,
\begin{equation}\label{eq:G_T1_t_hoeffding}
    \Pr(\bigl\|\widehat{\BFG}_{\ttR:t}\bigr\|\geq\tau)
    \leq 2q\exp(-\frac{\tau^2 (t-\ttR)}{2 q^2\bar{z}^2 \varsigma^2}).
\end{equation}
Thus, setting $\tau= \zeta_3 \coloneqq 2q \bar{z}\varsigma \sqrt{\frac{\log(T)}{t-\ttR}}$ yields
\begin{equation}\label{eq:prob-G_T1_to_t}
    \Pr(\|\widehat{\BFG}_{\ttR:t}\|> \zeta_3)\leq 2q\exp(-2\log(T)) = 2q T^{-2}.
\end{equation}

It follows from \eqref{eq:norm-alpha_est_t} and \eqref{eq:bound-on-part1'} that
\[
\bigl\{\bigl\|\widehat{\BFG}_{\ttR:t} \bigr\|\leq \zeta_3\bigr\} \cap \scrB_t\cap \scrC_t  \subseteq
\bigl\{
\|\widehat{\BFalpha}_t - \BFalpha \| \leq  \zeta_3 \bar{\kappa}
\bigr\}
\subseteq \biggl\{\|\widehat{\BFalpha}_t - \BFalpha \|\leq C_{\scrA}\sqrt{\frac{\log(T)}{t-\ttR}}\biggr\}
= \scrA_t,
\]
where the second step follows from
the assumption on $C_{\scrA}$ in \eqref{eq:condition-C_A} as well as the definitions of $\zeta_3$ and  $\bar{\kappa}$.
Hence,
\[
\bigl\{\bigl\|\widehat{\BFG}_{\ttR:t} \bigr\|\leq \zeta_3\bigr\}  \cap \scrB_t\cap \scrC_t \subseteq
\scrA_t \cap \scrB_t\cap \scrC_t,
\]
and thus
\[
\bigl\{\bigl\|\widehat{\BFG}_{\ttR:t} \bigr\|> \zeta_3\bigr\}  \cap \scrB_t\cap \scrC_t \supseteq
\scrA_t^\SFc \cap \scrB_t\cap \scrC_t.
\]
It follows that
\begin{align*}
    \Pr(\scrA_t^\SFc \cap \scrB_t\cap \scrC_t)
    \leq \Pr(\bigl\|\widehat{\BFG}_{\ttR:t}\bigr\|>\zeta_3)\leq 2qT^{-2},
\end{align*}
where the second inequality follows from \eqref{eq:prob-G_T1_to_t}.
\end{proof}

We now define for any $t>\ttU$ the following joint events:
\begin{gather*}
    \scrA_{\ttU:t}=\bigcap_{s=\ttU+1}^t \scrA_s,\quad
    \scrB_{\ttU:t}=\bigcap_{s=\ttU+1}^t \scrB_s,\quad
    \scrC_{\ttU:t}=\bigcap_{s=\ttU+1}^t \scrC_s, \quad
    \scrD_{\ttU:t}\coloneqq \bigcap_{s=\ttU+1}^t \scrD_s.
\end{gather*}

\begin{lemma}\label{lemma:D-bounds}
Suppose $C_{\ttU}\geq 6 q^2\bar{z}^4 \eta^{-2}$.
Then, for all $t=\ttU+1,\ldots,T$,
\begin{equation*}
 \Pr(\scrC_t)\geq 1-2q T^{-3} \qq{and} \Pr(\scrC_{\ttU:t})\geq 1-2q (t-\ttU)T^{-3}.
\end{equation*}
\end{lemma}

\begin{proof}[Proof.]
Note that by the definition of $\widehat{\BFSigma}_{zz,\ttR:t}$,
\[
\mathbb{E}[\widehat{\BFSigma}_{zz,\ttR:t}] = \frac{1}{\ttR-t}\sum_{s=\ttR+1}^t\mathbb{E}[\BFz_s \BFz_s^\intercal] =  \BFSigma_{zz}.
\]
In addition, $\|\BFz_s \BFz_s^\intercal \| \leq q \bar{z}^2$.
Hence, it follows from Lemma~\ref{lemma:Hoeffding_matrix} that for all $t \geq \ttU$,
\begin{align*}
\Pr(\scrC_t^{\SFc}) = \Pr(\|\widehat{\BFSigma}_{zz,\ttR:t}-\BFSigma_{zz}\| > \eta)
 \leq{}& 2q\exp(-\frac{\eta^2 (t-\ttR)}{2q^2\bar{z}^4})  \\
 \leq{}&  2q \exp(-\frac{\eta^2 (\ttU-\ttR)}{2q^2\bar{z}^4})
  = 2q\exp(-\frac{\eta^2 C_{\ttU} \log(T)}{2q^2\bar{z}^4}) \\
 \leq{}& 2q\exp(-3\log(T))=2q T^{-3},
\end{align*}
where the last inequality holds because $C_{\ttU}\geq  6q^2\bar{z}^4 \eta^{-2}$.
Therefore,
$\Pr(\scrC_t)\geq 1-2q T^{-3}$; further, by the Bonferroni inequality,
\[
\Pr(\scrC_{\ttU:t}) \geq  1 - \sum_{s=\ttU+1}^t \Pr(\scrC_s^\SFc) \geq 1-2q (t-\ttU)T^{-3}. \qquad \Box
\]
\end{proof}

\begin{lemma}\label{lemma:C-bounds}
Suppose
\begin{gather}
C_{\ttR} \geq 1024K^4 p^2q^3 \bar{v}^4\bar{z}^4 \varsigma^2 \breve{\kappa}^2 L^2 \eta^{-2}, \label{eq:condition-C_T1'}\\
C_{\ttU}\geq \eta^{-2}\max\bigl\{128(K-1)^2 pq\bar{v}^2\bar{z}^2, \; 256 K^2  L^2 p^2q \bar{v}^4\bar{z}^2 C_{\scrA}^2\bigr\}. \label{eq:condition-C_T2'}
\end{gather}
Suppose that  the constant $\eta$ in the definitions of $\scrB_t$ and
$\scrC_t$ satisfies \eqref{eq:condition-eta}.
Then, for all $t=\ttU+1,\ldots,T$,
\begin{equation*}
    \Pr(\scrB_t^{\SFc} \cap \scrA_{\ttR} \cap \scrA_{\ttU:t-1} )\leq  2(Kp+q+1)T^{-2}.
\end{equation*}
\end{lemma}

\begin{proof}[Proof.]

A subtlety in analyzing the event $\scrB_t^c = \{\|\widehat{\BFSigma}_{vz,\ttR:t}-\BFSigma_{vz}^*\| > \eta \}$ for $t>\ttR$ is that $\widehat{\BFSigma}_{vz,\ttR:t}$ is not an unbiased estimator of $\BFSigma_{vz}^*$---that is,
$\mathbb{E}[\widehat{\BFSigma}_{vz,\ttR:t}] \neq \BFSigma_{vz}^*$.
This is because $\BFSigma_{vz}^*$ is defined with respect to the (unknown) oracle policy (i.e., the optimal arm $a^*_s$ is taken for all $s=\ttR+1,\ldots,t$);
whereas $\widehat{\BFSigma}_{vz,\ttR:t}$ is defined with respect to the actual policy that is used to generate the data.
To address the issue, we define
\begin{equation*}
    \widehat{\BFSigma}_{vz,\ttR:t}^*\coloneqq \frac{1}{t-\ttR}\sum_{s=\ttR+1}^t \BFv_t^* \BFz_t^\intercal \in\Real^{(Kp)\times q},
\end{equation*}
and analyze separately the two components of the following decomposition:
\begin{equation}\label{eq:decomp-I1-I2}
\|\widehat{\BFSigma}_{vz,\ttR:t}-\BFSigma_{vz}^*\| \leq
\underbrace{
\|\widehat{\BFSigma}_{vz,\ttR:t} - \widehat{\BFSigma}_{vz,\ttR:t}^* \|}_{I_1} +
\underbrace{
\| \widehat{\BFSigma}_{vz,\ttR:t}^* -\BFSigma_{vz}^*\|}_{I_2}.
\end{equation}

For $I_2$, note that $\mathbb{E}[\BFv_t^* \BFz_t^\intercal] = \BFSigma_{vz}^*$
and that, by Lemma~\ref{lemma:matrix-norm},
\[
\|\BFv_t^* \BFz_t^\intercal - \BFSigma_{vz}^*\|
\leq \|\BFv_t^* \BFz_t^\intercal\| + \|\BFSigma_{vz}^*\|
\leq  2\sqrt{K pq}\bar{v}\bar{z}.
\]
It then follows from  Lemma~\ref{lemma:Hoeffding_matrix} that for all $t\geq \ttU$,
\begin{align}
    \Pr(\| \widehat{\BFSigma}_{vz,\ttR:t}^* - \BFSigma_{vz}^*\| > \frac{\eta}{2})
    \leq{}& 2(Kp+q)\exp(-\frac{(t-\ttR)\eta^2/4}{2(2\sqrt{Kpq}\bar{v}\bar{z})^2})  \nonumber \\
    \leq {}& 2(Kp+q) \exp(-\frac{\eta^2 C_{\ttU} \log(T)}{32K pq\bar{v}^2\bar{z}^2}) \nonumber \\
    \leq{}& 2(Kp+q)\exp(-2\log(T)) = 2(Kp+q) T^{-2},\label{eq:I_2-bound}
\end{align}
where the second inequality holds because $t-\ttR \geq \ttU-\ttR = C_{\ttU}\log(T)$,
and the third holds because of the condition \eqref{eq:condition-C_T2'}.

We now consider $I_1$. Note that
\begin{align}
    & \bigl\|\widehat{\BFSigma}_{vz,\ttR:t}-\widehat{\BFSigma}^*_{vz,\ttR:t}\bigr\|
    = \left\|\frac{1}{t-\ttR}\sum_{s=\ttR+1}^t \begin{pmatrix}
    (\ind(a_s = 1)-\ind(a_s^*=1)) \BFv_s \nonumber \\
    \vdots \\
    (\ind(a_s = K)-\ind(a_s^*= K)) \BFv_s
    \end{pmatrix}\BFz_s^\intercal \right\| \nonumber \\
    \leq {} & \frac{2}{t-\ttR}\sum_{s=\ttR+1}^t\ind(a_s\neq a_s^*) \| \BFv_s \BFz_s^\intercal \|
    \leq   \frac{2 \sqrt{pq}\bar{v}\bar{z}}{t-\ttR}\sum_{s=\ttR+1}^t\ind(a_s\neq a_s^*), \label{eq:I_1-bound}
\end{align}
where the first inequality holds because there are at most non-zero entries in the vector $(\ind(a_s = 1)-\ind(a_s^*=1),\ldots, \ind(a_s = K)-\ind(a_s^*=K))^\intercal$;
the last inequality  follows from Lemma~\ref{lemma:matrix-norm}.
We show next that conditional on some other events,
\begin{equation*}
\{a_s = j\}  \subseteq \biggl\{|\BFv_s^\intercal \BFdelta_{j,a^*_s}|\leq \frac{\eta}{8K L\sqrt{pq}\bar{v}\bar{z}} \biggr\},\quad \forall s\geq \ttR+1,\; \forall j\neq a^*_s,
\end{equation*}
where $\BFdelta_{i,j} = \BFalpha_i - \BFalpha_j$ and $L$ is the constant defined in the margin condition of Assumption~\ref{assump:IV}.
We consider two cases: (i) $s = \ttR + 1,\ldots, \ttU$ and (ii) $s=\ttU+1,\ldots,T$.
Let $\widehat{\BFdelta}_{i,j,s} \coloneqq \widehat{\BFalpha}_{i,s} - \widehat{\BFalpha}_{j,s}$ for each $s$.

\textbf{Case (i): $s = \ttR + 1,\ldots, \ttU$.}
Recall that for $s=\ttR+1, \ldots, \ttU$, the arm selected  at time $s$ is
\[a_s = \argmax_{i=1,\ldots,K} \bigl\{\BFv_s^\intercal \widehat{\BFalpha}_{i,\ttR}\bigr\}.\]
Hence,
for any $j\neq a^*_s$,
if $a_s = j$, then $\BFv_s^\intercal \widehat{\BFdelta}_{j,a^*_s,\ttR} = \BFv_s^\intercal \widehat{\BFalpha}_{j,\ttR} - \BFv_s^\intercal \widehat{\BFalpha}_{a^*_s,\ttR} > 0$.
Thus,
\begin{align}
    0\geq \BFv_s^\intercal \BFdelta_{j,a^*_s}
    > \BFv_s^\intercal \BFdelta_{j,a^*_s} - \BFv_s^\intercal \widehat{\BFdelta}_{j,a^*_s,\ttR}
    ={}& \BFv_s^\intercal [(\BFalpha_j - \widehat{\BFalpha}_{j,\ttR}) - (\BFalpha_{a^*_s} - \widehat{\BFalpha}_{a^*_s,\ttR})] \nonumber  \\
    \geq{}& - 2 \| \BFv_s\| \|\BFalpha - \widehat{\BFalpha}_{\ttR} \|
    \geq -2\sqrt{p}\bar{v} \|\BFalpha - \widehat{\BFalpha}_{\ttR} \|   . \label{eq:v_trans_delta}
\end{align}
Further, conditional on $\scrA_{\ttR} =  \bigl\{\|\widehat{\BFalpha}_{\ttR} - \BFalpha \| \leq  2K q\bar{z}\varsigma \breve{\kappa}  C_{\ttR}^{-\frac{1}{2}} \bigr\}$ defined in \eqref{eq:def-A_T1}, we have
\[
|\BFv_s^\intercal \BFdelta_{j,a^*_s}   | \leq 2 \sqrt{p}\bar{v} \|\BFalpha - \widehat{\BFalpha}_{\ttR} \| \leq
 4K \sqrt{p}  q \bar{v} \bar{z}\varsigma \breve{\kappa}  C_{\ttR}^{-\frac{1}{2}}
 \leq \frac{\eta}{8K L\sqrt{pq} \bar{v}\bar{z}},
\]
where the last inequality follows from the condition \eqref{eq:condition-C_T1'}.
Hence,
\begin{equation}\label{eq:margin-event-1}
\scrA_{\ttR}\cap \{a_s = j\}  \subseteq \biggl\{|\BFv_s^\intercal \BFdelta_{j,a^*_s}  |\leq \frac{\eta}{8K L\sqrt{pq}\bar{v}\bar{z}} \biggr\},\quad \forall s= \ttR+1,\ldots,\ttU,\; \forall j\neq a_s^*.
\end{equation}

\textbf{Case (ii): $s= \ttU+1,\ldots,T$.}
Recall that for $s\geq \ttU+1$, the arm selected at time $s$ is
\[
a_s=\argmax_{i=1,\ldots,K} \left\{\BFv_s^\intercal\widehat{\BFalpha}_{i,s-1} \right\}.
\]
It follows from the same argument as \eqref{eq:v_trans_delta} that for any $j\neq a^*_s$, if $a_s = j$, then
\begin{align}
| \BFv_s^\intercal \BFdelta_{j,a_s^*} |  \leq{}&  2 \sqrt{p}\bar{v} \|\BFalpha - \widehat{\BFalpha}_{s-1} \|   \label{eq:v_trans_delta'} \\
\leq{}&  2\sqrt{p} \bar{v} C_{\scrA} \sqrt{\frac{\log(T)}{s-1-\ttR}}
\leq    2\sqrt{p} \bar{v}C_{\scrA}C_{\ttU}^{-\frac{1}{2}}  \nonumber \\
\leq{}&
 \frac{\eta}{8KL\sqrt{pq}\bar{v}\bar{z}}, \nonumber
\end{align}
conditional on $\scrA_{s-1} = \Bigl\{ \|\widehat{\BFalpha}_{s-1} - \BFalpha \|\leq C_{\scrA} \sqrt{\frac{\log(T)}{s-1-\ttR}} \Bigr\}$,
where the last inequality holds because of the condition \eqref{eq:condition-C_T2'}.
Therefore,
\begin{equation}\label{eq:margin-event-2}
\begin{aligned}
\scrA_{s-1}\cap  \{a_s = j\}  \subseteq{}&  \biggl\{|\BFv_s^\intercal \BFdelta_{j, a^*_s}|\leq \frac{\eta}{8K L\sqrt{pq}\bar{v}\bar{z}} \biggr\},
\quad \forall s= \ttU+1,\ldots,T,\; \forall j\neq a^*_s.
\end{aligned}
\end{equation}

Consequently, combining \eqref{eq:margin-event-1} and \eqref{eq:margin-event-2} yields that
\begin{align*}
& \scrA_{\ttR} \cap  \scrA_{s-1} \cap \{a_s = j\}  \\
\subseteq{}&  \biggl\{|\BFv_s^\intercal \BFdelta_{j, a^*_s}|\leq \frac{\eta}{8K L\sqrt{pq}\bar{v}\bar{z}} \biggr\},
\quad \forall s= \ttR+1,\ldots,T, \; \forall j\neq a^*_s.
\end{align*}

Now, let us return the task of bounding $I_1$, which is defined in \eqref{eq:decomp-I1-I2}.
By \eqref{eq:I_1-bound}, conditional on $\scrA_{\ttR} \cap  \scrA_{\ttU:t-1} $, we have
\begin{align*}
\|\widehat{\BFSigma}_{vz,\ttR:t}-\widehat{\BFSigma}^*_{vz,\ttR:t}\|
\leq \frac{2 \sqrt{pq}\bar{v}\bar{z}}{t-\ttR}\sum_{s=\ttR+1}^t\ind(a_s\neq a_s^*)
\leq \frac{2 \sqrt{pq}\bar{v}\bar{z}}{t-\ttR}\sum_{s=\ttR+1}^t \sum_{j\neq a^*_s}\xi_{j,s},
\end{align*}
for all $t=\ttU+1,\ldots,T$,
where $ \xi_{j,s}\coloneqq \ind\bigl( |\BFv_s^\intercal \BFdelta_{j,a^*_s}|\leq \frac{\eta}{8K L\sqrt{pq}\bar{v}\bar{z}} \bigr)$.
Hence,
\begin{align}
& \Pr(\Bigl\{\|\widehat{\BFSigma}_{vz,\ttR:t}-\widehat{\BFSigma}^*_{vz,\ttR:t}\| \geq \frac{\eta}{2}\Bigr\} \cap \scrA_{\ttR} \cap  \scrA_{\ttU:t-1} ) \nonumber \\
\leq{}& \Pr\biggl(\frac{1}{t-\ttR}\sum_{s=\ttR+1}^t \sum_{j\neq a^*_s}\xi_{j,s} \geq \frac{\eta}{4 \sqrt{pq}\bar{v}\bar{z}}\biggr) \nonumber \\
\leq{}& \Pr\biggl(\biggl|\frac{1}{t-\ttR}\sum_{s=\ttR+1}^t \sum_{j\neq a^*_s}(\xi_{j,s}-\mathbb{E}[\xi_{j,s}]) \biggr|\geq \frac{\eta}{4 \sqrt{pq}\bar{v}\bar{z}}-\sum_{j\neq a^*_s}\mathbb{E}[\xi_{j,s}]\biggr) \nonumber \\
\leq{}& \Pr\biggl(\biggl|\frac{1}{t-\ttR}\sum_{s=\ttR+1}^t \sum_{j\neq a^*_s} (\xi_{j,s}-\mathbb{E}[\xi_{j,s}]) \biggr|\geq \frac{\eta}{8 \sqrt{pq}\bar{v}\bar{z}}\biggr) \label{eq:pre-hoeffding}
\end{align}
where the last step holds because
\begin{equation*}
    \sum_{j\neq a^*_s}\mathbb{E}[\xi_{j,s}] = \sum_{j\neq a^*_s} \Pr(|\BFv_s^\intercal \BFdelta_{j,a_s^*}| \leq \frac{\eta}{8K L\sqrt{pq} \bar{v}\bar{z}})\leq  \sum_{j\neq a^*_s} \frac{\eta}{8K \sqrt{pq}\bar{v}\bar{z}}\leq  \frac{\eta}{8\sqrt{pq}\bar{v}\bar{z}},
\end{equation*}
by the margin condition in Assumption~\ref{assump:IV}.

Note that $\{\sum_{j\neq a^*_s}\xi_{j,s}:s = \ttR+1,\ldots,T\}$  are i.i.d. bounded random variables taking values on $[0,K-1]$.
Thus, they are sub-Gaussian with variance proxy $\frac{K-1}{4}$; see, e.g., \citet[page~24]{Wainwright19}.
Then, setting $\zeta_6 \coloneqq \frac{\eta}{8\sqrt{pq} \bar{v}\bar{z}}$ and
applying Lemma~\ref{lemma:Hoeffding_SG} leads to
\begin{align}
     \Pr\biggl(\biggl|\frac{1}{t-\ttR}\sum_{s=\ttR+1}^{t}(\xi_s-\mathbb{E}[\xi_s])\biggl|\geq \zeta_6\biggr)
    \leq{}& 2\exp(\frac{-2\zeta_6^2 (t-\ttR)}{(K-1)^2}) \nonumber \\
    \leq{}&  2\exp(\frac{-\eta^2  C_{\ttU}\log(T) }{32(K-1)^2 pq\bar{v}^2\bar{z}^2})
    \leq 2T^{-4},
\end{align}
for all $t = \ttU+1,\ldots,T$, where the last inequality follows from the condition \eqref{eq:condition-C_T2'}.
It then follows from \eqref{eq:pre-hoeffding} that
\begin{align}
& \Pr(\Bigl\{\|\widehat{\BFSigma}_{vz,\ttR:t}-\widehat{\BFSigma}^*_{vz,\ttR:t}\| \geq \frac{\eta}{2}\Bigr\}  \cap \scrA_{\ttR} \cap  \scrA_{\ttU:t-1}) \nonumber \\
\leq{}&  2T^{-4},\quad \forall t = \ttU+1,\ldots,T. \label{eq:I_1-bound'}
\end{align}

Combining \eqref{eq:decomp-I1-I2}, \eqref{eq:I_2-bound}, and \eqref{eq:I_1-bound'} results in
\begin{align*}
& \Pr(\Bigl\{\|\widehat{\BFSigma}_{vz,\ttR:t}-\BFSigma^*_{vz}\| \geq \eta\Bigr\} \cap  \scrA_{\ttR} \cap  \scrA_{\ttU:t-1} )    \\
\leq{}&
\Pr(\Bigl\{\|\widehat{\BFSigma}_{vz,\ttR:t}-\widehat{\BFSigma}^*_{vz,\ttR:t}\| \geq \frac{\eta}{2}\Bigr\} \cap \scrA_{\ttR} \cap  \scrA_{\ttU:t-1}  ) \\
& + \Pr(\| \widehat{\BFSigma}_{vz,\ttR:t}^* - \BFSigma_{vz}^*\| > \frac{\eta}{2}) \\
\leq{}& 2(Kp+q) T^{-2}  + 2T^{-4} \leq  2(Kp+q+1) T^{-2},
\end{align*}
for all $t = \ttU+1,\ldots,T$, which concludes the proof.
\end{proof}

We now summarize the conditions of Lemmas~\ref{lemma:RanP}--\ref{lemma:C-bounds} as follows.

\begin{assumption}\label{assump:constants}
The constants $C_{\ttR}$  and $C_{\ttU}$  satisfy
\begin{align*}
C_{\ttR} \geq{}&
\max\Biggl\{\left(\frac{40K\max\{\sqrt{pq}\bar{v}\bar{z},\; q\bar{z}^2\}}{\textstyle \min\{\phi_{\min}(\breve{\BFSigma}_{vz}),\phi_{\min}(\breve{\BFSigma}_{zz})\}}\right)^2,\; 1024K^4 p^2q^3 \bar{v}^4\bar{z}^4 \varsigma^2 \breve{\kappa}^2 L^2 \eta^{-2}\Biggr\}, \\
C_{\ttU}\geq{}& \max\big\{
6 q^2\bar{z}^4 \eta^{-2},
128(K-1)^2  pq\bar{v}^2\bar{z}^2\eta^{-2}, \; 256 K^2  L^2 p^2 q \bar{v}^4\bar{z}^2 C_{\scrA}^2\eta^{-2} \big\},
\end{align*}
where $C_{\scrA} \geq 2q \bar{z}\varsigma \bar{\kappa}$,
$\eta\leq \frac{1}{10}\min\bigl\{\phi_{\min}(\BFSigma_{vz}^*), \; \phi_{\min}(\BFSigma_{zz})\bigr\}$,
and $\breve{\kappa}$ and
$\bar{\kappa}$ are defined in \eqref{eq:kappa-breve} and \eqref{eq:kappa-bar},  respectively.
\end{assumption}

\begin{lemma}
\label{lemma:iter-bounds}
Suppose  $C_{\ttR}$ and  $C_{\ttU}$ satisfy Assumption~\ref{assump:constants}.
Then, for all $t=\ttU+1,\ldots,T$,
\begin{gather*}
\Pr(\scrA_{\ttU:t})\geq  1-  [6Kp+(10K+8)q+1]T^{-1},\\
    \Pr(\scrB_{\ttU:t})\geq 1-  [6Kp+(10K+8)q+1] T^{-1}.
\end{gather*}
\end{lemma}

\begin{proof}[Proof.]
We first summarize the results from Lemmas~\ref{lemma:RanP}--\ref{lemma:C-bounds}:
for all $t=\ttU+1,\ldots,T$,
\begin{align}
&\Pr(\scrA_{\ttR})\geq 1-2K(2p+5q) T^{-2}, \label{eq:A_T1}\\
&\Pr({A}_t^{\SFc}\cap \scrB_t\cap \scrC_t) \leq 2q T^{-2},\label{eq:ABC}\\
&\Pr(\scrC_t)\geq 1-2q T^{-3},\label{eq:C}\\
&\Pr(\scrC_{\ttU:t})\geq 1-2q(t-\ttU)T^{-3} \geq 1- 2q T^{-2}\label{eq:C-1}\\
&\Pr(\scrB_t^{\SFc}\cap \scrA_{\ttR}\cap \scrA_{\ttU:t-1})\leq 2(Kp+q+1) T^{-2}.\label{eq:B}
\end{align}

By \eqref{eq:ABC} and \eqref{eq:C}, we have
\begin{align}
    \Pr(\scrA_t^{\SFc}\cap \scrB_t)
    ={}& \Pr(\scrA_t^{\SFc}\cap \scrB_t \cap \scrC_t) + \Pr(\scrA_t^{\SFc}\cap \scrB_t\cap \scrC_t^\SFc) \nonumber \\
    \leq{}& \Pr(\scrA_t^{\SFc}\cap \scrB_t \cap \scrC_t) + \Pr(\scrC_t^\SFc) \leq  2q(T^{-2}+T^{-3})\leq 4q T^{-2}.\label{eq:A-2}
\end{align}

By   \eqref{eq:A_T1} and \eqref{eq:B}, we have
\begin{align}
    \Pr(\scrB_t^{\SFc}\cap \scrA_{\ttU:t-1})
    \leq{}&
   \Pr(\scrB_t^{\SFc}\cap \scrA_{\ttR}\cap  \scrA_{\ttU:t-1})
   + \Pr(\scrA_{\ttR}^\SFc) \nonumber \\
    \leq{}& 2(Kp+q+1)T^{-2} + 2K(2p+5q)  T^{-2} \nonumber \\
    ={}& [6Kp + (10K+2)q+1]T^{-2}.\label{eq:J-2}
\end{align}

Define $\scrG_t\coloneqq \scrA_{\ttU:t}\cap \scrB_{\ttU:t}$ for all $t=T_{\ttU}+1,\ldots,T$. Let $\Pr(\scrG_{\ttU})=1$ for convenience.
Moreover, note that
\begin{align}
   \Pr(\scrB_t^{\SFc}\cap \scrG_{t-1})
    ={}& \Pr(\scrB_t^{\SFc}\cap \scrA_{\ttU:t-1}\cap \scrB_{\ttU:t-1})\nonumber \\
    ={}& \Pr(\scrB_t^{\SFc}\cap \scrA_{\ttU:t-1}\cap \scrB_{\ttU:t-1}\cap \scrC_{\ttU:t-1}) +  \Pr(\scrB_t^{\SFc}\cap \scrA_{\ttU:t-1}\cap \scrB_{\ttU:t-1}\cap \scrC_{\ttU:t-1}^\SFc)\nonumber\\
    \leq{}& \Pr(\scrB_t^{\SFc}\cap \scrA_{\ttU:t-1} )
     +  \Pr(\scrC_{\ttU:t-1}^\SFc)\nonumber\\
    \leq{}&  [6Kp+(10K+4)q + 1]T^{-2},\label{eq:C-2}
\end{align}
where the second inequality follows from \eqref{eq:C-1} and \eqref{eq:J-2}.
Therefore,
\begin{align}
    \Pr(\scrG_t) ={}& \Pr(\scrG_{t-1}\cap \scrA_t \cap \scrB_t )
    = \Pr(\scrG_{t-1}) - \Pr(\scrG_{t-1}\cap (\scrA_t ^{\SFc}\cup \scrB_t ^{\SFc}))\nonumber\\
    ={}& \Pr(\scrG_{t-1}) -\Pr(\scrG_{t-1}\cap  \scrA_t ^{\SFc}\cap \scrB_t) - \Pr(\scrG_{t-1}\cap  \scrB_t ^{\SFc}) \nonumber\\
    \geq{}& \Pr(\scrG_{t-1}) - \Pr(\scrA_t^{\SFc}\cap \scrB_t)- \Pr(\scrG_{t-1}\cap  \scrB_t ^{\SFc}) \nonumber\\ \geq{}& \Pr(\scrG_{t-1}) - [6Kp+(10K+8)q+1]T^{-2},\label{eq:K-1}
\end{align}
where the last inequality follows from \eqref{eq:A-2} and \eqref{eq:C-2}.

Note that $\Pr(\scrG_{\ttU})=1$, applying \eqref{eq:K-1} iteratively, we have that for all $t>\ttU$,
\begin{equation*}
    \Pr(\scrG_t)\geq 1 - (t-\ttU)[6Kp+(10K+8)q+1]T^{-2} \geq  1-  [6Kp+(10K+8)q+1]T^{-1}.
\end{equation*}
The proof is concluded by noting that
$\Pr(\scrA_{\ttU:t})\geq \Pr(\scrG_t)$ and $\Pr(\scrB_{\ttU:t})\geq \Pr(\scrG_t)$.
\end{proof}

\section{Proof of Theorem~\ref{theo:IV-Greedy-regret}}

Let $\ttR= C_{\ttR}\log(T)$ and $\ttU= (C_{\ttR}+C_{\ttU})\log(T)$ for some constants $C_{\ttR}$ and $C_{\ttU}$ that satisfy Assumption~\ref{assump:constants}.
Let $\bar{\delta} \coloneqq  \max_{1\leq i \neq j\leq K} \|\BFdelta_{i,j}\| $,
where $\BFdelta_{i,j} \coloneqq \BFalpha_i - \BFalpha_j$.
We show in the following that for all $T>\ttU$,
\begin{align*}
 \Reg(T)
\leq {}&
[\sqrt{p}\bar{v}\bar{\delta}(C_{\ttR} + C_{\ttU})  + 16 (K-1) L p q^3 \bar{v}^2 \bar{z}^2\varsigma^2 \bar{\kappa}^2] \log(T)  \\
& + \sqrt{p}\bar{v} \bar{\delta} (K-1) [12Kp + (20K+20)q+3].
\end{align*}

Note that for all $s\geq 1$,
\begin{align}
R_s^{\pi^*} - R_s^\pi ={}& \BFv_s^\intercal \BFalpha_{a^*_s} - \BFv_s^\intercal \BFalpha_{a_s}
= \sum_{j\neq a^*_s} \ind(a_s = j) |\BFv_s^\intercal \BFdelta_{j, a^*_s}|.
\label{eq:single-regret}
\end{align}

We first focus on the case that $s=\ttU+2,\ldots,T$.
It follows from \eqref{eq:v_trans_delta'}  that for any $j\neq a^*_s$,
conditional on the event $\{a_s = j\}$,
\begin{align}
|\BFv_s^\intercal \BFdelta_{j,a^*_s} |
\leq 2\sqrt{p}\bar{v} \|\widehat{\BFalpha}_{s-1} - \BFalpha\|. \label{eq:v_delta}
\end{align}
Further, recall from \eqref{eq:main-2sls} that
\[
\widehat{\BFalpha}_{s-1} - \BFalpha=
\Bigl(\widehat{\BFSigma}_{vz,{\ttR:s-1}}\widehat{\BFSigma}_{zz,{\ttR:s-1}}^{-1} \widehat{\BFSigma}_{vz,{\ttR:s-1}}^\intercal\Bigr)^{-1}\widehat{\BFSigma}_{vz,\ttR:s-1}\widehat{\BFSigma}_{zz,{\ttR:s-1}}^{-1} \widehat{\BFG}_{{\ttR:s-1}},
\]
and from \eqref{eq:bound-on-part1'} in the  Appendix that
\[
\scrB_{s-1}\cap \scrC_{s-1} \subseteq
\Bigl\{\Bigl\| \left(\widehat{\BFSigma}_{vz,\ttR:s-1}\widehat{\BFSigma}_{zz,\ttR:s-1}^{-1} \widehat{\BFSigma}_{vz,\ttR:s-1}^\intercal\right)^{-1}\widehat{\BFSigma}_{vz,\ttR:s-1}\widehat{\BFSigma}_{zz,\ttR:s-1}^{-1}\Bigr\| \leq \bar{\kappa}\Bigr\}.
\]
It then follows from \eqref{eq:v_delta} that for any $j\neq a^*_s$,
\begin{align}\label{eq:a-v-1}
\{a_s = j\} \cap  \scrB_{s-1}\cap \scrC_{s-1} \subseteq
\Bigl\{ |\BFv_s^\intercal \BFdelta_{j, a^*_s} | \leq
2\sqrt{p}\bar{v}\bar{\kappa} \|  \widehat{\BFG}_{{\ttR:s-1}}\| \Bigr\}.
\end{align}
Hence, for any $j\neq a^*_s$,
\begin{align*}
    & \mathbb{E}[\ind(a_s = j) |\BFv_s^\intercal \BFdelta_{j, a^*_s}| ] \nonumber \\
    ={}& \mathbb{E}[(\ind(\{a_s =j \}\cap \scrB_{s-1}\cap \scrC_{s-1} ) |\BFv_s^\intercal \BFdelta_{j, a^*_s}|]
    + \mathbb{E}[\ind(\{a_s =j \}\cap (\scrB_{s-1}\cap \scrC_{s-1})^\SFc) |\BFv_s^\intercal \BFdelta_{j, a^*_s}| ]   \nonumber \\
    \leq{}& \mathbb{E}[(\ind(\{a_s = j\}\cap \scrB_{s-1}\cap \scrC_{s-1} ) |\BFv_s^\intercal \BFdelta_{j, a^*_s}|]
    + \mathbb{E}[\ind( (\scrB_{s-1}\cap \scrC_{s-1})^\SFc) |\BFv_s^\intercal \BFdelta_{j, a^*_s}| ] \nonumber\\
    \leq{}&
    \mathbb{E}\Bigl[\ind\Bigl(|\BFv_s^\intercal \BFdelta_{j, a^*_s} | \leq
2\sqrt{p}\bar{v}\bar{\kappa} \|  \widehat{\BFG}_{{\ttR:s-1}}\| \Bigr) |\BFv_s^\intercal \BFdelta_{j, a^*_s}| \Bigr]
+ 2\sqrt{p}\bar{v} \| \BFdelta_{j, a^*_s}\|  \Pr( \scrB_{s-1}^\SFc\cup \scrC_{s-1}^\SFc )  \nonumber \\
    \leq {} & 2\sqrt{p}\bar{v}\bar{\kappa}
    \underbrace{\mathbb{E}\Bigl[\ind\Bigl(|\BFv_s^\intercal \BFdelta_{j, a^*_s} | \leq
2\sqrt{p}\bar{v}\bar{\kappa} \|  \widehat{\BFG}_{{\ttR:s-1}}\| \Bigr) \|  \widehat{\BFG}_{{\ttR:s-1}}\| \Bigr] }_{U_{1,j}}
+ 2\sqrt{p}\bar{v} \bar{\delta} \underbrace{\Pr( \scrB_{s-1}^\SFc\cup \scrC_{s-1}^\SFc )}_{U_2},
\end{align*}
where
the second inequality follows from \eqref{eq:a-v-1}, and  that $|\BFv_s^\intercal \BFdelta_{j, a^*_s}| \leq \|\BFv_s\| \| \BFdelta_{j, a^*_s}\| \leq 2\sqrt{p}\bar{v} \bar{\delta} $.
By applying \eqref{eq:single-regret}, we have
\begin{align}
     \mathbb{E}[R_s^{\pi^*}-R_s^\pi]     = \sum_{j\neq a^*_s }\mathbb{E}[\ind(a_s = j) |\BFv_s^\intercal \BFdelta_{j, a^*_s}| ]
     \leq \sum_{j\neq a^*_s } (U_{1,j} + U_2).
     \label{eq:reg-bound-1}
\end{align}

For $U_2$, we may apply Lemmas~\ref{lemma:D-bounds} and \ref{lemma:iter-bounds} to conclude that
\begin{align}
U_2 \leq{}& \Pr(\scrB_{s-1}^c) +\Pr(\scrC_{s-1}^\SFc)
\leq \Pr(\scrB_{\ttU:{s-1}}^\SFc)+\Pr(\scrC_{s-1}^\SFc)  \nonumber \\
\leq{}& [6Kp+(10K+8)q+1]T^{-1} + 2q T^{-3} \leq [6K p+(10K+10) q+1]T^{-1}. \label{eq:reg-bound-p}
\end{align}

For $U_{1,j}$, we have
\begin{align}
U_{1,j} = {}& \mathbb{E}\biggl[\E\Bigl[\ind\Bigl(|\BFv_s^\intercal \BFdelta_{j, a^*_s} | \leq
2\sqrt{p}\bar{v}\bar{\kappa} \|  \widehat{\BFG}_{{\ttR:s-1}}\| \Bigr) \|  \widehat{\BFG}_{{\ttR:s-1}}\| \Bigr]  \,\Big|\,  \|  \widehat{\BFG}_{{\ttR:s-1}}\| \biggr]
\nonumber \\
\leq {}&  \mathbb{E}\biggl[ \Pr\Bigl(|\BFv_s^\intercal \BFdelta_{j, a^*_s} | \leq
2\sqrt{p}\bar{v}\bar{\kappa} \|  \widehat{\BFG}_{{\ttR:s-1}}\| \,\Big|\,  \|  \widehat{\BFG}_{{\ttR:s-1}}\|  \Bigr) \|  \widehat{\BFG}_{{\ttR:s-1}}\| \biggr]
\nonumber \\
\leq{}&   \mathbb{E}\Bigl[ 2L \sqrt{p}\bar{v}\bar{\kappa} \|  \widehat{\BFG}_{{\ttR:s-1}}\|^2  \Bigr],\label{eq:U_1-bound}
\end{align}
where the second inequality follows from the
margin condition of Assumption~\ref{assump:IV},

Let $\SFp(w)$ denote the probability density function of $\|\widehat{\BFG}_{\ttR:s-1}\|$.
Then,
\begin{align}
    \mathbb{E}\Bigl[\|\widehat{\BFG}_{\ttR:s-1}\|^2\Bigr] ={}& \int_{0}^\infty  w^2 \SFp(w)\dd{w}
    = \int_{0}^\infty  \SFp(w) \int_{0}^{w} 2 \tau \dd{\tau} \dd{w} \nonumber \\
    ={}&\int_{0}^\infty 2\tau  \int_{\tau}^\infty  \SFp(w) \dd{w} \dd{\tau}
    = \int_{0}^\infty 2\tau \Pr(\|\widehat{\BFG}_{\ttR:s-1}\|\geq \tau) \dd{\tau},\label{eq:integral}
\end{align}
where the third equality follows from Fubini's theorem.
We then apply \eqref{eq:G_T1_t_hoeffding}:
\begin{gather}
    \int_{0}^\infty 2\tau \Pr(\|\widehat{\BFG}_{\ttR:s-1}\|\geq \tau) \dd{\tau }
    \leq \int_{0}^\infty 4q\tau \exp(-\frac{\tau^2 (s-\ttU-1)}{2 q^2\bar{z}^2 \varsigma^2}) \dd{\tau}
    =\frac{4q^3\bar{z}^2\varsigma^2 }{s-\ttU-1}.\label{eq:pr-bound-final}
\end{gather}

Combining \eqref{eq:U_1-bound},  \eqref{eq:integral}, and  \eqref{eq:pr-bound-final}, we have that for any $j\neq a^*_s$,
\begin{equation}\label{eq:reg-v-bound-final}
    U_{1,j} \leq  \frac{8L \sqrt{p} q^3 \bar{v} \bar{z}^2\varsigma^2 \bar{\kappa}}{s-\ttU-1}.
\end{equation}

Plugging in \eqref{eq:reg-bound-p} and \eqref{eq:reg-v-bound-final} into \eqref{eq:reg-bound-1}, we have
\begin{equation}
    \mathbb{E}[R_s^{\pi^*}-R_s^\pi]
    \leq  (K-1)\biggl[\frac{16 L p q^3 \bar{v}^2 \bar{z}^2\varsigma^2 \bar{\kappa}^2}{s-\ttU-1} + 2\sqrt{p}\bar{v} \bar{\delta} [6Kp+(10K+10)q+1]T^{-1}\biggr].\label{eq:reg-s}
\end{equation}
Note that
$
\sum_{s=\ttU+2}^T \frac{1}{s-\ttU-1} \leq \int_1^T \frac{\dd{u}}{u} = \log(T)
$.
Then,
summing \eqref{eq:reg-s} over $s=\ttU+2,\ldots,T$ yields
\begin{align}
    & \sum_{s=\ttU+2}^T \mathbb{E}[R_s^{\pi^*}-R_s^\pi]  \nonumber \\
    \leq{}& (K-1) \bigl[16 L p q^3 \bar{v}^2 \bar{z}^2\varsigma^2 \bar{\kappa}^2 \log(T) + 2\sqrt{p}\bar{v} \bar{\delta} [6Kp+(10K+10)q+1] \bigr]. \label{eq:regret-stage-3}
\end{align}

Moreover, for the first $\ttU+1$ periods,
we can simply take advantage of the fact that each reward is bounded and $\ttU$ is of order $\mathcal{O}(\log(T))$.
Specifically, by \eqref{eq:single-regret},
\begin{align}
\sum_{s=1}^{\ttU+1} \mathbb{E}[R_s^{\pi^*} - R_s^\pi] ={}& \sum_{s=1}^{\ttU+1} \mathbb{E}[\ind(a_s\neq a_s^*) |\BFv_s^\intercal \BFdelta_{a_s,a^*_s}|]
\leq \sum_{s=1}^{\ttU+1} \mathbb{E}[\|\BFv_s\|\|\BFdelta_{a_s,a^*_s}\| ] \nonumber \\
\leq{}& \sqrt{p}\bar{v}\bar{\delta} (\ttU +1)
=\sqrt{p}\bar{v} \bar{\delta}\bigl[(C_{\ttR} + C_{\ttU}) \log(T) + 1\bigr]. \label{eq:regret-pre-T2}
\end{align}

Combining \eqref{eq:regret-stage-3} and \eqref{eq:regret-pre-T2} completes the proof.
\hfill$\Box$

\section{Proof of Theorem~\ref{theorem:asymp}}

Let $\ttR= C_{\ttR}\log(T)$ and $\ttU= (C_{\ttR}+C_{\ttU})\log(T)$ for some constants $C_{\ttR}$ and $C_{\ttU}$ that satisfy Assumption~\ref{assump:constants}.
We show in the following that
\begin{gather}
\sqrt{T-\ttR} (\widehat{\BFalpha}_{T} - \BFalpha )
\rightsquigarrow \calN\Bigl(\BFzero,\sigma^2\bigl(\BFSigma_{vz}^*\BFSigma_{zz}^{-1}(\BFSigma_{vz}^*)^\intercal\bigr)^{-1}\Bigr),
\end{gather}
as $T\to\infty$, where $\BFSigma_{vz}^* = \mathbb{E}[\BFv_t^* \BFz_t^\intercal]$ and $\BFSigma_{zz}=\mathbb{E}[\BFz_t \BFz_t^\intercal]$.

By \eqref{eq:main-2sls}, for all $T>\ttR$,
\begin{align*}
\widehat{\BFalpha}_{T} - \BFalpha = \Bigl(\widehat{\BFSigma}_{vz,{\ttR:T}}\widehat{\BFSigma}_{zz,{\ttR:T}}^{-1} \widehat{\BFSigma}_{vz,{\ttR:T}}^\intercal\Bigr)^{-1}\widehat{\BFSigma}_{vz,\ttR:T}\widehat{\BFSigma}_{zz,{\ttR:T}}^{-1} \widehat{\BFG}_{{\ttR:T}}.
\end{align*}

It follows from the definition of $\widehat{\BFG}_{\ttR:T}$ and
the central limit theorem that as $T\to\infty$,
\begin{gather*}
\sqrt{T-\ttR}\widehat{\BFG}_{{\ttR:T}}
= \frac{1}{\sqrt{T-\ttR}}\sum_{s=\ttR+1}^{T} \BFz_s \varepsilon_s
\rightsquigarrow \mathcal{N}(\BFzero,\sigma^2
    \BFSigma_{zz}).
\end{gather*}

By Lemmas~\ref{lemma:D-bounds} and \ref{lemma:iter-bounds},
for arbitrarily small $\tau>0$, for $T$ large enough (dependent on $\tau$), we have that:
\begin{align*}
    &\Pr(\|\widehat\BFSigma_{vz,\ttR:T}-\BFSigma_{vz}^*\|\leq \tau)\geq 1-[6Kp+(10K+8)q+1]T^{-1},\\
    &\Pr(\|\widehat{\BFSigma}_{zz,\ttR:T}-\BFSigma_{zz}\| \leq \tau)\geq 1-2q T^{-3}.
\end{align*}
Therefore, as $T\rightarrow \infty$,
\begin{gather*}
    \widehat\BFSigma_{vz,\ttR:T} - \BFSigma_{vz}^*\rightarrow_p \BFzero \qq{and}
    \widehat\BFSigma_{zz,\ttR:T}-\BFSigma_{zz}\rightarrow_p \BFzero.
\end{gather*}
It then follows from the  continuous mapping theorem that, as $T\to\infty$,
\begin{align*}
\sqrt{T-\ttR}
(\widehat\BFalpha_{T} - \BFalpha)
\rightsquigarrow {}& \bigl(\BFSigma_{vz}^*\BFSigma_{zz}^{-1}(\BFSigma_{vz}^*)^\intercal\bigr)^{-1}\BFSigma_{vz}^*\BFSigma_{zz}^{-1}\cdot \mathcal{N}(\BFzero,\sigma^2 \BFSigma_{zz}) \\
\stackrel{d}{=} {}& \calN\Bigl(\BFzero,\sigma^2\bigl(\BFSigma_{vz}^*\BFSigma_{zz}^{-1}(\BFSigma_{vz}^*)^\intercal\bigr)^{-1}\Bigr).
\end{align*}

\section{Proof of Corollary~\ref{coro:inf}}

\begin{lemma}
\label{lemma:sigma}
Suppose
\begin{equation}\label{eq:condition-C_T2}
C_{\ttU}\geq \max\Biggl\{\frac{3}{\textstyle  \Cbern \min\Bigl\{  \frac{\Delta^2}{4K_\varepsilon^2}, \frac{\Delta}{2K_\varepsilon}\Bigr\}}, \;
\frac{6K pq^2 \bar{v}^2 \bar{z}^2\varsigma^2 \bar{\kappa}^2 }{\textstyle  \min \Bigl\{\frac{\Delta^2}{128\sigma^2}, \frac{\Delta}{64} \Bigr\}}\Biggr\},
\end{equation}
where $\Cbern$ is the constant in Lemma~\ref{lemma:Bernstein} and
$K_\varepsilon = \| \varepsilon_1^2 - \sigma^2\|_{\psi_1} < \infty$.
Suppose that  the constant $\eta$ in the definitions of $\scrB_t$ and
$\scrC_t$ satisfies
\begin{equation}\label{eq:condition-eta''}
\eta\leq \frac{1}{10}\min\Bigl\{\phi_{\min}(\BFSigma_{vz}^*),\; \phi_{\min}(\BFSigma_{zz})\Bigr\}.
\end{equation}
Then, for all $t=\ttU+1,\ldots,T$,
\begin{equation*}
    \Pr(\scrB_t \cap \scrC_t \cap \scrD_t^{\SFc})\leq (2q+2) T^{-3}.
\end{equation*}
\end{lemma}

\begin{proof}[Proof.]
Denote the estimated residual  by $\widehat{\varepsilon}_t \coloneqq R_t - \widetilde{\BFv}_t^\intercal \widehat{\BFalpha}_t$, and
define for all $t>\ttR$,
\begin{align}
\widehat{\BFvarepsilon}_{\ttR:t}\coloneqq{}&
\begin{pmatrix}
\widehat{\varepsilon}_{\ttR} \\
\vdots \\
\widehat{\varepsilon}_t
\end{pmatrix}
= \BFR_{\ttR: t} -  \widetilde{\BFV}_{\ttR: t}
\widehat{\BFalpha}_t
= \BFvarepsilon_{\ttR:t} + \widetilde{\BFV}_{\ttR: t}
\BFalpha  - \widetilde{\BFV}_{\ttR: t}
\widehat{\BFalpha}_t   \nonumber \\
={}& \BFvarepsilon_{\ttR: t} - \widetilde{\BFV}_{\ttR:t}\BFM_{\ttR:t}  \widehat{\BFG}_{{\ttR:t}}, \label{eq:varepsilon-hat}
\end{align}
where $\BFM_{\ttR:t} \coloneqq \Bigl(\widehat{\BFSigma}_{vz,{\ttR:t}}\widehat{\BFSigma}_{zz,{\ttR:t}}^{-1} \widehat{\BFSigma}_{vz,{\ttR:t}}^\intercal\Bigr)^{-1}\widehat{\BFSigma}_{vz,\ttR:t}\widehat{\BFSigma}_{zz,{\ttR:t}}^{-1}$,
and the last equality follows from  \eqref{eq:main-2sls}.
Then, it follows from \eqref{eq:variance-estimator} that
\begin{align}
\widehat{\sigma}_t^2 ={}& \frac{1}{t-\ttR} \| \widehat{\BFvarepsilon}_{\ttR:t}\|^2
= \frac{1}{t-\ttR}  \|\BFvarepsilon_{\ttR: t} - \widetilde{\BFV}_{\ttR:t}\BFM_{\ttR:t}  \widehat{\BFG}_{{\ttR:t}} \|^2 \nonumber \\
={}& \frac{1}{t-\ttR} \Bigl(  \|\BFvarepsilon_{\ttR: t}\|^2 - 2 \BFvarepsilon_{\ttR: t}^\intercal  \widetilde{\BFV}_{\ttR:t}\BFM_{\ttR:t}  \widehat{\BFG}_{{\ttR:t}} + \|\widetilde{\BFV}_{\ttR:t}\BFM_{\ttR:t}  \widehat{\BFG}_{{\ttR:t}}\|^2 \Bigr) \nonumber \\
\leq{}& \frac{1}{t-\ttR} \Bigl(  \|\BFvarepsilon_{\ttR: t}\|^2 + 2 \|\BFvarepsilon_{\ttR: t}\|   \|\widetilde{\BFV}_{\ttR:t}\BFM_{\ttR:t}  \widehat{\BFG}_{{\ttR:t}}\| + \|\widetilde{\BFV}_{\ttR:t}\BFM_{\ttR:t}  \widehat{\BFG}_{{\ttR:t}}\|^2 \Bigr). \label{eq:sigma-hat-square}
\end{align}
Therefore, to examine the event $\scrD_t^\SFc = \{|\widehat{\sigma}_t^2 - \sigma^2| \leq \Delta \}$, it suffices to analyze the following two events separately: (i) $\frac{1}{t-\ttR}\|\BFvarepsilon_{\ttR: t}\|^2$ is in the proximity of $\sigma^2$, and (ii) $\frac{1}{t-\ttR} \|\widetilde{\BFV}_{\ttR:t}\BFM_{\ttR:t}  \widehat{\BFG}_{{\ttR:t}}\|^2$ is small.

We first analyze the former. Define
$
\scrE_t\coloneqq \left\{ \left| \frac{1}{t-\ttR}\|\BFvarepsilon_{\ttR: t}\|^2 - \sigma^2 \right| \leq \frac{\Delta}{2}\right\}
$.
Note that
\[
\frac{1}{t-\ttR}\|\BFvarepsilon_{\ttR: t}\|^2 - \sigma^2 = \frac{1}{t-\ttR}\sum_{s=\ttR+1}^t (\varepsilon_s^2-\sigma^2),
\]
and $\mathbb{E}[\varepsilon_s^2] = \sigma^2$.
Moreover, since $\varepsilon_s$ is sub-Gaussian, it follows from
Lemma~2.7.6 of \cite{Vershynin18} that
$\varepsilon_s^2 - \sigma^2$  is sub-exponential.
Hence, by Lemma~\ref{lemma:Bernstein},
\begin{align}
\Pr( \scrE_t^\SFc )
\leq{}& 2\exp(-(t-\ttR) \Cbern \min\Bigl\{ \frac{\Delta^2}{4K_\varepsilon^2}, \frac{\Delta}{2K_\varepsilon} \Bigr\}) \nonumber \\
\leq{}& 2\exp(-C_{\ttU} \log(T) \Cbern \min\Bigl\{ \frac{\Delta^2}{4K_\varepsilon^2}, \frac{\Delta}{2K_\varepsilon} \Bigr\}) \nonumber \\
\leq{}& 2 \exp(-3\log(T)) = 2T^{-3}, \label{eq:applying-Bernstein}
\end{align}
where the last inequality holds because of the condition \eqref{eq:condition-C_T2}.

Now, we analyze the term $\frac{1}{t-\ttR} \|\widetilde{\BFV}_{\ttR:t}\BFM_{\ttR:t}  \widehat{\BFG}_{{\ttR:t}}\|^2$.
Note that by Lemma~\ref{lemma:matrix-norm},
\begin{equation}\label{eq:V-tilde-norm-square}
\frac{1}{t-\ttR} \|\widetilde{\BFV}_{\ttR:t}\|^2 \leq \frac{1}{t-\ttR}  \bigl(\sqrt{Kp(t-\ttR)}\|\widetilde{\BFV}_{\ttR:t}\|_{\max}\bigr) ^2
\leq K p \bar{v}^2.
\end{equation}
It follows from \eqref{eq:step4'} that conditional on the event $\scrB_t\cap\scrC_t$ with  $\eta$ satisfying \eqref{eq:condition-eta''},
\begin{equation}\label{eq:M-norm-square}
\| \BFM_{\ttR:t} \|^2
= \left\| \left(\widehat{\BFSigma}_{vz,\ttR:t}\widehat{\BFSigma}_{zz,\ttR:t}^{-1} \widehat{\BFSigma}_{vz,\ttR:t}^\intercal\right)^{-1}\widehat{\BFSigma}_{vz,\ttR:t}\widehat{\BFSigma}_{zz,\ttR:t}^{-1}\right\|^2
\leq \bar{\kappa}^2.
\end{equation}

Note that
\begin{align*}
\frac{1}{t-\ttR} \|\widetilde{\BFV}_{\ttR:t}\BFM_{\ttR:t}  \widehat{\BFG}_{{\ttR:t}}\|^2
\leq \frac{1}{t-\ttR} \|\widetilde{\BFV}_{\ttR:t}\|^2 \| \BFM_{\ttR:t} \|^2 \|\widehat{\BFG}_{{\ttR:t}}\|^2.
\end{align*}
It follows from  \eqref{eq:V-tilde-norm-square} and \eqref{eq:M-norm-square} that
conditional on the event $\scrB_t\cap\scrC_t \cap \{ \|\widehat{\BFG}_{\ttR:t}\| \leq  \zeta_4 \} $ with $\zeta_4 \coloneqq \sqrt{6} q \bar{z}\varsigma \sqrt{\frac{\log(T)}{t-\ttR}}$,
we have
\begin{align}
\frac{1}{t-\ttR}  \|\widetilde{\BFV}_{\ttR:t}\BFM_{\ttR:t}  \widehat{\BFG}_{{\ttR:t}}\|^2
\leq{}&  K p\bar{v}^2   \zeta_4^2  \bar{\kappa}^2
= 6K pq^2 \bar{v}^2 \bar{z}^2\varsigma^2 \bar{\kappa}^2\frac{\log(T)}{t-\ttR}  \nonumber \\
\leq{}& \frac{6K pq^2 \bar{v}^2\bar{z}^2\varsigma^2 \bar{\kappa}^2 }{C_{\ttU}}
\leq \zeta_5
\end{align}
where $\zeta_5 \coloneqq \min \bigl\{\frac{\Delta^2}{128\sigma^2}, \frac{\Delta}{64} \bigr\}$. Note that the second inequality holds because $t-\ttR \geq \ttU- \ttR \geq C_{\ttU}\log(T)$, and the last inequality because of the condition \eqref{eq:condition-C_T2}.
Consequently,
\begin{equation}\label{eq:VMG-small}
\bigl\{ \|\widehat{\BFG}_{\ttR:t}\| \leq  \zeta_4 \bigr\} \cap \scrB_t\cap\scrC_t
\subseteq \biggl\{ \frac{1}{t-\ttR}  \|\widetilde{\BFV}_{\ttR:t}\BFM_{\ttR:t}  \widehat{\BFG}_{{\ttR:t}}\|^2 \leq \zeta_5 \biggr\} \coloneqq \scrF_t.
\end{equation}

Note that conditional on the event $\scrE_t \cap \scrF_t$, we have
\begin{align}
    |\widehat\sigma_t^2-\sigma^2|
    \leq{}& \biggl|\frac{1}{t-\ttR}\|\BFvarepsilon_{\ttR:t}\|^2-\sigma^2\biggr| + \biggl| \widehat{\sigma}^2_t-\frac{1}{t-\ttR}\|\BFvarepsilon_{\ttR:t} \|^2 \biggr| \nonumber  \\
    \leq{}& \biggl|\frac{1}{t-\ttR}\|\BFvarepsilon_{\ttR:t}\|^2-\sigma^2\biggr| + 2 \biggl(\frac{1}{t-\ttR} \|\BFvarepsilon_{\ttR:t}\|^2\cdot \frac{1}{t-\ttR}\|\widetilde{\BFV}_{\ttR:t}\BFM_{\ttR:t}  \widehat{\BFG}_{{\ttR:t}}\|^2\biggr)^{1/2} \nonumber \\
    & + \frac{1}{t-\ttR} \|\widetilde{\BFV}_{\ttR:t}\BFM_{\ttR:t}  \widehat{\BFG}_{{\ttR:t}}\|^2 \nonumber
    \\
    \leq{}& \frac{\Delta}{2} + 2\sqrt{\Bigl(\sigma^2+\frac{\Delta}{2}\Bigr)\zeta_5}+\zeta_5,\label{eq:sigma-hat-bound}
\end{align}
where the second inequality follows from \eqref{eq:sigma-hat-square}.
Further, by the definition of $\zeta_5$, we have
\begin{align*}
\Bigl(\sigma^2+\frac{\Delta}{2}\Bigr)\zeta_5
\leq \sigma^2 \cdot \frac{\Delta^2}{128\sigma^2} + \frac{\Delta}{2}\cdot \frac{\Delta}{64} =  \frac{\Delta^2}{64}.
\end{align*}
It then follows from \eqref{eq:sigma-hat-bound} that $|\widehat\sigma_t^2-\sigma^2| \leq    \frac{\Delta}{2} + \frac{\Delta}{4} + \frac{\Delta}{64} < \Delta $.
Hence, $\scrE_t \cap \scrF_t \subseteq \scrD_t$, which in conjunction with \eqref{eq:VMG-small} implies that
\[
\scrE_t \cap \bigl\{ \|\widehat{\BFG}_{\ttR:t}\| \leq  \zeta_4 \bigr\} \cap  \scrB_t\cap\scrC_t
\subseteq
\scrE_t  \cap \scrF_t \subseteq
\scrD_t,
\]
and thus $\scrE_t \cap  \bigl\{ \|\widehat{\BFG}_{\ttR:t}\| \leq  \zeta_4 \bigr\} \cap \scrB_t\cap\scrC_t   \subseteq \scrB_t\cap\scrC_t \cap \scrD_t
$.
Hence,
\begin{align}
\scrB_t\cap\scrC_t \cap \scrD_t^{\SFc}
\subseteq{}&
\Bigl( \scrE_t \cap \bigl\{ \|\widehat{\BFG}_{\ttR:t}\| \leq  \zeta_4 \bigr\} \Bigr)^\SFc \cap \scrB_t\cap\scrC_t
\subseteq  \scrE_t^\SFc \cup \bigl\{ \|\widehat{\BFG}_{\ttR:t}\| >  \zeta_4 \bigr\}. \label{eq:BCD}
\end{align}

Note that by \eqref{eq:G_T1_t_hoeffding}, we have
\begin{equation}\label{eq:G_T1_t_hoeffding'}
\Pr(\|\widehat{\BFG}_{\ttR:t}\| > \zeta_4) \leq 2q\exp(-\frac{\zeta_4^2(t-\ttR)}{2q^2\bar{z}^2\varsigma^2}) \leq   2q T^{-3}.
\end{equation}
Combining \eqref{eq:applying-Bernstein}, \eqref{eq:BCD},  and  \eqref{eq:G_T1_t_hoeffding'} yields
\[
\Pr(\scrB_t\cap\scrC_t \cap \scrD_t^{\SFc}) \leq
\Pr(\scrE_t^\SFc) +
\Pr(\|\widehat{\BFG}_{\ttR:t}\| > \zeta_4)
\leq  (2q+2)T^{-3}.
\]
\end{proof}

\begin{proof}[Proof of Corollary~\ref{coro:inf}.]

By Lemmas~\ref{lemma:D-bounds} and \ref{lemma:iter-bounds},
for arbitrarily small $\tau>0$, for $T$ large enough (dependent on $\tau$), we have that:
\begin{align*}
    &\Pr(\|\widehat\BFSigma_{vz,\ttR:T}-\BFSigma_{vz}^*\|\leq \tau)\geq 1-[6Kp+(10K+8)q+1]T^{-1},\\
    &\Pr(\|\widehat{\BFSigma}_{zz,\ttR:T}-\BFSigma_{zz}\| \leq \tau)\geq 1-2q T^{-3}.
\end{align*}
Therefore, as $T\rightarrow \infty$,
\begin{gather*}
    \widehat\BFSigma_{vz,\ttR:T} - \BFSigma_{vz}^*\rightarrow_p \BFzero \qq{and}
    \widehat\BFSigma_{zz,\ttR:T}-\BFSigma_{zz}\rightarrow_p \BFzero.
\end{gather*}
It implies that
\begin{equation*}
    \widehat{\BFOmega}_T =\Bigl(\widehat\BFSigma_{vz,\ttR:T}\widehat\BFSigma_{vz,\ttR:T}^{-1}(\widehat\BFSigma_{vz,\ttR:T})^\intercal \Bigr)^{-1}\rightarrow_p \BFOmega^*.
\end{equation*}

Similarly, by Lemma \ref{lemma:sigma}, for arbitrarily small $\eta$ and large enough $T$ (dependent on $\eta$), we have
\begin{equation*}
   \Pr(\mathscr{B}_T\cap \mathscr{C}_T \cap \mathscr{D}_T^\SFc)\leq (2q+2)T^{-3}.
\end{equation*}
By Lemma \ref{lemma:D-bounds}  and Lemma \ref{lemma:iter-bounds}, we know that for $T$ large enough,
\begin{align*}
    \Pr(\mathscr{B}_T\cap \mathscr{C}_T)\geq{}& 1 - \Pr(\scrB_T^\SFc) - \Pr(\scrC_T^\SFc)\\
    \geq{}& 1- [6Kp+(10K+8)q+1]T^{-1}-2qT^{-3}.
\end{align*}
Therefore,
\begin{align*}
    \Pr(\mathscr{D}_T^\SFc )={}& \Pr(\mathscr{B}_T\cap \mathscr{C}_T \cap \mathscr{D}_T^\SFc) + \Pr((\mathscr{B}_T\cap \mathscr{C}_T)^\SFc \cap \mathscr{D}_T^\SFc)     \\
    \leq{} & (2q+2)T^{-3} + [6Kp+(10K+8)q+1]T^{-1}-2qT^{-3}.
\end{align*}
Hence,     $\Pr(\mathscr{D}_t )\rightarrow 1$ as $T\rightarrow \infty$.
Since $\eta$ can be arbitrarily small, it implies that $\widehat\sigma^2_T\rightarrow_p \sigma^2$.
Combining all the above, we have that
\begin{equation*}
    \widehat\sigma^2_T \widehat{\BFOmega}_T\rightarrow_p \sigma^2 \BFOmega^*.
\end{equation*}
By Theorem \ref{theorem:asymp}, we have that
$\sqrt{T-T_1}(\widehat{\BFalpha}_T-\BFalpha)\rightsquigarrow \mathcal{N}(0,\sigma^2 \BFOmega^*)$ as $T\rightarrow \infty$. Since $\widehat\sigma^2_T \widehat{\BFOmega}_T$ is a consistent estimator of $\sigma^2 \BFOmega^*$, the statements in Corollary \ref{coro:inf} hold.
\end{proof}

\section{Proof of Proposition \ref{prop:Non-Convergence}}

\begin{proof}[Proof of Proposition \ref{prop:Non-Convergence}]
WLOG., we can assume that $\BFv_t \geq 0$ for all $\BFv_t\in \mathcal{V}$. Otherwise we can always apply a constant shift to $\BFv_t$ to make such condition to hold.
    Given that $R_i(\BFv_t) = \BFalpha_i^\intercal \BFv_t + \epsilon_t$. WLOG., suppose $a_t=1$ for $t=T_1+1$, i.e., suppose arm 1 is pulled at time $T_1+1$. Consider the event that:
    \begin{equation*}
        \mathcal{A} := \{\epsilon_{T_1+1}<\bar{\epsilon}\}\cap \{\epsilon_{t}\in [-\bar{\epsilon}_1,\bar{\epsilon}_1],t=T_1+2,...,T_2\}\cap \{v_{T_1+1} > \bar{v}\},
    \end{equation*}
    with some $\bar{\epsilon}<0$, $\bar{\epsilon}_1>0$ and $\bar{v}>0$.

    It can be shown that for arbitrarily fixed $\bar{\epsilon}$ and $\bar{\epsilon}_1$, and fixed finite number of $T_2-T_1=C$, we have that $$\Pr(\mathcal{A})>2\underline{p}$$ for some fixed positive constant $\underline{p}>0$.

     For $\bar{\epsilon}$ negative enough and $\bar{\epsilon}_1$ close enough to $0$, we can conclude that $\hat\BFalpha_{1,t}<\BFalpha_{1} - \bar{\alpha}$ with $\bar{\alpha}>0$ being an arbitrarily large positive vector, while $|\hat\BFalpha_{2,t}-\BFalpha_2|< C \bar{\epsilon}_1$ is arbitrarily close to $0$.

    Therefore, if $\bar{\epsilon}$ is negative enough, then, we can have that for all $v\in \mathcal{V}$, where $\mathcal{V}$ is a bounded compact set, $\hat\BFalpha_1^\intercal v < \hat\BFalpha_2^\intercal v$ for $t=T_2+1$. Therefore, the arm 2 will be pulled for sure under event $\mathcal{A}$ at $t=T_2+1$.

    Then, we show that there exists a non-trivial probability such that arm 2 will be pulled at all time $t=T_2+2,...,T$. It suffices to show that $\hat{\BFalpha}_{2,t}\rightarrow {\BFalpha}_{2}$ uniformly as $t\rightarrow \infty$.

  By definition, we have that:
\begin{equation} %
\|\widehat{\BFalpha}_{t} - \BFalpha\| \leq \Bigl\|
\Bigl(\widehat{\BFSigma}_{vz,{\ttR:t}}\widehat{\BFSigma}_{zz,{\ttR:t}}^{-1} \widehat{\BFSigma}_{vz,{\ttR:t}}^\intercal\Bigr)^{-1}\widehat{\BFSigma}_{vz,\ttR:t}\widehat{\BFSigma}_{zz,{\ttR:t}}^{-1} \Bigr\| \cdot \bigl\| \widehat{\BFG}_{{\ttR:t}} \bigr\|.
\end{equation}
For some arbitrarily small positive constant $C_A>0$, define the events that:
\begin{gather*}
    \scrA_t\coloneqq \Biggl\{\|\widehat{\BFalpha}_t - \BFalpha \|\leq C_A \Biggr\}, \quad
    \scrB_t\coloneqq \bigl\{\|\widehat{\BFSigma}_{vz,\ttR:t}-\BFSigma_{vz}^*\|\leq \eta\bigr\}, \\
    \scrC_t\coloneqq \bigl\{\|\widehat{\BFSigma}_{zz,\ttR:t}-\BFSigma_{zz}\|\leq \eta\bigr\},  %
\end{gather*}

First, similar to Lemma \ref{lemma:conditional-A}, given $\mathcal{B}_t$ and $\mathcal{C}_t$ hold, we have that:
\begin{equation}
\begin{gathered}
\phi_{\min}(\widehat{\BFSigma}_{vz,\ttR:t})> \phi_{\min}(\BFSigma_{vz}^*)-\eta>0, \quad
\phi_{\max}(\widehat{\BFSigma}_{vz,\ttR:t}) < \phi_{\max}(\BFSigma_{vz}^*)+\eta, \\
\phi_{\min}(\widehat{\BFSigma}_{zz,\ttR:t}) >  \phi_{\min}(\BFSigma_{zz})-\eta>0, \quad
\phi_{\max}(\widehat{\BFSigma}_{zz,\ttR:t})<  \phi_{\max}(\BFSigma_{zz})+\eta.
\end{gathered}
\end{equation}
This implies that
\begin{align*}
    & \left\| \left(\widehat{\BFSigma}_{vz,\ttR:t}\widehat{\BFSigma}_{zz,\ttR:t}^{-1} \widehat{\BFSigma}_{vz,\ttR:t}^\intercal\right)^{-1}\widehat{\BFSigma}_{vz,\ttR:t}\widehat{\BFSigma}_{zz,\ttR:t}^{-1}\right\| \nonumber \\
    ={}& \phi_{\max}\left(\left(\widehat{\BFSigma}_{vz,\ttR:t}\widehat{\BFSigma}_{zz,\ttR:t}^{-1} \widehat{\BFSigma}_{vz,\ttR:t}^\intercal\right)^{-1}\widehat{\BFSigma}_{vz,\ttR:t}\widehat{\BFSigma}_{zz,\ttR:t}^{-1}\right) \nonumber \\
    \leq {}&  \frac{\phi_{\max}(\widehat{\BFSigma}_{vz,\ttR:t})\phi_{\max}(\widehat{\BFSigma}_{zz,\ttR:t})}{(\phi_{\min}(\widehat{\BFSigma}_{vz,\ttR:t}))^2 \phi_{\min}(\widehat{\BFSigma}_{zz,\ttR:t})}  \\
    \leq{}& \frac{(\phi_{\max}({\BFSigma}_{vz}^*)+\eta)(\phi_{\max}({\BFSigma}_{zz})+\eta)}{(\phi_{\min}({\BFSigma}_{vz}^*)-\eta)^2 (\phi_{\min}({\BFSigma}_{zz})-\eta)} \\
    \leq{}& \biggl(\frac{11}{10}\biggr)^2\biggl(\frac{10}{9}\biggr)^3 \frac{\phi_{\max}(\BFSigma_{vz}^*)\phi_{\max}(\BFSigma_{zz})}{(\phi_{\min}(\BFSigma_{vz}^*))^2 \phi_{\min}(\BFSigma_{zz})}  \\
    \leq{}&  \bar{\kappa}.
\end{align*}

Therefore,
\begin{equation}
\scrB_t\cap \scrC_t \subseteq
\Bigl\{\Bigl\| \left(\widehat{\BFSigma}_{vz,\ttR:t}\widehat{\BFSigma}_{zz,\ttR:t}^{-1} \widehat{\BFSigma}_{vz,\ttR:t}^\intercal\right)^{-1}\widehat{\BFSigma}_{vz,\ttR:t}\widehat{\BFSigma}_{zz,\ttR:t}^{-1}\Bigr\| \leq \bar{\kappa}\Bigr\}.
\end{equation}

By Corollary \ref{coro:Hoeffding_vector}, we have that for all $\tau\geq 0$ and $t>\ttU$,
\begin{equation*}
    \Pr(\bigl\|\widehat{\BFG}_{\ttU:t}\bigr\|\geq\tau)
    \leq 2q\exp(-\frac{\tau^2 (t-\ttU)}{2 q^2\bar{z}^2 \varsigma^2}).
\end{equation*}
Note that
$ \widehat{\BFG}_{\ttR:t} = \frac{t-\ttU}{t-\ttR}\widehat{\BFG}_{\ttU:t} + \frac{\ttU-\ttR}{t-\ttR}\widehat{\BFG}_{\ttR:\ttU}$, and by construction $\|\frac{\ttU-\ttR}{t-\ttR}\widehat{\BFG}_{\ttR:\ttU}\|\leq C\bar{\epsilon}_1$ can be arbitrarily small when $\epsilon_1$ is small enough, we have that:
\begin{equation*}
    \Pr(\{\|\widehat{G}_{T_1:t}\|\geq \tau\}\cap \mathcal{A})\leq 2q\underline{p}\exp\left(-\frac{(\tau - C\bar{\epsilon}_1)^2(t-\ttR)}{2q^2\bar{z}\varsigma^2}\right).
\end{equation*}
Let $ 2C\bar{\epsilon}_1 = \tau$, we have that:
\begin{equation*}
    \Pr(\{\|\widehat{G}_{T_1:t}\|\geq \tau\}\cap \mathcal{A})\leq 2q\underline{p}\exp\left(-\frac{\tau ^2(C+t-\ttU)}{8q^2\bar{z}\varsigma^2}\right).
\end{equation*}
Let $C_A \geq 4q \bar{z}\varepsilon C\bar{\epsilon}_1$, We have that
\begin{equation}\label{eq:component-1}
    \Pr(\mathscr{A}_t^c \cap \mathscr{B}_t\cap \mathscr{C}_t \cap \mathcal{A}) \leq 2q\bar{p}\exp(-(C+t-T_2)).
\end{equation}
Similar to Lemma \ref{lemma:C-bounds}, we have that
it follows from Lemma~\ref{lemma:Hoeffding_matrix} that for all $t \geq \ttU$,
\begin{align*}
\Pr(\scrC_t^{\SFc}) = \Pr(\|\widehat{\BFSigma}_{zz,\ttR:t}-\BFSigma_{zz}\| > \eta)
 \leq{}& 2q\exp(-\frac{\eta^2 (t-\ttR)}{2q^2\bar{z}^4})  \\
\leq {} & 2q\exp(-\frac{\eta^2 (C+t-\ttU)}{2q^2\bar{z}^4}).
\end{align*}
By Bonferroni inequality, we have that:
\begin{equation*}
    \Pr(\mathscr{C}_{T_2,t}) \geq 1-\sum_{s=\ttU+1}^t \Pr(\mathscr{C}_{s}^c) \geq 1-2q^3\bar{z}^4/\eta^2 \exp(-\frac{\eta^2 C}{2q^2\bar{z}^4}),
\end{equation*}
for all $t> \ttU$. Next, similar to Lemma \ref{lemma:D-bounds}, for all $t\geq \ttU$, we define
\begin{equation*}
    \widehat{\BFSigma}_{vz,\ttR:t}^*\coloneqq \frac{1}{t-\ttR}\sum_{s=\ttR+1}^t \BFv_t^* \BFz_t^\intercal \in\Real^{(Kp)\times q},
\end{equation*}
and analyze separately the two components of the following decomposition:
\begin{equation}
\|\widehat{\BFSigma}_{vz,\ttR:t}-\BFSigma_{vz}^*\| \leq
\underbrace{
\|\widehat{\BFSigma}_{vz,\ttR:t} - \widehat{\BFSigma}_{vz,\ttR:t}^* \|}_{I_1} +
\underbrace{
\| \widehat{\BFSigma}_{vz,\ttR:t}^* -\BFSigma_{vz}^*\|}_{I_2}.
\end{equation}

For $I_2$, note that $\mathbb{E}[\BFv_t^* \BFz_t^\intercal] = \BFSigma_{vz}^*$
and that, by Lemma~\ref{lemma:matrix-norm},
\[
\|\BFv_t^* \BFz_t^\intercal - \BFSigma_{vz}^*\|
\leq \|\BFv_t^* \BFz_t^\intercal\| + \|\BFSigma_{vz}^*\|
\leq  2\sqrt{K pq}\bar{v}\bar{z}.
\]
It then follows from  Lemma~\ref{lemma:Hoeffding_matrix} that for all $t\geq \ttU$,
\begin{align}
    \Pr(\| \widehat{\BFSigma}_{vz,\ttR:t}^* - \BFSigma_{vz}^*\| > \frac{\eta}{2})
    \leq{}& 2(Kp+q)\exp(-\frac{(t-\ttR)\eta^2/4}{2(2\sqrt{Kpq}\bar{v}\bar{z})^2})  \nonumber \\
    \leq {}& 2(Kp+q) \exp(-\frac{\eta^2 (C+t-\ttU)}{32K pq\bar{v}^2\bar{z}^2}) \nonumber.
\end{align}
And
\begin{align}
    \Pr(\| \widehat{\BFSigma}_{vz,\ttR:t}^* - \BFSigma_{vz}^*\| > \frac{\eta}{2} \cap \mathcal{A}\cap \mathscr{A}_{\ttU:t-1})
    \leq{}& 2\underline{p}\exp\left(-\frac{\eta^2 (t-\ttR)}{32 pq\var{v}^2\bar{z}^2 (K-1)^2} \right) \nonumber\\
    \leq{} & 2\underline{p}\exp\left(-\frac{\eta^2 (C+t-\ttU)}{32 pq\var{v}^2\bar{z}^2 (K-1)^2} \right).
\end{align}
Therefore, we have that:
\begin{equation*}
    \Pr(\mathscr{B}_t^c \cap \mathcal{A}\cap \mathscr{A}_{\ttU:t-1}) \leq 2\underline{p}(\exp\left(-\frac{\eta^2 (C+t-\ttU)}{32 pq\var{v}^2\bar{z}^2 (K-1)^2} \right) + (Kp+q) \exp(-\frac{\eta^2 (C+t-\ttU)}{32K pq\bar{v}^2\bar{z}^2}) ).
\end{equation*}

Similar to the proof of Lemma \ref{lemma:iter-bounds}, we can show that by induction, we have:
\begin{equation}
    \Pr(A\cap A_{\ttU:T}^c)\leq  \bar{C}T^{-1}
\end{equation}
for some fixed constant $\bar{C}>0$, and hence, $\Pr(A\cap A_{\ttU:T})\geq \Pr(A) - \Pr(A\cap A_{\ttU:T}^c) \geq 2\underline{p} - \bar{C} T^{-1}\geq \underline{p}$ for $T$ large enough. It implies that there exists probability of $\underline{p}>0$ such that arm 2 is pulled forever in Phase 3, and the reward of arm 1 is always underestimated, leading to linear regret of the algorithm. Therefore, without Phase 2, the algorithm could be non-converging with a non-trivial positive probability $\geq \underline{p}$.
\end{proof}

\end{appendix}

\end{document}